\documentclass[aps,prb,preprint,onecolumn,citeautoscript,superscriptaddress]{revtex4-1} 
\usepackage{float}
\usepackage{amsmath,amssymb,mathrsfs,bm,feynmf,setspace}
\usepackage{graphicx}
\usepackage[tight]{subfigure}  
\usepackage{color} 
\usepackage[papersize={8.5in,11in}]{geometry}
\usepackage[colorlinks=true]{hyperref} 
\hypersetup{
    bookmarks=true,         
    unicode=false,          
    pdftoolbar=true,        
    pdfmenubar=true,        
    pdffitwindow=false,     
    pdfstartview={FitH},    
    pdftitle={My title},    
    pdfauthor={Author},     
    pdfsubject={Subject},   
    pdfcreator={Creator},   
    pdfproducer={Producer}, 
    pdfkeywords={keyword1} {key2} {key3}, 
    pdfnewwindow=true,      
    colorlinks=true,       
    linkcolor=red,          
    citecolor=blue,        
    filecolor=magenta,      
    urlcolor=cyan           
} 
\usepackage{bm}
\newcommand{\beq}{\begin{equation}}
\newcommand{\eeq}{\end{equation}}
\newcommand{\be}{\begin{equation}}
\newcommand{\ee}{\end{equation}}
\newcommand{\ba}{\begin{array}{ccc}}
\newcommand{\ea}{\end{array}}
\newcommand{\nn}{\nonumber \\}

\newcommand{\bk}{{\bm k}}

\def\bea{\begin{eqnarray}}
\def\eea{\end{eqnarray}}
\newcommand{\tr}{\textrm{tr}}

\begin{document}

\title{Quantum quenches and competing orders
}
 \author{Wenbo Fu}
 \affiliation{Department of Physics, Harvard University, Cambridge, MA 02138, USA}
 \author{Ling-Yan Hung}
 \affiliation{Department of Physics, Harvard University, Cambridge, MA 02138, USA}
 \author{Subir Sachdev}
 \affiliation{Department of Physics, Harvard University, Cambridge, MA 02138, USA}
 \affiliation{Perimeter Institute for Theoretical Physics,
Waterloo, Ontario N2L 2Y5, Canada}
 \date{\today\\
 \vspace{1.6in}}
\begin{abstract}
We study the non-equlibrium dynamics of an electronic model of competition between an unconventional
charge density wave (a bond density wave) 
and $d$-wave superconductivity. In a time-dependent Hartree-Fock+BCS approximation, the dynamics reduces to the equations of
motion of operators realizing the generators of SU(4) at each pair of momenta, $(\bk, -\bk)$, in the Brillouin zone. 
We also study the non-equilibrium dynamics of a quantum generalization of a O(6)
non-linear sigma model of competing orders in the underdoped cuprates
(Hayward {\em et al.}, Science {\bf 343}, 1336 (2014)). We obtain results, in the large $N$ limit
of a O($N$) model, on the time-dependence of correlation
functions following a pulse disturbance.
We compare our numerical studies 
with recent picosecond optical experiments.
We find that, generically, the oscillatory responses in our models share
various qualitative features with the experiments. 

\end{abstract}
\maketitle

\section{Introduction}
\label{sec:intro}

A remarkable series of recent optical experiments \cite{Orenstein,Gedik,andrea1,andrea2} have explored time-dependent
non-equilibrium physics in the cuprate superconductors at the picosecond time scale. 
Our work is specifically motivated by the observations of Ref.~\onlinecite{Orenstein}: these experiments observed terahertz 
oscillations in the reflectivity
of underdoped of YBCO in a time-domain, pump-probe experiment. The onset temperature of the reflectivity oscillations was the same
as the onset temperature of charge ordering in the recent X-ray measurements,\cite{keimer,chang,hawthorn} and so the oscillations
were interpreted \cite{Orenstein} as an oscillation in the amplitude of the charge order. The reflectivity oscillations also showed
a remarkable $\pi$ phase shift, and a temperature-dependent frequency, across the superconducting critical temperature $T_c$. The
authors interpreted these phenomena in a classical phenomenological model of competition between superconductivity
and charge order. 

Our purpose here is to develop a more microscopic and quantum model of these oscillations. We will do this
by examining two distinct models. 

The first is a simple `hot spot' electronic model for the competition between unconventional charge
density wave (CDW) order (a bond density wave) and superconductivity (SC) which was proposed in Ref.~\onlinecite{jay}.
The CDW ordering wavevectors of this first model are $(\pm Q_0, \pm Q_0)$, where $Q_0$ is determined by the positions
of the hot spots.
 We will extend
the equilibrium results to time-dependent phenomena using a time-dependent 
Hartree-Fock-BCS theory similar to that used in Ref.~\onlinecite{levitov} for the quench dynamics of BCS superconductors.
This model has the advantage of dealing directly with the underlying fermionic degrees of freedom. 
However, our analysis has the disadvantage that the spatial correlations of the order parameter are treated in a mean-field
manner. Our results for this model appear in Sections~\ref{sec:hotspot}--\ref{sec:pulse}.

The second model, described in Sections~\ref{sec:intro2}--\ref{sec:pulseII},
has a more complete treatment of spatial fluctuations of the CDW and SC orders, but works instead with an effective
model for these bosonic order parameters alone.
Also, the CDW wavevectors can now be either along the cardinal directions or the diagonals: so they can also take the
experimentally observed values of $(\pm Q_0, 0)$, $(0, \pm Q_0)$.
This  model for the competing order parameters has an energy functional which is drawn directly from 
recent work by Hayward {\em et al.}\cite{o6} They argued for a non-linear sigma model for a 6-component order parameter:
two of the components, $\vec{\Psi}$, represented $d$-wave SC, while the remaining four, $\vec{\Phi}$, represented
the complex order parameters for CDWs along the $x$ and $y$ directions; we will implicitly assume that $\vec{\Psi}$ ($\vec{\Phi}$)
is a 2 (4) component real vector. The thermal fluctuations in Ref.~\onlinecite{o6} were restricted to be on the space constrained
by $\vec{\Psi}^2 + \vec{\Phi}^2 = 1$, and we will also impose this constraint.
However, we need to extend the model of Ref.~\onlinecite{o6} to include a kinetic energy term to describe
the dynamic questions of interest here. In Sections~\ref{sec:hotspot}--\ref{sec:pulse}, the dynamics 
is derived from the equations of motion
of the underlying electrons, and so for the non-linear sigma model 
the analogous procedure is to integrate out the fermionic degrees in a path-integral
formulation of the Hamiltonian. While integrating out fermions is a delicate matter in a metal due to the presence of Fermi surface, we argue that for our purposes the consequences are simple. The key point is the observation that {\em both\/} the $\vec{\Psi}$
and $\vec{\Phi}$ gap out the {\em same\/} important portion of the Fermi surface in the anti-nodal region (near the ``hot spots''). As our study is restricted to the manifold $\vec{\Psi}^2 + \vec{\Phi}^2 = 1$, we can always
assume that the antinodal Fermi surface is gapped. Consequently, integrating out the fermions only induces analytic time derivative
terms in the effective action for $\vec{\Psi}$ and $\vec{\Phi}$, and we will only keep terms containing upto 2 time derivatives.
We will ignore the small damping that can be induced by the gapless fermions in the nodal regions because both the CDW and SC orders
do not couple strongly to these fermions.

We will conclude the paper in Section~\ref{sec:conc} 
with a summary of our results, and a comparison of the distinct methodologies employed in the paper.  

\section{Hot spot model}
\label{sec:hotspot}

\begin{figure}[h]
\centering
\includegraphics[width=2.7in]{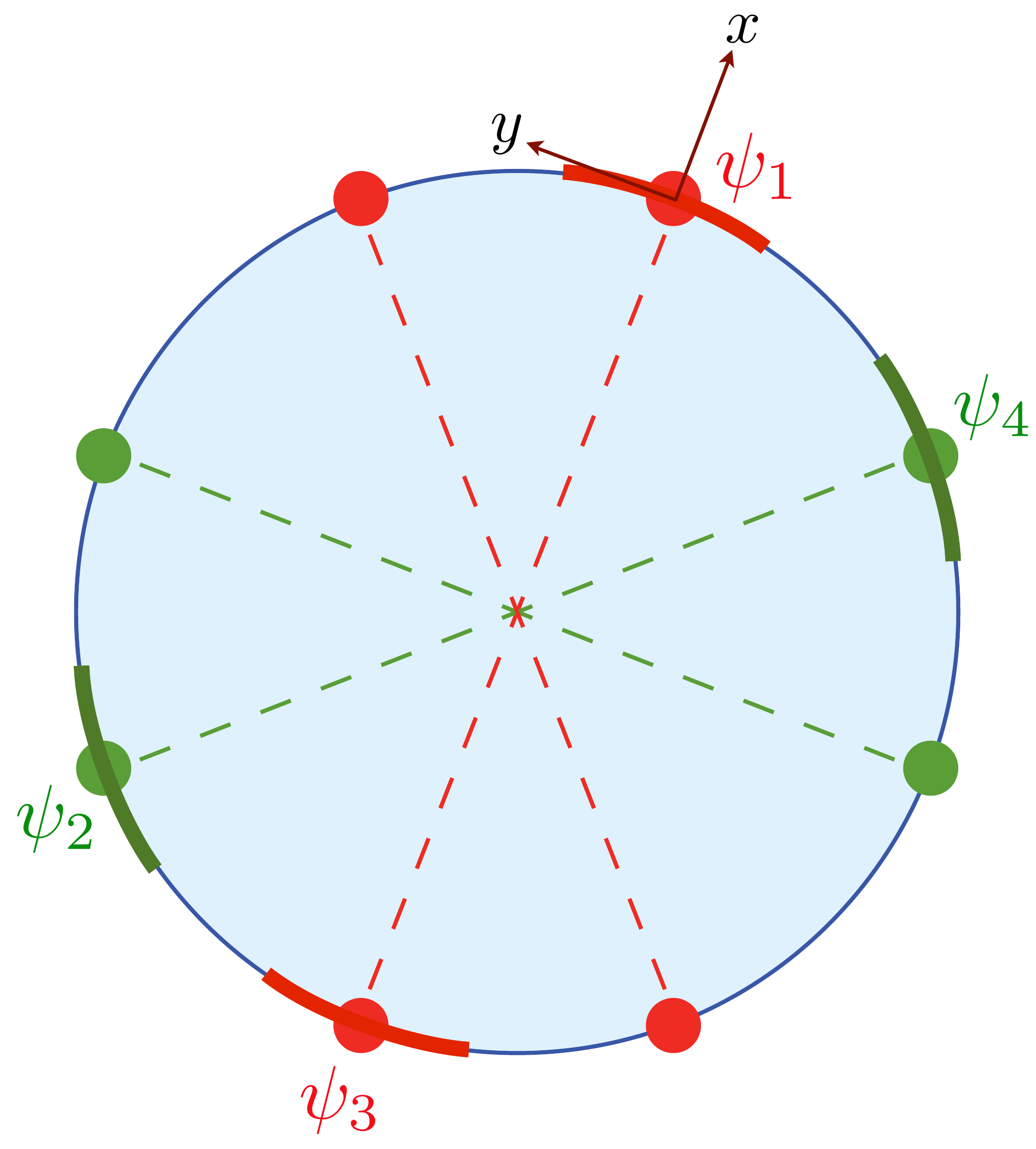}
\caption{Definitions of the $\psi_{1,2,3,4}$ fermions around the Fermi surface. Each fermion resides around a curved patch
of the Fermi surface shown by the thick lines. 
The red (green) hot spots are where the superconducting and bond density wave orders are positive (negative).}
\label{fig:hotspots}
\end{figure}

We begin by reviewing the equilibrium properties of the simple ``hot spot'' model of competing orders presented
in Ref.~\onlinecite{jay}. The model is defined in terms of 4 species of fermions
$\psi_{a \alpha}$, $a = 1 \ldots 4$, $\alpha = \uparrow, \downarrow$ located near ``hotspots'' on the Fermi surface
as shown in Fig.~\ref{fig:hotspots}. 
Their kinetic energy is given by
\bea
&& H_0 =  \sum_{\bk} \Biggl[ \epsilon_1 (\bk)  \, \psi_{1 \alpha}^\dagger (\bk)  \psi_{1 \alpha}^{\vphantom \dagger} (\bk)
+ \epsilon_2 (\bk)\, \psi_{2 \alpha}^\dagger (\bk)  \psi_{2 \alpha}^{\vphantom \dagger} (\bk) +\nn
&& \epsilon_1 (- \bk) \, \psi_{3 \alpha}^\dagger (\bk)  \psi_{3 \alpha}^{\vphantom \dagger} (\bk)
+ \epsilon_2 (- \bk) \, \psi_{4 \alpha}^\dagger (\bk)  \psi_{4 \alpha}^{\vphantom \dagger} (\bk) \Biggr].
\eea
We take the origin of momentum space at the hot spots, and orient the $x$-axis orthogonal to the Fermi surface
for the $\psi_{1,3}$ fermions; so we can write
\beq
\epsilon_1 (\bk) = k_x + \gamma k_y^2 . \label{eps1}
\eeq
We have taken the Fermi velocity to be unity, while $\gamma$ measures the curvature of the Fermi surface. 
The dispersion $\epsilon_2 (\bk)$ has the form obtained by rotating $\epsilon_1 (\bk)$ so that the direction
orthogonal to the Fermi surfaces of the $\psi_{2,4}$ has a linear dispersion.
After rescaling momenta appropriately, we can choose the convenient momentum space 
cutoffs $- \pi < k_x, k_y < \pi $, and the value $\gamma = 1/\pi$.  

Next, we add interactions between these fermions. 
The microscopic exchange ($J$) interactions and Coulomb repulsion ($V$) when projected onto the hot spots lead to
\bea
H_1 &=& \int d^2 x \Biggl[
 - J \left(  \psi_{1 \alpha}^\dagger \vec{\sigma}_{\alpha\beta} \psi_{2 \beta}^{\vphantom \dagger} 
+  \psi_{2 \alpha}^\dagger \vec{\sigma}_{\alpha\beta} \psi_{1 \beta}^{\vphantom \dagger} \right)  \nn
&~& \quad\quad\quad\quad \cdot
\left(  \psi_{3 \gamma}^\dagger \vec{\sigma}_{\gamma\delta} \psi_{4 \delta}^{\vphantom \dagger} 
 + \psi_{4 \gamma}^\dagger \vec{\sigma}_{\gamma\delta} \psi_{3 \delta}^{\vphantom \dagger}  \right) \\
&~&  - V \left(  \psi_{1 \alpha}^\dagger \psi_{2 \alpha}^{\vphantom \dagger} 
+  \psi_{2 \alpha}^\dagger  \psi_{1 \alpha}^{\vphantom \dagger} \right)  
\left(  \psi_{3 \beta}^\dagger \psi_{4 \beta}^{\vphantom \dagger} 
 + \psi_{4 \beta}^\dagger  \psi_{3 \beta}^{\vphantom \dagger}  \right) \Biggr] \nonumber
\eea
The full Hamiltonian $H_0+H_1$ has an exact SU(2)$\times$SU(2) pseudospin rotation symmetry \cite{metlitski10-2} when $\gamma=0$ and $V=0$.

Next, we review the Hartree-Fock-BCS theory of the hotspot model $H_0 + H_1$. 
The superconducting (SC) order parameter, $\Delta$, involves pairing of particles on antipodal points on the Fermi surface,
while the charge density wave (CDW) order, $\Pi$, involves pairing of particles with holes on the antipodal point \cite{jay};
consequently, the CDW ordering wavevector has the values $(\pm Q_0, \pm Q_0)$, as is clear from Fig.~\ref{fig:hotspots}.
\bea
\Delta_1 (\bk) &=& \left\langle \varepsilon_{\alpha\beta} \psi_{1\alpha}^\dagger (\bk) \psi_{3\beta}^{\dagger} (-\bk) \right\rangle 
\quad;\quad \Delta_1 \equiv \sum_\bk \Delta_1 (\bk) \nn
\Delta_2 (\bk) &=& \left\langle \varepsilon_{\alpha\beta} \psi_{2\alpha}^\dagger (\bk) \psi_{4\beta}^{\dagger} (-\bk) \right\rangle 
\quad;\quad \Delta_2 \equiv \sum_\bk \Delta_2 (\bk)\nn
 \Pi_1 (\bk) &=& \left\langle  \psi_{1\alpha}^\dagger (\bk) \psi_{3\alpha}^{\vphantom \dagger} (\bk) \right\rangle
 \quad;\quad \Pi_1 \equiv \sum_\bk \Pi_1 (\bk)\nn
 \Pi_2 (\bk) &=& \left\langle  \psi_{2\alpha}^\dagger (\bk) \psi_{4\alpha}^{\vphantom \dagger} (\bk) \right\rangle
\quad;\quad \Pi_2 \equiv \sum_\bk \Pi_2 (\bk) \label{orders}
\eea
It was found\cite{jay} that optimal state has a $d$-wave signature for both the superconducting and charge orders, with 
$\Delta_1 = - \Delta_2$ and $\Pi_1 = - \Pi_2$. 
For the charge order, this $d$-wave structure implies that the charge modulation is primarily on the {\em bonds\/} of the underlying 
lattice.\cite{rolando}
With the above orders, 
the mean field Hamiltonian is
\bea
&& H_{MF} = H_0 +\frac{(3J-V)}{2} \left( - \Delta_1 \, \varepsilon_{\alpha\beta} \psi_{2\alpha}^{\vphantom \dagger} (\bk) \psi_{4\beta}^{\vphantom \dagger} (-\bk) \right. \nn
&& + \Delta_2^\ast \, \varepsilon_{\alpha\beta} \psi_{1\alpha}^{ \dagger} (\bk) \psi_{3\beta}^{ \dagger} (-\bk) 
 - \Delta_2 \, \varepsilon_{\alpha\beta} \psi_{1\alpha}^{\vphantom \dagger} (\bk) \psi_{3\beta}^{\vphantom \dagger} (-\bk) \nn
&& \left. + \Delta_1^\ast \, \varepsilon_{\alpha\beta} \psi_{2\alpha}^{ \dagger} (\bk) \psi_{4\beta}^{ \dagger} (-\bk) 
 \right) \nn
 && + \frac{(3J+V)}{2} \left( \Pi_1 \, \psi_{4 \alpha}^\dagger (\bk) \psi_{2 \alpha}^{\vphantom \dagger} (\bk) 
 + \Pi_2^\ast \, \psi_{1 \alpha}^\dagger (\bk) \psi_{3 \alpha}^{\vphantom \dagger} (\bk) \right. \nn
&& \left. + \Pi_2 \, \psi_{3 \alpha}^\dagger (\bk) \psi_{1 \alpha}^{\vphantom \dagger} (\bk) 
 + \Pi_1^\ast \, \psi_{2 \alpha}^\dagger (\bk) \psi_{4 \alpha}^{\vphantom \dagger} (\bk) 
 \right) . \label{hmf}
\eea

\begin{figure}
\includegraphics[width=4in]{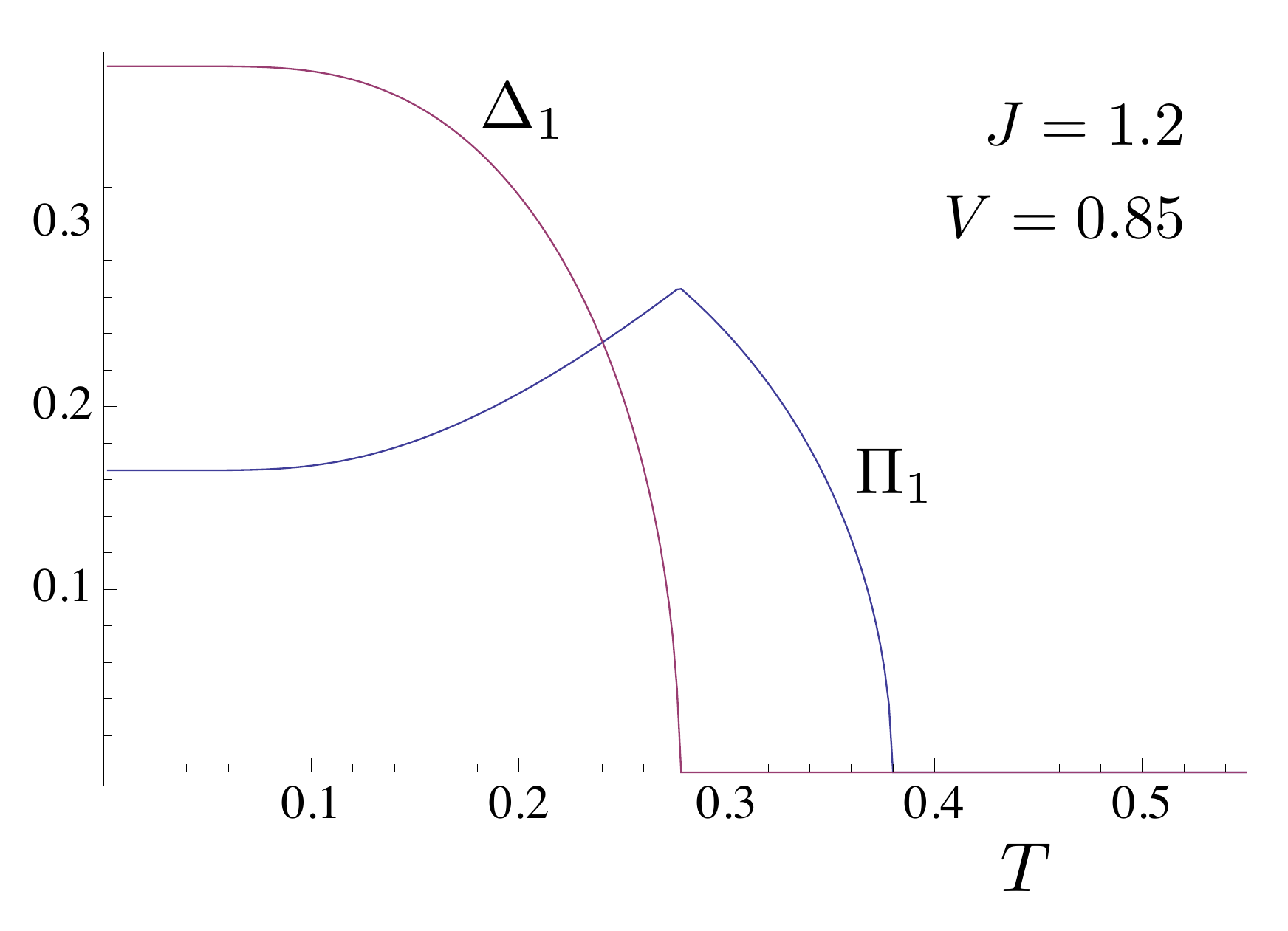}
\caption{Superconducting ($\Delta_1$) and bond ($\Pi_1$) orders in the hot spot model as a function of $T$.}
\label{fig:phase}
\end{figure}

Ref.~\onlinecite{jay} presented the solution of the equilibrium properties of the Hartree-Fock-BCS equations for a variety of values
of $J$ and $V$. Here, we reproduce in Fig.~\ref{fig:phase} 
the solution at one set of parameter values to illustrate the basic temperature dependence 
of the mean-field order parameters. 
Note that the CDW order, $\Pi_1$ has an onset at a higher $T$. However, at the superconducting $T_c$, it starts `competing'
for the Fermi surface with the SC order $\Delta_1$, and so decreases with decreasing $T$.

\section{Equations of motion}
\label{sec:motion}

We will follow the same general strategy as in Ref.~\onlinecite{levitov}: we will work with Heisenberg equations of motion from the
Hamiltonian $H_{MF}$, where the mean field order parameters $\Delta_1$ and $\Pi_1$ take their instantaneous average values.

An important feature of this method for the present model is that that commutators of the operators $\Delta_1 (\bk)$ and 
$\Pi_1 (\bk) $ with $H_{MF}$ do not close among themselves: they produce additional operators whose equations of motion we have
to also consider. By repeatedly evaluating commutators of the operators so generated, we find that we also have to consider the operators
\beq
N_i (\bk) = \psi_{i \alpha}^\dagger (\bk)  \psi_{i \alpha}^{\vphantom \dagger} (\bk) \quad, \quad P_i (\bk) = \varepsilon_{\alpha\beta} \psi_{i\alpha}^{\dagger} (\bk) \psi_{i\beta}^{\dagger} (-\bk)
\eeq
where $i=1 \ldots 4$; note 
\beq
P_i (-\bk) = P_i (\bk) \quad, \quad N_i^\dagger (\bk) = N_i (\bk). 
\eeq
Among all the operators introduced so far, the operator
\beq
N_1 (\bk) + N_3 (\bk) - N_1 (-\bk) - N_3 (-\bk)
\eeq
commutes with all other operators. The remaining 15 operators
\bea
&& N_1 (\bk) + N_3 (-\bk), N_1 (\bk) + N_1 (-\bk) -1,N_3 (\bk) + N_3 (-\bk) -1,\nn
&& \Delta_1 (\bk), \Delta_1 (-\bk), \Delta_1^\dagger (\bk), \Delta_1^\dagger (-\bk),\nn 
&& \Pi_1 (\bk), \Pi_1 (-\bk), \Pi_1^\dagger (\bk), \Pi_1^\dagger (-\bk),\nn
&&  P_1 (\bk), P_1^\dagger (\bk), P_3 (\bk), P_3^\dagger (\bk) \label{su4}
\eea
form the Lie algebra of SU(4). This is to be compared with the SU(2) algebra of Ref.~\onlinecite{levitov} of the operators
$P_1 (\bk)$, $P_1^\dagger (\bk)$, $N_1 (\bk) + N_1 (-\bk) - 1$.

It is now a straightforward, but tedious, exercise to evaluate the commutators of this SU(4) algebra, and so generate the equations
of motion associated with $H_{MF}$. We display the explicit form of these equations of motion in Appendix~\ref{app:motion}.

\section{Quench}

\begin{figure}
\begin{center}
\includegraphics[width=3in]{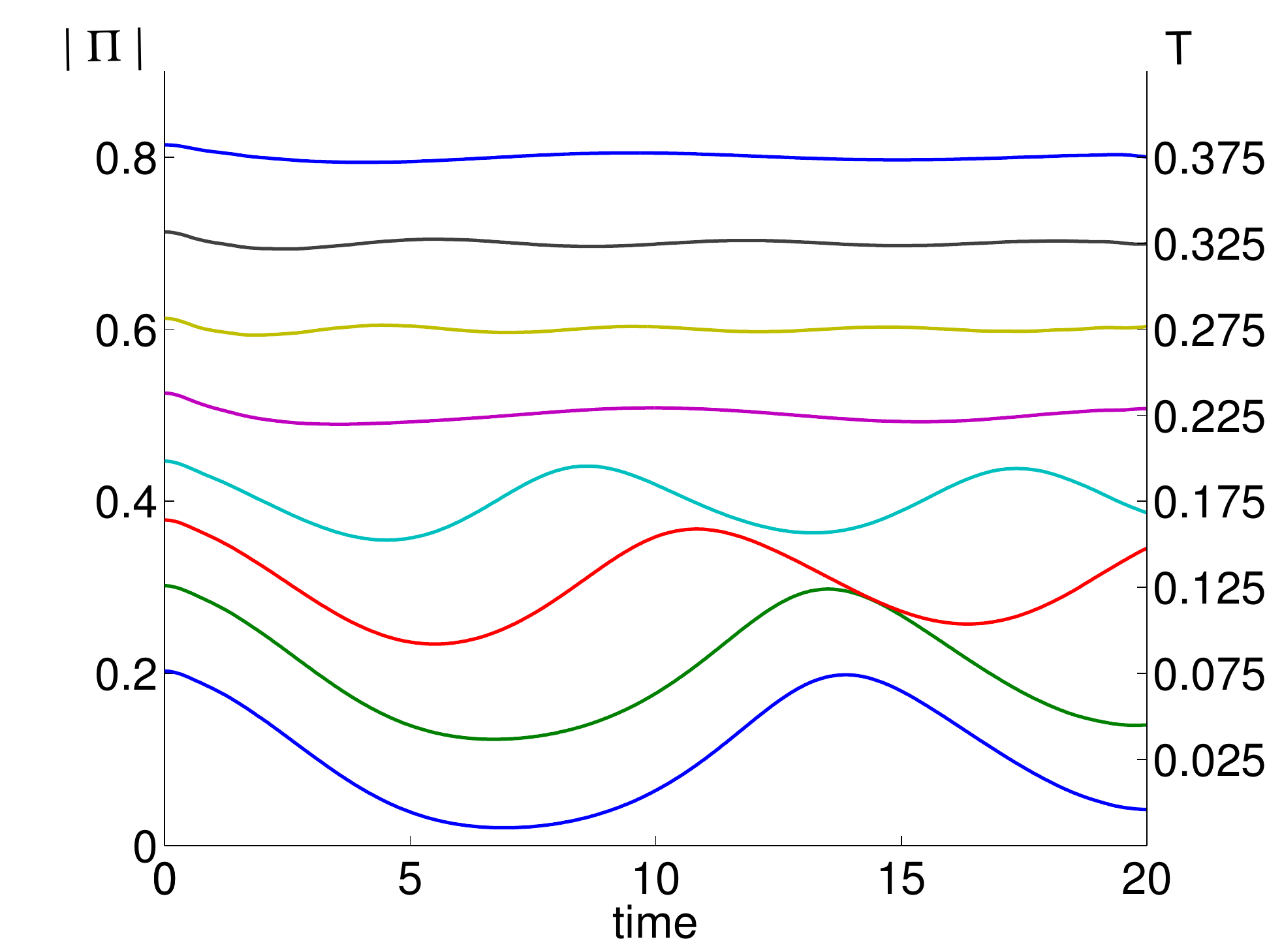}
\includegraphics[width=3in]{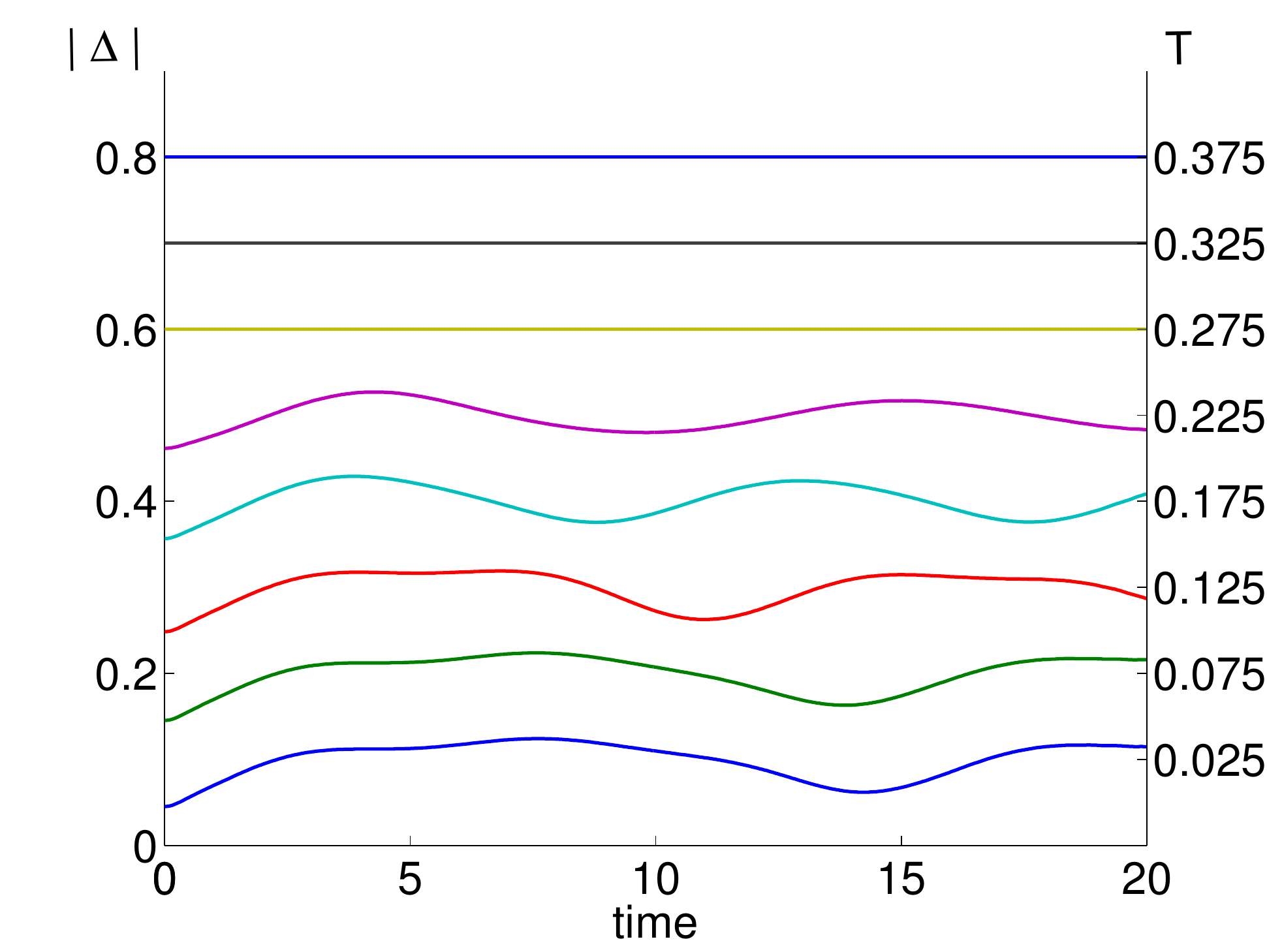}
\caption{Oscillation of CDW order parameter $\Pi$(left) and SC order parameter $\Delta$(right) as a function of time in the quench case, from low temperature at the bottom to high temperature at the top, temperatures are taken from 0.025 to 0.375 with 0.05 step. Note that here we are plotting the absolute value of the order parameters. Also, we have added constants to the curves to make them evenly spaced.
The initial value  $J_0=1.2$, $V_0=0.9$, the quench is taken as $\Delta J=0, \Delta V=-0.1$.  The initial $T_c$ at equilibrium can be computed to be 0.25, the final $T_c$ to be 0.33.}
\label{qv-01}
\end{center}
\end{figure}

First let us consider the quench case. By quench, we mean the coupling changes abruptly, {\em i.e.\/} 
\beq
V(t)=V_0+\Delta V \, \theta(t) \quad , \quad  J(t)=J_0+\Delta J \, \theta(t)
\eeq
where $\theta (t)$ is the step function, and $\Delta V$ and $\Delta J$ are the sizes of the steps.
Similar problems have been considered in the BCS system.\cite{levitov} 
We take the system to be at equilibrium at the beginning with both CDW and SC order,
at a fixed temperature which can be both below and above superconducting critical temperature $T_c$ in Fig.~\ref{fig:phase}. The evolutions of order parameters can be obtained using Heisenberg equations of motions. We obtained oscillations of the 
CDW order parameter $\Pi$, and the SC order parameter $\Delta$ as a function of time at different temperatures, as shown in Fig.~\ref{qv-01} for the parameters $J_0=1.2,V_0=0.9$, $\Delta J=0,\Delta V=-0.1$. We find that at high temperature which is larger than $T_c$, $\Delta$ stays zero, while the oscillation of $\Pi$ is suppressed.
 
\begin{figure}
\begin{center}
\includegraphics[width=4.5in]{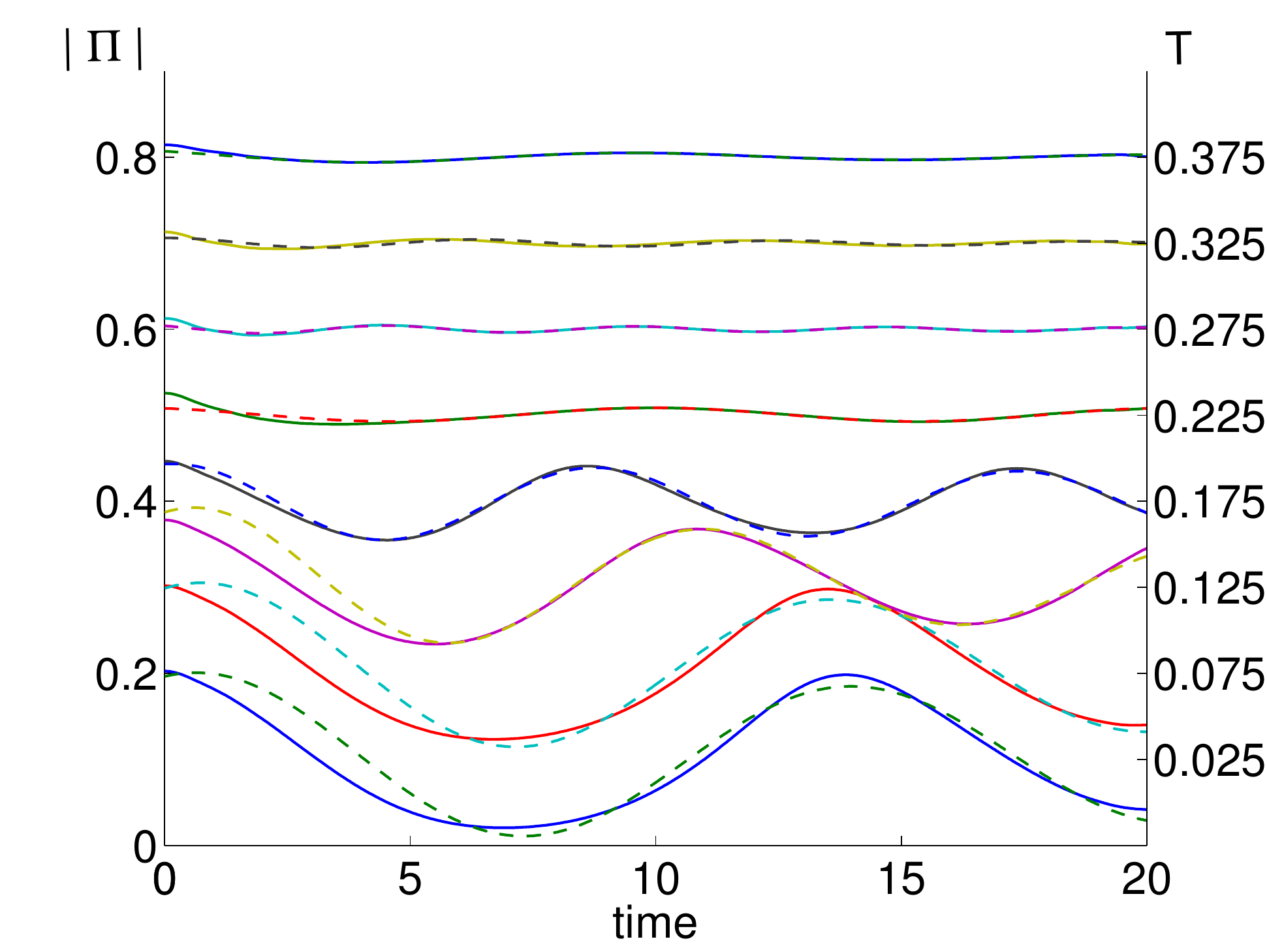}
\caption{Fitting of CDW order parameters $\Pi$ in the left panel of Fig.~\ref{qv-01}, the dashed lines are fitting lines using Eq.~\ref{fit}. We used the data after $time=5$ and later fittings also obey this rule. }
\label{qv-01fit}
\end{center}
\end{figure}

As in Ref.~\onlinecite{Orenstein}, we use a decaying sinusoidal function to fit the data. 
Unlike the experimental data, the amplitude of the oscillation does not decay to zero at long times in the present mean-field model. 
Later we will analyze the fourier spectrum of the oscillation, but for now we proceed with a naive fitting to the following function 
\beq
f(t)=a e^{-b (t-t_0)}\sin(c (t-t_0)+2\pi d)+e
\label{fit}
\eeq
which works quite well over the time window studied.
The fit is shown in Fig.~\ref{qv-01fit}. We have set an onset $t_0$ in the fitting function, the fitting phase $d$ will depend on the choice of $t_0$, since the frequency $c$ does not stay constant over the temperature range. In all of the fits we choose $t_0=5.5$, which will give nearly $\pi$ phase shift in the later pulse case. The fitting phase $d$ here just helps us to better see the phase shift rather than simply estimating by eye.
 
\begin{figure}
\begin{center}
\includegraphics[width=2in]{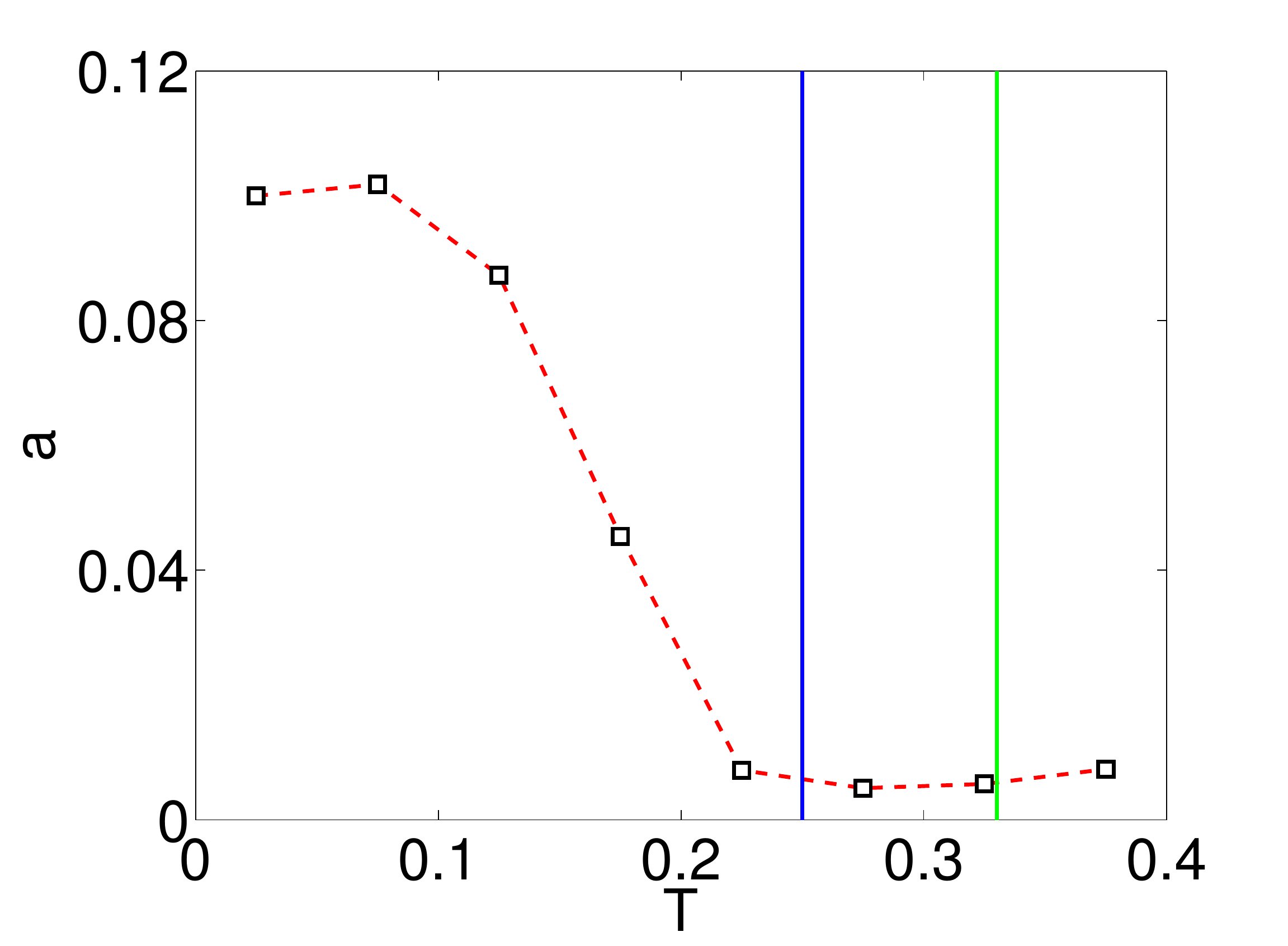}
\includegraphics[width=2in]{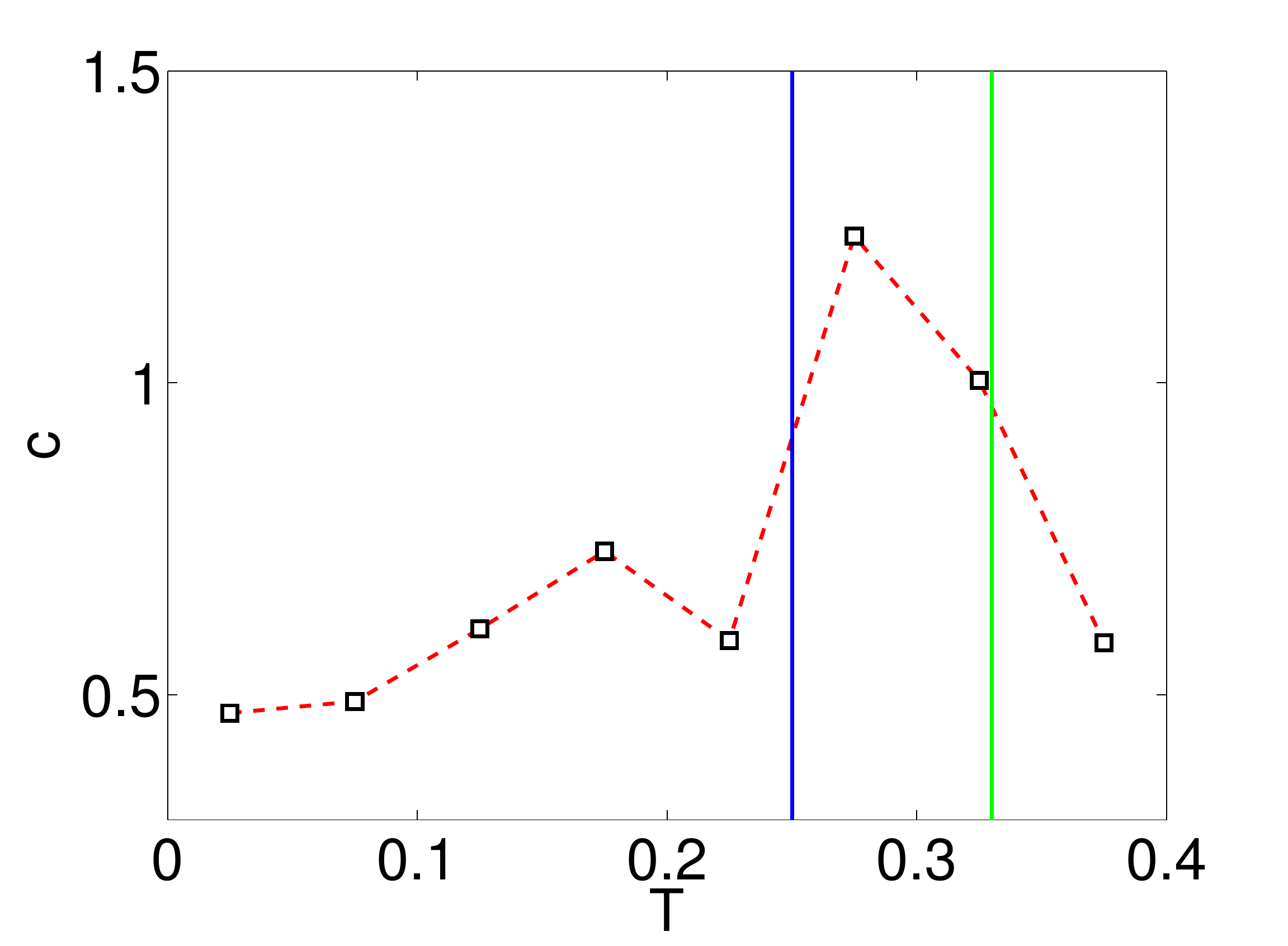}
\includegraphics[width=2in]{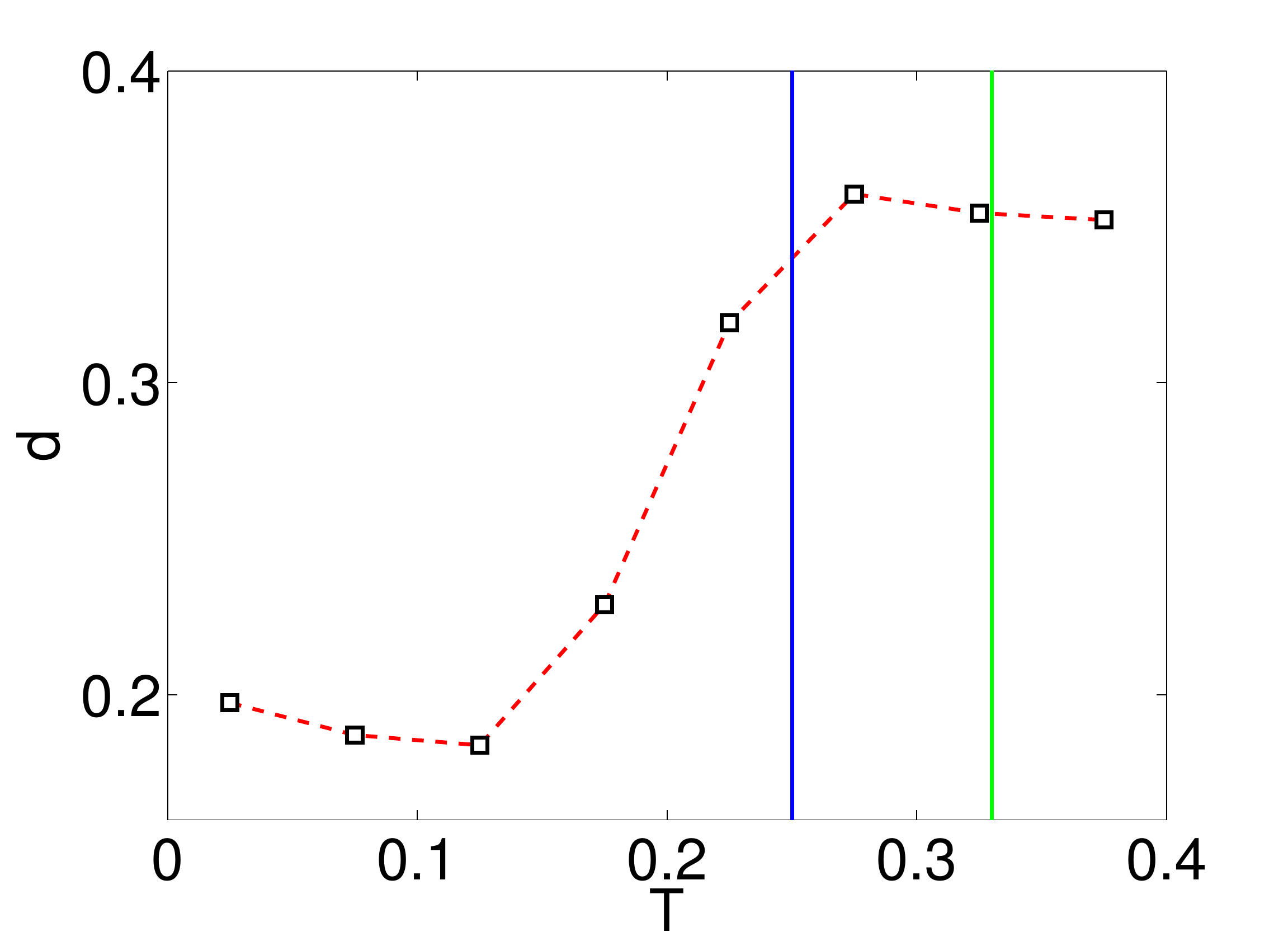}
\caption{Left to right: amplitude a, frequency c and phase d of the fit Eq.~\ref{fit} fitting the data in Fig.~\ref{qv-01fit} as a function of temperature. The blue line denotes initial equilibrium $T_c=0.25$, the green line denotes after quench, the equilibrium $T_c=0.33$.}
\label{qv-01fitpara}
\end{center}
\end{figure}

In Fig.~\ref{qv-01fitpara}, we show the variation of the amplitude $a$, frequency $c$, phase $d$ in Eq.~\ref{fit} as a function of temperature. The most important feature is that the amplitude $a$ is enhanced below the initial $T_c$ at equilibrium, which is also a key feature in the $O(6)$ field theory description that will be discussed in later sections. This resembles the oscillatory behavior in the experiment Ref.~\onlinecite{Orenstein}. Also the frequency varies against temperature, and there is a phase shift in the oscillations upon crossing $T_c$. Here the phase shift is smaller than $\pi$ by our choice of $t_0=5.5$.

\begin{figure}
\begin{center}
\includegraphics[width=3in]{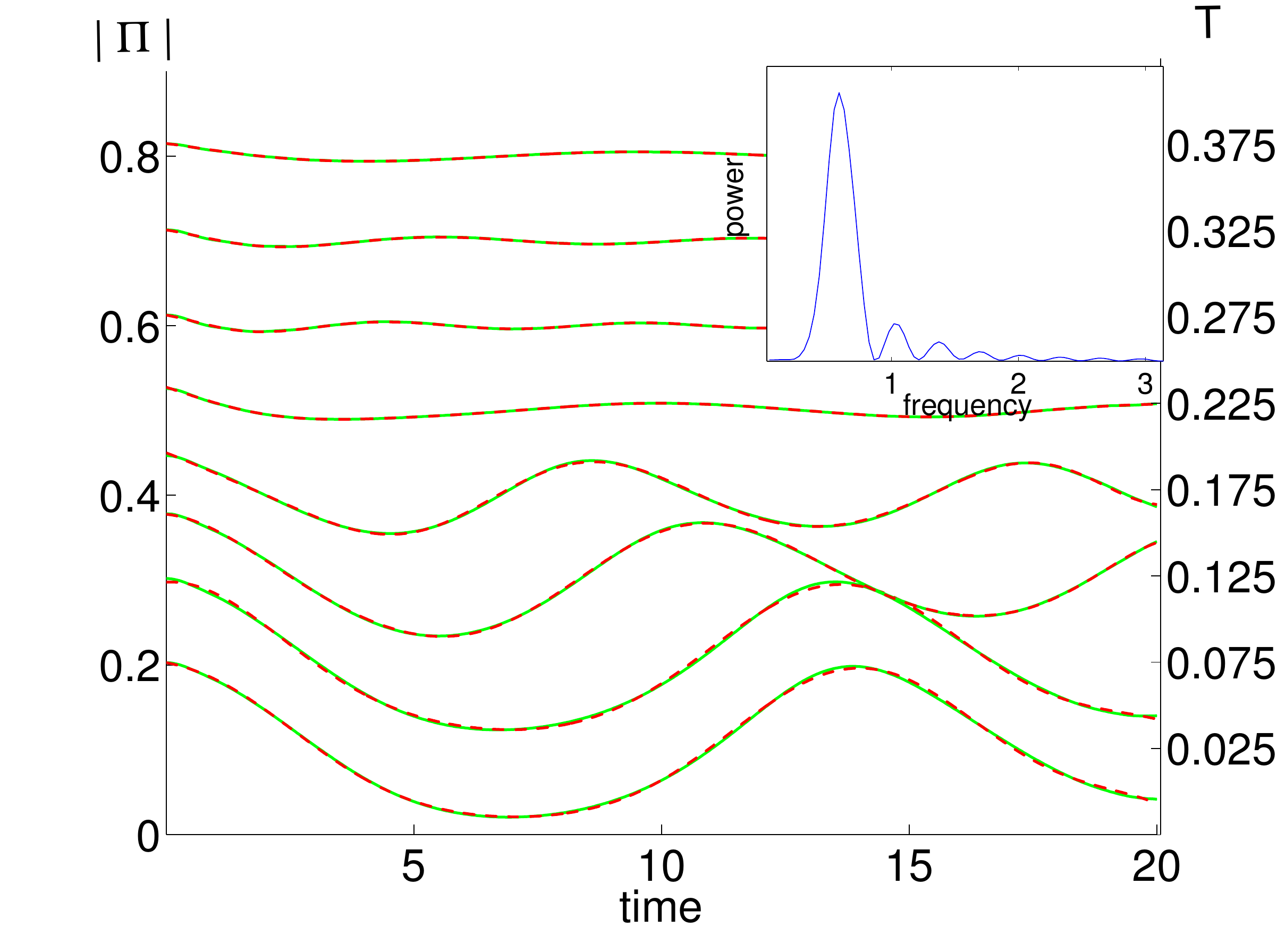}
\includegraphics[width=3in]{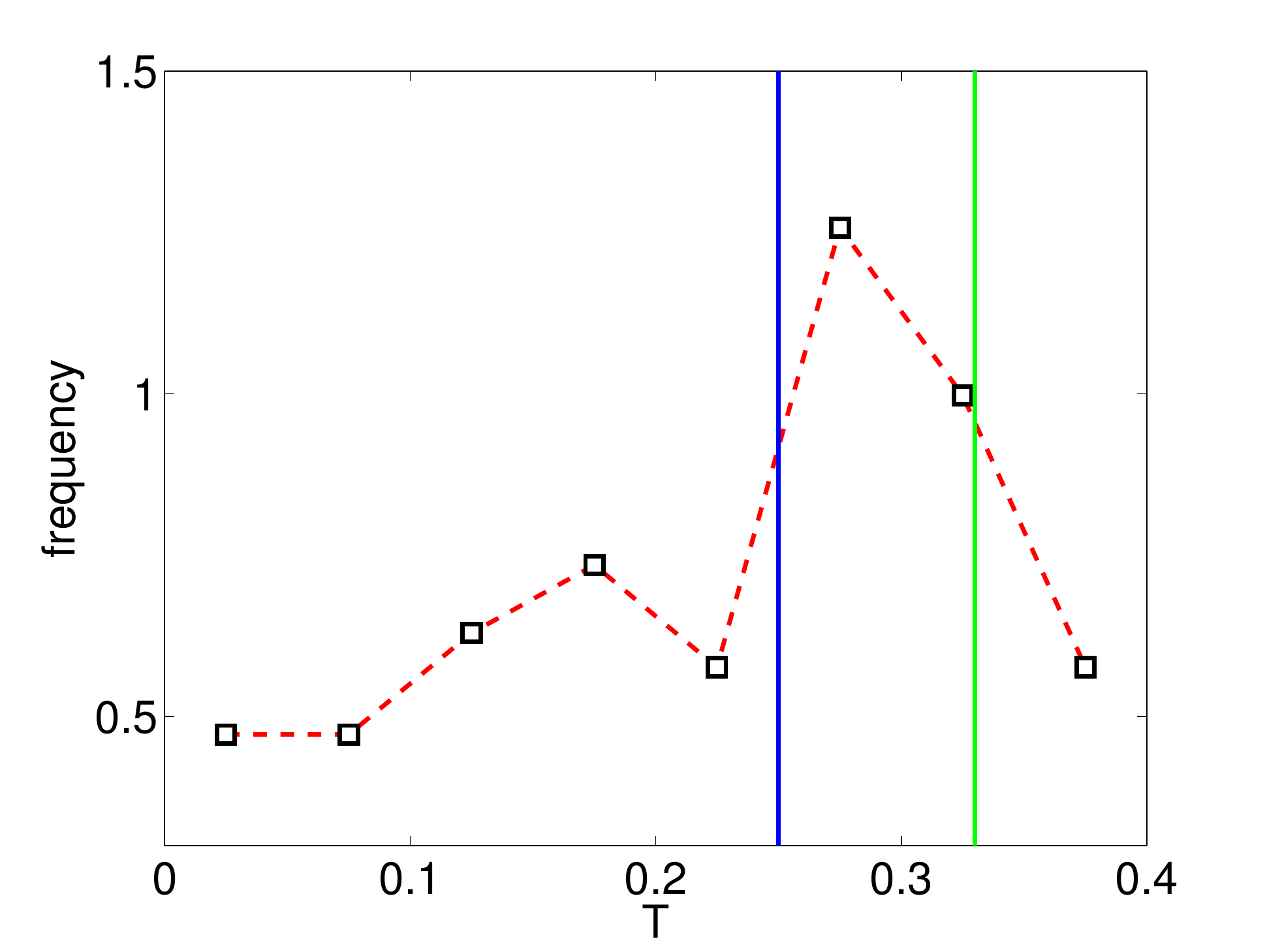}
\caption{Left: spectral fitting of the left panel in Fig.~\ref{qv-01}. The inserted figure is the power spectrum at $T=0.225$ near $T_c$. Right: peak frequencies from the left panel as a function of temperature. }
\label{qv-01fitspec}
\end{center}
\end{figure}

We can further use spectral analysis to fit the simulation data. Using the Lomb-Scargle algorithm, we can find frequencies' spectral power as shown, for example, in the inserted figure in Fig.~\ref{qv-01fitspec}. We choose frequencies which have  spectral power larger than ten percent of the largest power to fit the oscillation. Including frequencies from main peak also large side peaks, the fit result is quite good. However the peak frequencies are quite similar as the fitting number $c$ that we found using the simple decaying sinusoidal function as shown in the middle panel of Fig.~\ref{qv-01fitpara}.

To understand the fitting frequency behavior above $T_c$, we have $\Delta=0$, and the mean-field Hamiltonian Eq.(\ref{hmf}) reduces to
\bea
&& H_{MF} = H_0 + \frac{(3J+V)}{2} \left( \Pi_1 \, \psi_{4 \alpha}^\dagger (\bk) \psi_{2 \alpha}^{\vphantom \dagger} (\bk) 
 + \Pi_2^\ast \, \psi_{1 \alpha}^\dagger (\bk) \psi_{3 \alpha}^{\vphantom \dagger} (\bk) \right. \nn
&& \left. + \Pi_2 \, \psi_{3 \alpha}^\dagger (\bk) \psi_{1 \alpha}^{\vphantom \dagger} (\bk) 
 + \Pi_1^\ast \, \psi_{2 \alpha}^\dagger (\bk) \psi_{4 \alpha}^{\vphantom \dagger} (\bk) 
 \right) . \label{hmf2}
\eea
From the commutation relations in Eq.~(\ref{commutation}), we can make the identification so that these operators satisfy the $SU(2)$ algebra:
\beq
\Pi_1\rightarrow S_+, \Pi_1^{\dagger}\rightarrow S_-, \frac{N_1-N_3}{2}\rightarrow S_z
\eeq
Furthermore if we ignore the curvature of the Fermi surface, then $\epsilon_1(\bk)=-\epsilon_1(-\bk)$, the above Hamiltonian becomes
\beq
H_{MF}=\sum_{\bk}\Biggl[2\epsilon_1(\bk)S_z(\bk)+\frac{3J+V}{2}\Biggl(-\left\langle S_{-}\right\rangle S_{+}(\bk)-\left\langle S_{+}\right\rangle S_{-}(\bk)\Biggr)\Biggr]
\label{hmf3}
\eeq
here we have used $\Pi_2=-\Pi_1$ and only considered $1,3$ hotspot field (the $2,4$ channel would be similar). And this resembles the well-known pseudospin formulation of the BCS system, as studied in Ref.~\onlinecite{levitov}. For a small deviation, the frequency would be proportional to the order parameter $\Pi$. This explains the fact that the oscillation frequency decrease rapidly above $T_c$.

We have also computed the positive quench case in Fig.~\ref{qv01} of Appendix~\ref{app:simulation}, where $J_0=1.2,V_0=0.9$, $\Delta J=0,\Delta V=0.1$. We also get amplitude enhancement below $T_c$, although the effect is not that big.

\section{Pulse}
\label{sec:pulse}

Since in the experiment\cite{Orenstein,Gedik,andrea1,andrea2}, the disturbance is a short-time optical pulse, it should be more reasonable to consider a 
pulse in our time-dependent Hamiltonian, {\em i.e.}
\beq
J(t)=J_0+\Delta J \left( 1 - \tanh^2 (\omega t) \right) \quad , \quad  V(t)=V_0+\Delta V \left( 1 - \tanh^2(\omega t) \right)
\label{pulsefunction}
\eeq

\begin{figure}
\begin{center}
\includegraphics[width=3in]{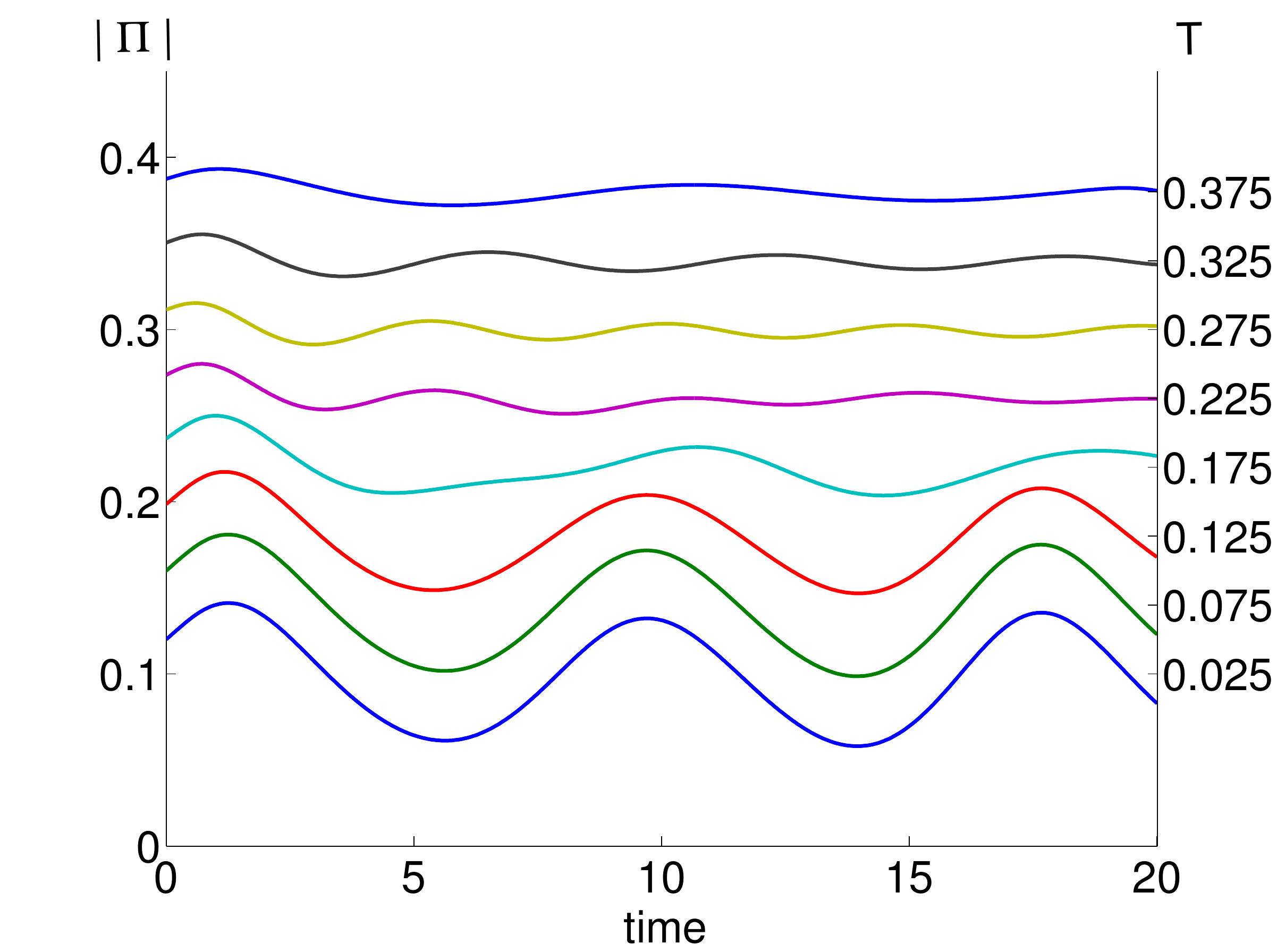}
\includegraphics[width=3in]{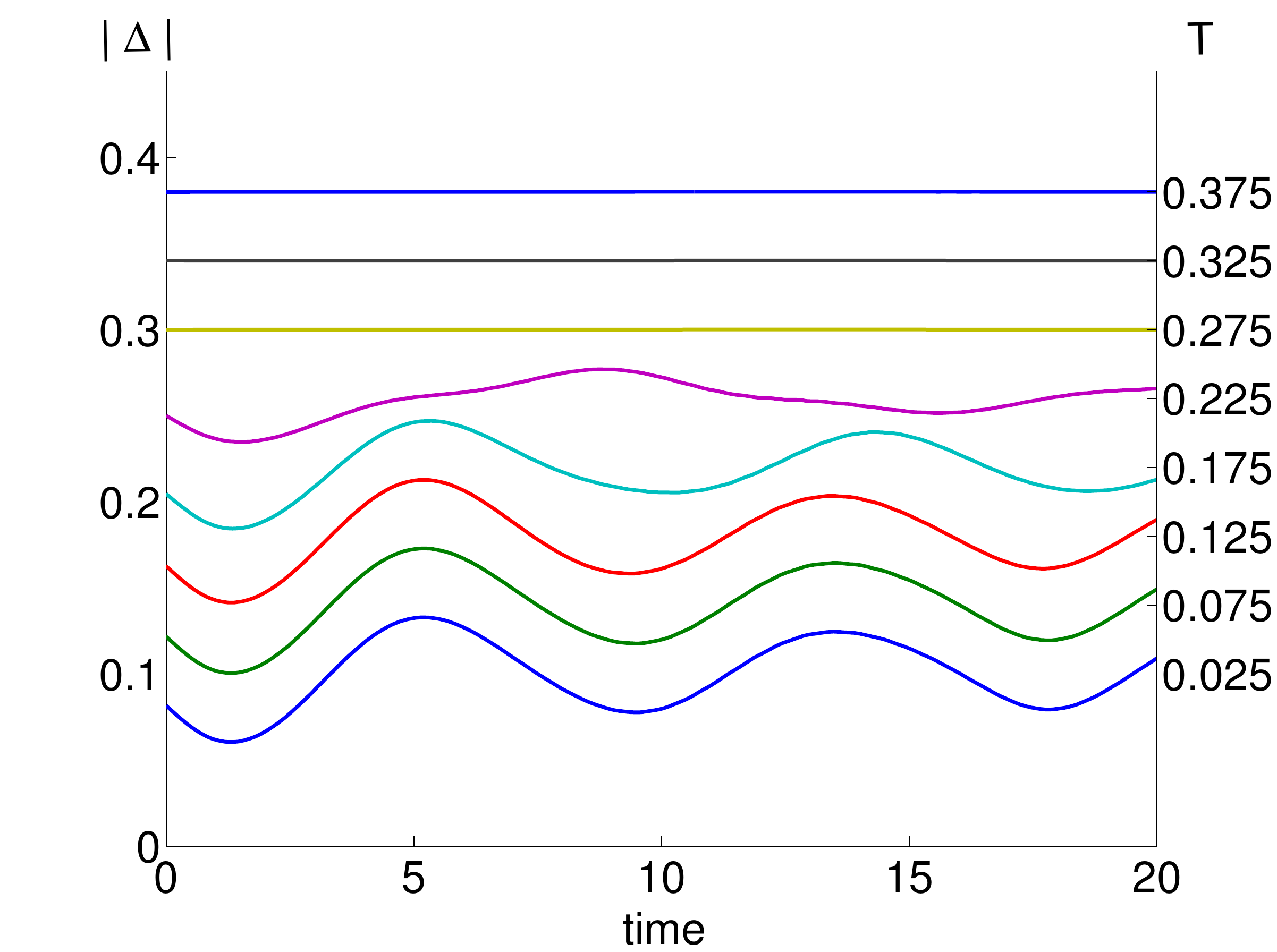}
\caption{Oscillation of CDW order parameter $\Pi$(left) and SC order parameter $\Delta$(right) as a function of time in the pulse case, from low temperature at the bottom to high temperature at the top, temperatures are taken from 0.025 to 0.375 with 0.05 step. The initial value  $J_0=1.2$, $V_0=0.9$, the pulse is taken as $\Delta J=0, \Delta V=0.1$, $\omega=1$. The initial $T_c$ at equilibrium can be computed to be 0.25, at the largest derivation $V=V_0+\Delta V$, the corresponding equilibrium $T_c$ to be 0.2.}
\label{pv01}
\end{center}
\end{figure}

\begin{figure}
\begin{center}
\includegraphics[width=4.5in]{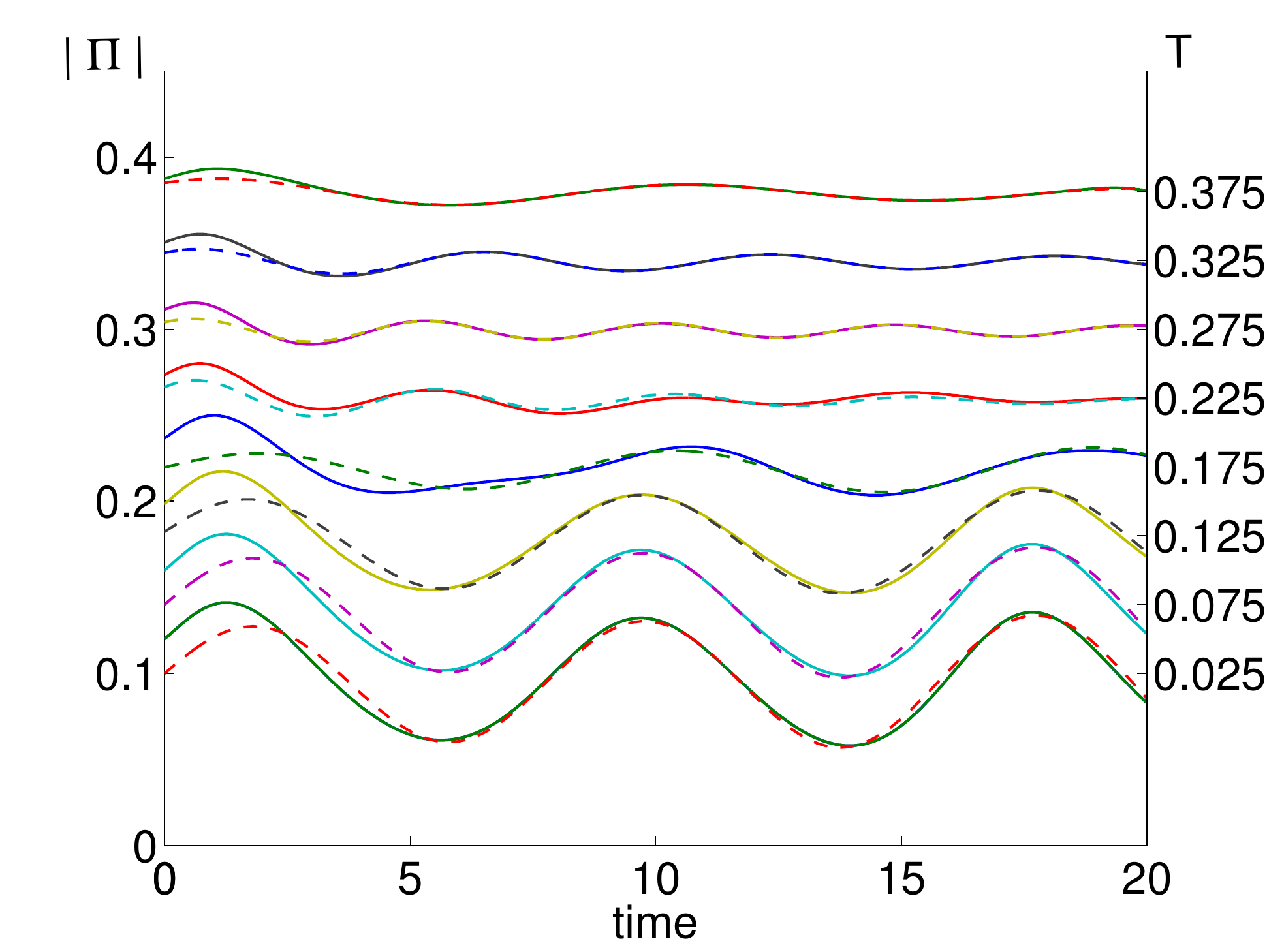}
\caption{Fitting of CDW order parameters $\Pi$ in the left panel of Fig.~\ref{pv01}, the dashed lines are fitting lines using Eq.~\ref{fit}.}
\label{pv01fit}
\end{center}
\end{figure}

\begin{figure}
\begin{center}
\includegraphics[width=2in]{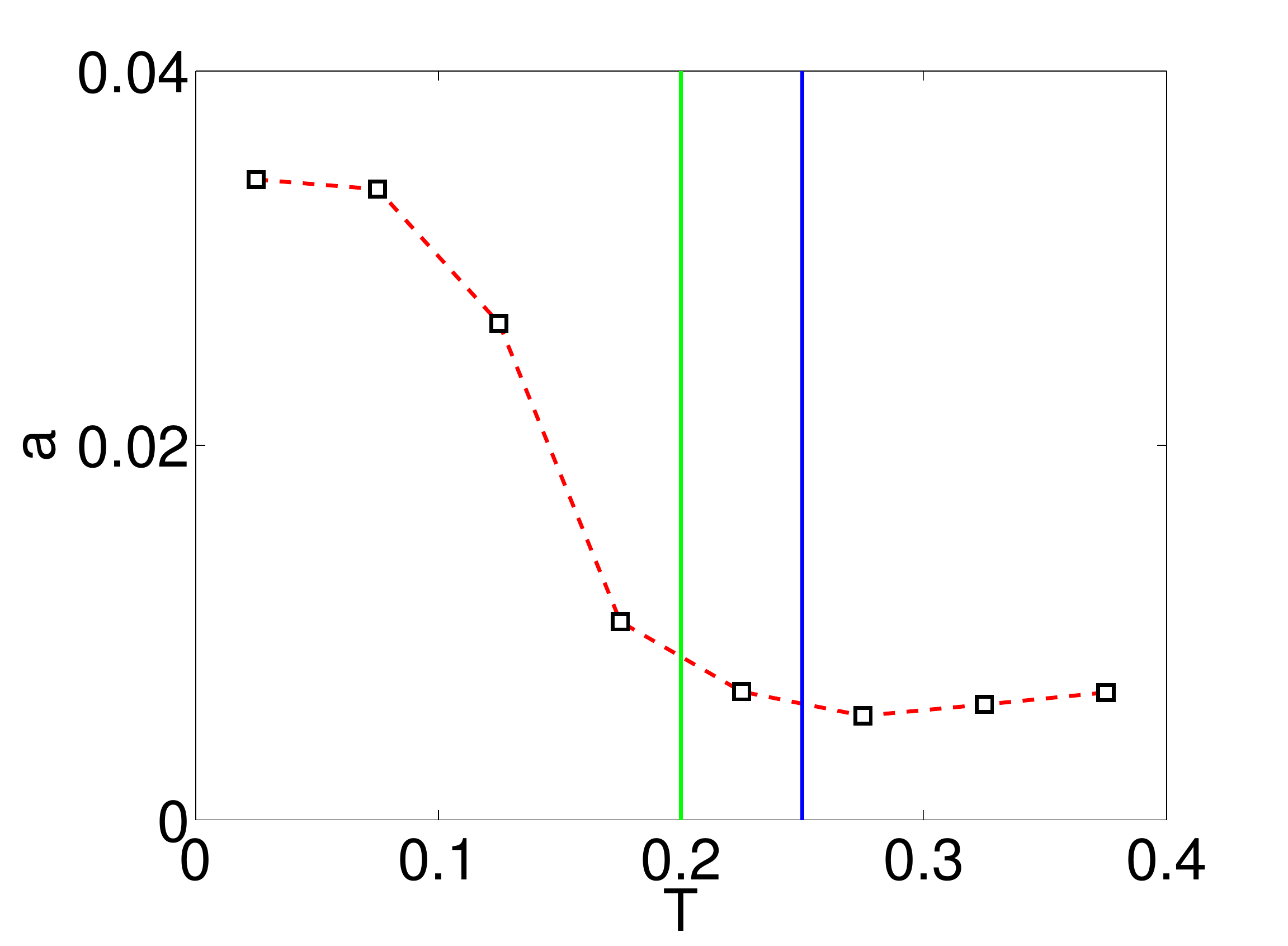}
\includegraphics[width=2in]{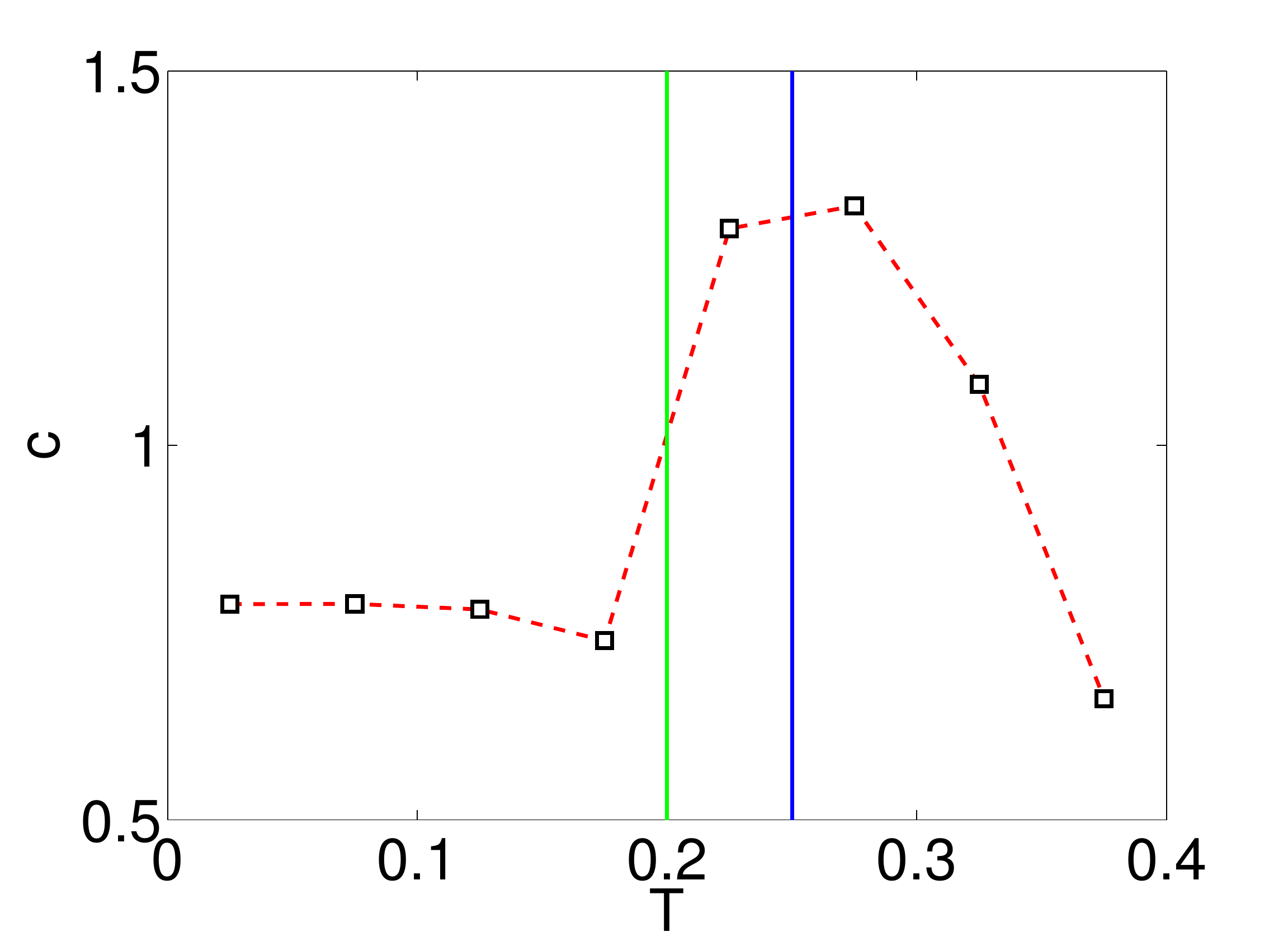}
\includegraphics[width=2in]{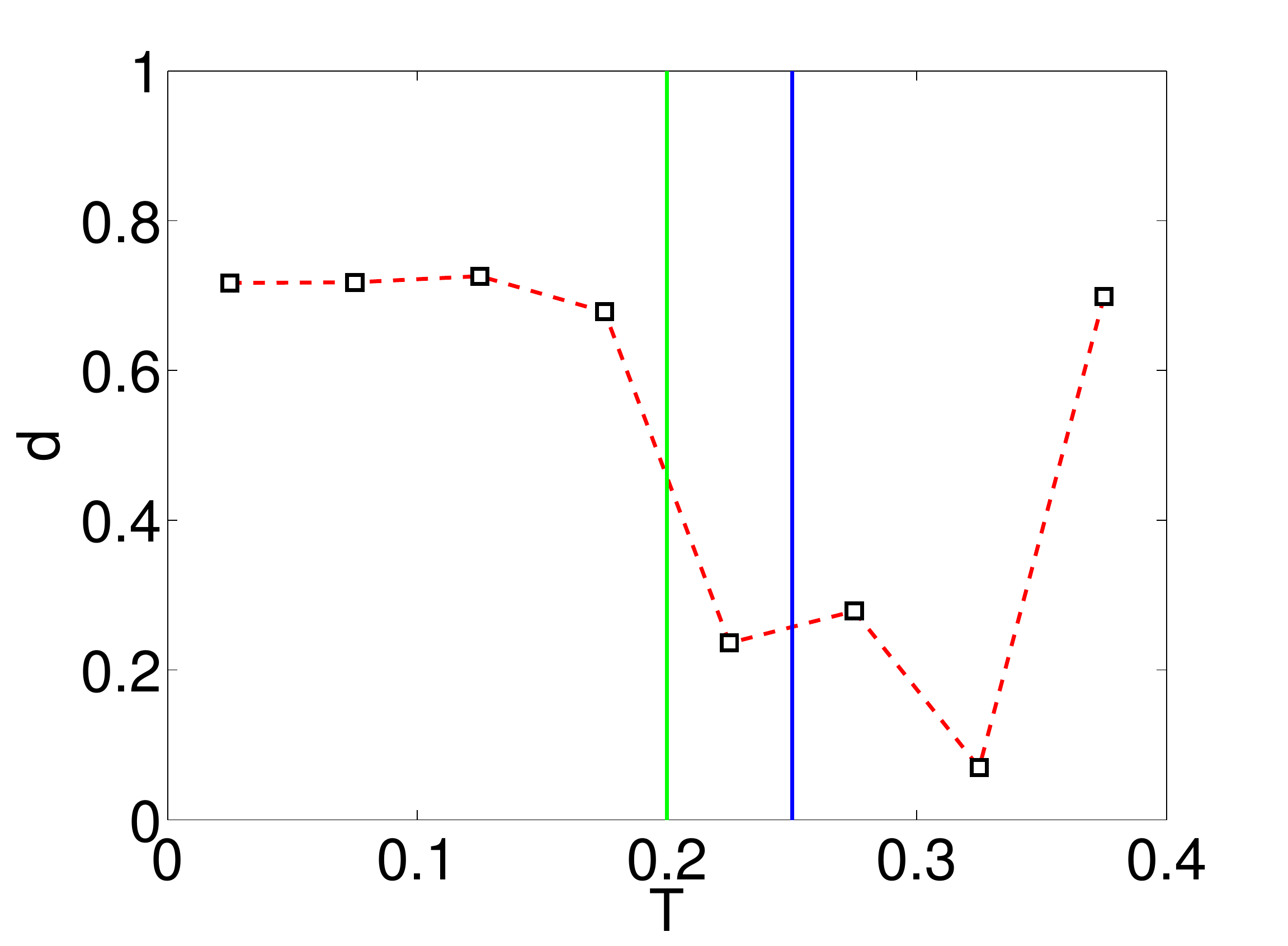}
\caption{Left to right: amplitude a, frequency c and phase d of the fit Eq.~\ref{fit} fitting the data in Fig.~\ref{pv01fit} as a function of temperature. The blue line denotes initial equilibrium $T_c=0.25$, the green line denotes at the largest derivation $V=V_0+\Delta V$, the equilibrium $T_c=0.2$.}
\label{pv01fitpara2}
\end{center}
\end{figure}

\begin{figure}
\begin{center}
\includegraphics[width=3in]{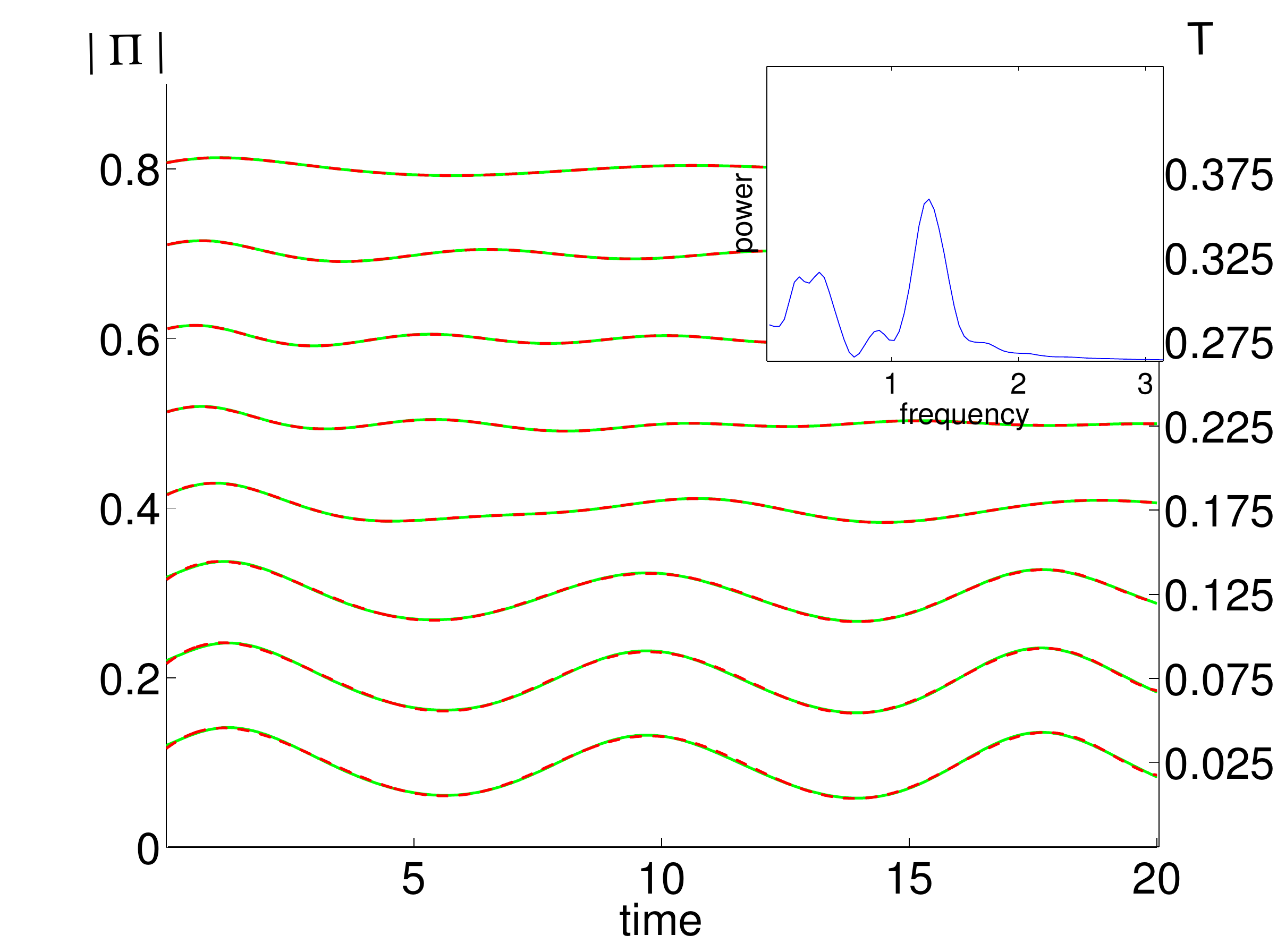}
\includegraphics[width=3in]{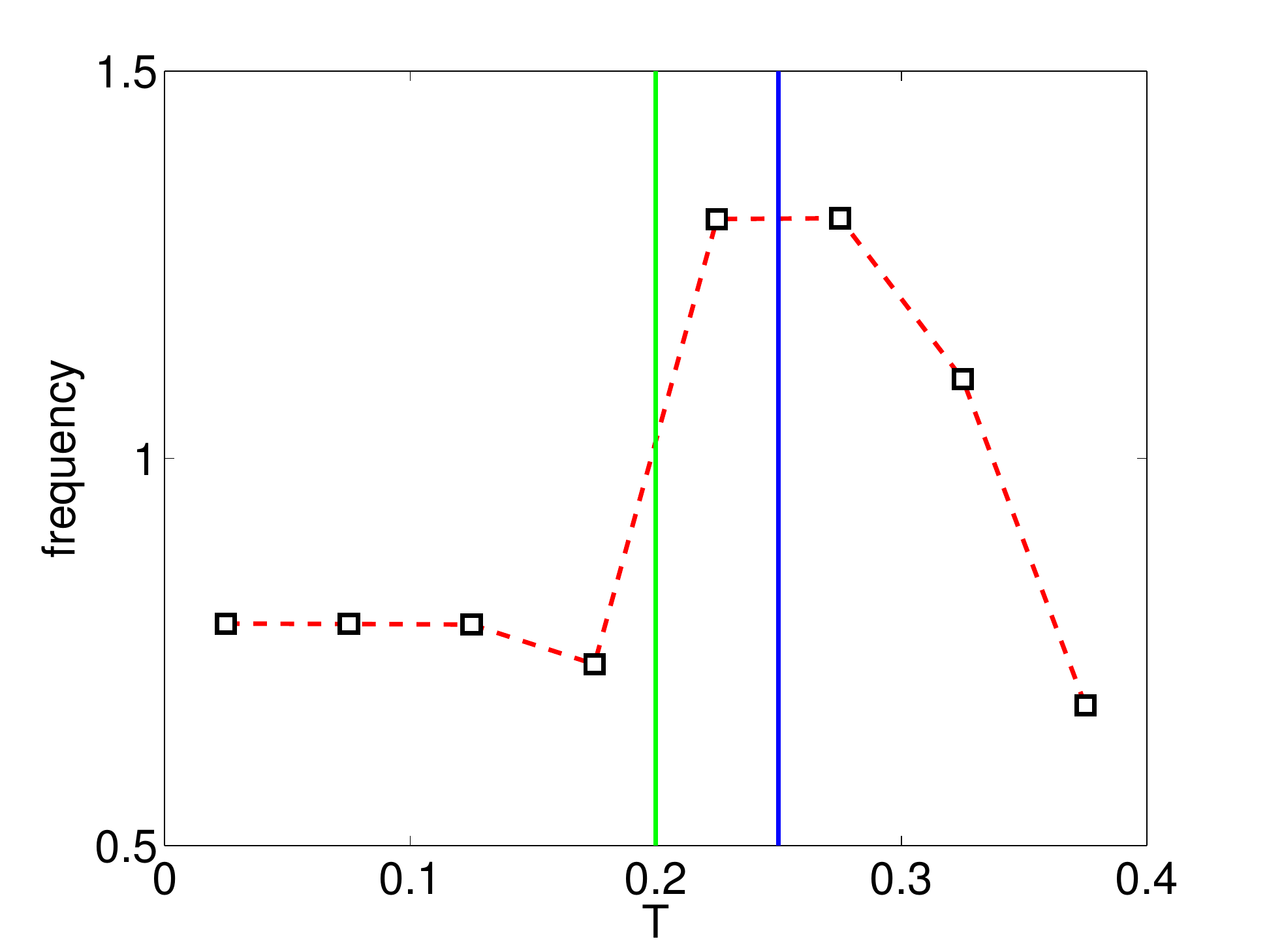}
\caption{Left: spectral fitting of the left panel in Fig.~\ref{pv01}. The inserted figure is the power spectrum at $T=0.225$ near $T_c$. Right: peak frequencies from the left panel as a function of temperature.}
\label{pv01fitspec}
\end{center}
\end{figure}

We have chosen $\omega=1, \Delta V=0.1$ as shown in Fig.~\ref{pv01}. Fig.~\ref{pv01fit} shows the fit using Eq.~\ref{fit} and the fitting parameters are shown in Fig.~\ref{pv01fitpara2}. We have similar amplitude enhancement effect and temperature dependent frequencies, however, simply by eye, 
the first valley at low temperature becomes a peak upon crossing $T_c$. In Fig.~\ref{pv01fitpara2}, we can see that there is a nearly $\pi$ (or $-\pi$) phase shift crossing $T_c$. The spectral analysis is shown in Fig.~\ref{pv01fitspec}, the peak frequencies are quite closed to the fitting frequencies. From the inserted figure in the left panel, near $T_c$ side peak will grow then becomes the main peak. This reflects the frequency abrupt change near $T_c$ in the right panel.

\begin{figure}
\begin{center}
\includegraphics[width=1.6in]{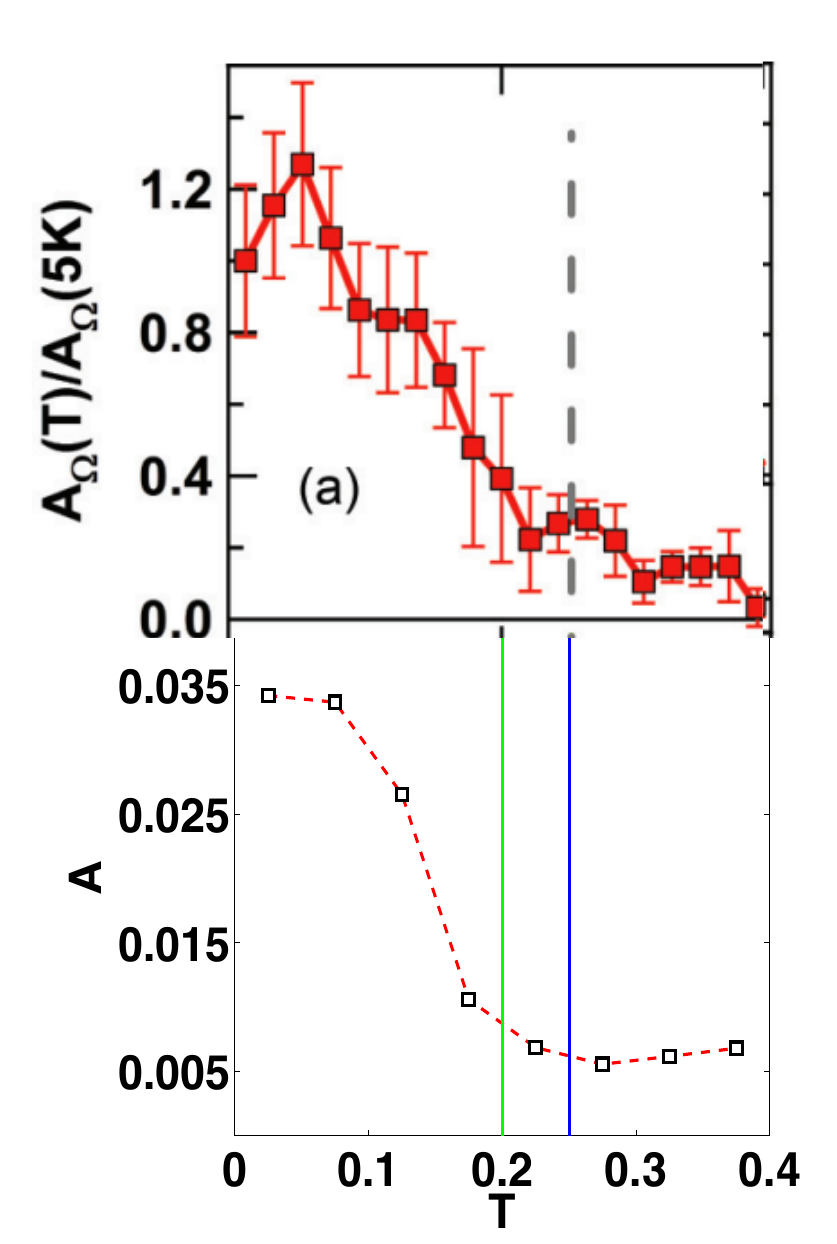}
\includegraphics[width=3in]{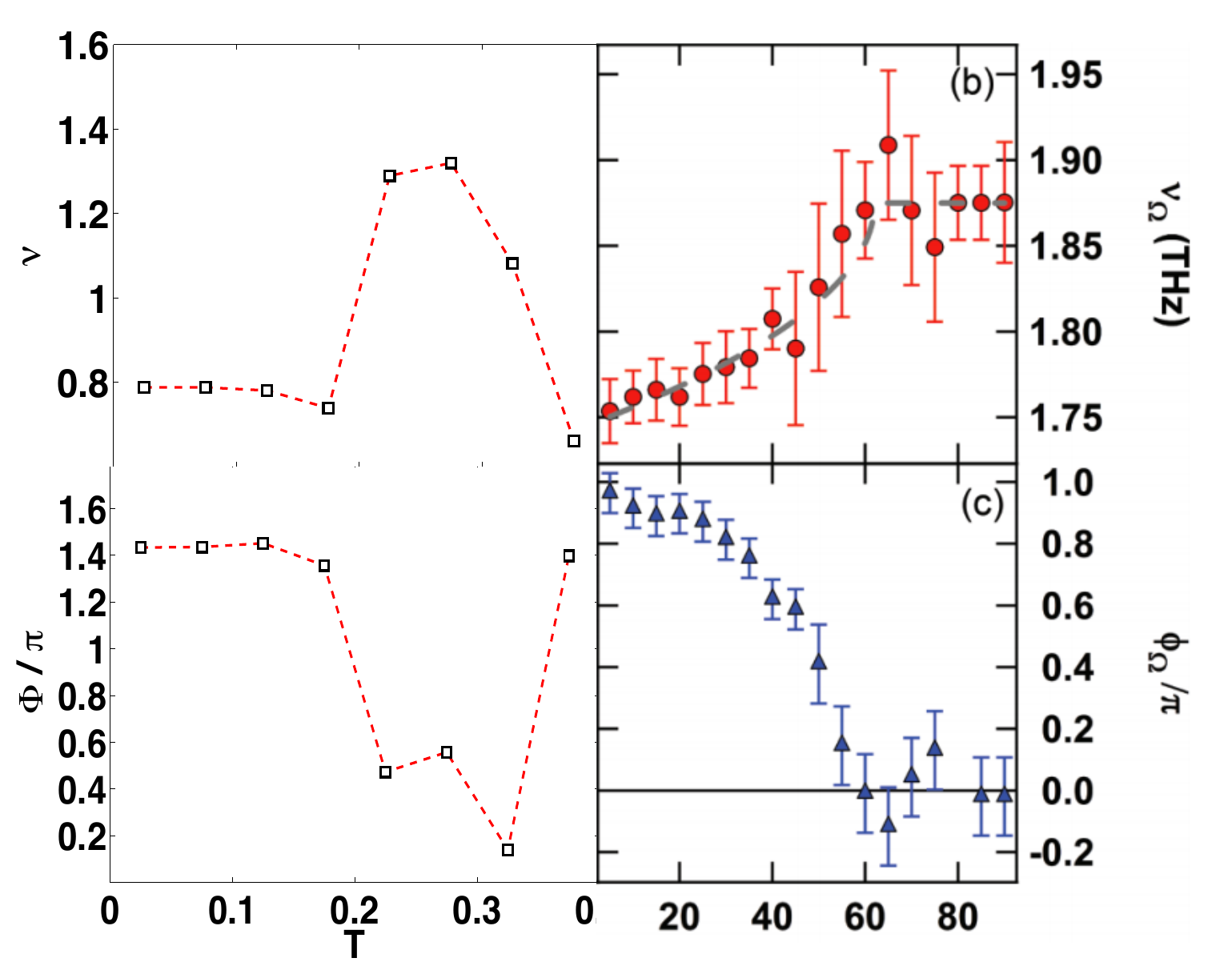}
\caption{Comparison of the numerics (dashed red line) from Fig.~\ref{pv01fitpara2} and experiment data\cite{Orenstein}(red and blue dots). Left panel is amplitude against temperature, where dashed grey line denotes initial $T_c$ before perturbation in the experiment,  the blue line denotes initial $T_c$ in the numerics, the green line denotes the equilibrium $T_c$ at the largest derivation $V=V_0+\Delta V$ in the numerics. The right panel shows the comparison in frequency and phase.}
\label{compare}
\end{center}
\end{figure}

Fig.~\ref{compare} shows a direct comparison of our simulation and the experimental data. Both data indicate enhancement of CDW oscillation below $T_c$. If we put back the real energy scale $T_c\approx 75K$ \cite{Orenstein} in the simulation, the fitting frequency is of the order THz, which is also the same order found in Ref.~\onlinecite{Orenstein}. The difference is that the frequency is more dependent on temperature in the numerics than in the experiment. 

We also calculated the case $\Delta V=-0.1$ as shown in Figs.~\ref{pv-01}--\ref{pv-01fitspec} of Appendix~\ref{app:simulation}. Other than the phase shift to the opposite direction, the other results are quite similar. In the experiment~\cite{Orenstein}, the pump duration is 60 fs. In the simulation above, if we put back the energy scale, the choice of $\omega=1$ means that the pulse duration is of the order 100 fs. We checked the case when we set $\omega=4$ (duration 25fs) as shown in  Figs.~\ref{pv01fre4}--\ref{pv01fre4fitspec} of Appendix~\ref{app:simulation}. Because the duration is shorter, the oscillation amplitude gets smaller, however the frequency and phase shift do not change much from the $\omega=1$ case.

Finally, we examined the quench or pulse $J$ case, i.e. $\Delta J\neq0$. In this case, our 
numerics showed no clear enhancement of oscillation below $T_c$. We showed one case when pulse $\Delta J=0.1, \Delta V=0$ in Fig.~\ref{pj01}: no obvious enhancement or phase shift is observed crossing $T_c$. In the experiment, the pump light suppresses the superconductivity condensation, which enhances the CDW as mentioned in Ref.~\onlinecite{Orenstein}. In our model the nearest neighbor repulsion $J$ favors CDW and suppresses the condensation, while anti-ferromagnetic coupling $J$ favor both. Therefore,  disturbing in $V$ mimics the pump effect in the experiment.


\section{Quantum non-linear sigma model}
\label{sec:intro2}

In the sections above, we worked with a model of electrons with
an underlying Fermi surface, and then examined the dynamics implied by the electron dispersion on order parameters
consisting of fermion bilinears. In the following, we will work directly with the competing order parameters, via an effective
Hamiltonian for the bosonic order parameters themselves. This approach will allow for spatial fluctuations of the order
parameters, and hence goes beyond the mean-field treatment of the previous sections. However, our analysis will be limited here
to a ``collisionless'' large $N$ limit in which true thermalization does not take place in the long-time limit.

As we noted in Section~\ref{sec:intro}, the CDW of the present model can have the experimentally
observed wavevectors of $(\pm Q_0, 0)$, $(0, \pm Q_0)$.

Our model for the competing order parameter has an energy functional which is drawn directly from 
recent work by Hayward {\em et al.}\cite{o6} as we discussed in Section~\ref{sec:intro}.
And this is supplemented with a ``relativistic'' time derivative term, as also discussed in Section~\ref{sec:intro},
thus ignoring any damping that may arise from the gapless fermions at the nodes of the $d$-wave superconductor.

In Lorentzian signature, the path integral of such an O(6) model takes the following form:
\beq
\int D\Psi \, D\Phi \, D\sigma \, e^{i S},
\eeq
where
\beq
S = \frac{N\rho_s}{2} \int d^3x \bigg( (\partial \vec{\Psi})^2 + \lambda (\partial \vec{\Phi})^2- ( g \vec{\Phi}^2 - \mu (\vec{\Phi}. \vec{\Phi})^2 + \sigma (\vec{\Phi}^2+ \vec{\Psi}^2 - 1))
\bigg),
\ee
where  the $\sigma$ integral imposes the condition
\beq
\vec{\Phi}^2 + \vec{\Psi}^2 =1,
\ee
and $\vec{\Psi}$ is an $N/3$ dimensional vector, whereas $\vec{\Phi}$ a $2N/3$ dimensional vector.
The symmetries of the problem also allow a linear time-derivative term $\Psi_1 \partial_t \Psi_2$, which is allowed by the absence of particle-hole symmetry about the Fermi surface. However, the particle-hole asymmetry is small and we will ignore it in our analysis.
Also we have chosen the velocity of `light' in our relativistic formulation to be unity by rescaling the time co-ordinate.

The subsequent procedure we follow is very much the same as in Ref~\onlinecite{cardy}, 
which is further elaborated and extended in Ref.~\onlinecite{gubser}, accommodating
more general dynamical evolution beyond a strict quantum quench.
The models considered in these previous works, however, focus on the linear sigma
model. Our path-integral treatment parallels that in Ref.~\onlinecite{hung} and particularly  
Ref.~\onlinecite{das} 
where the Schwinger Keldysh formalism is employed. We note however that this
is identical to the approach taken elsewhere\cite{cardy, gubser}, and that one can show that the 
self-consistent mean-field equation of a general linear sigma model with a $\phi^4$ coupling,
considered for example in Ref.~\onlinecite{gubser},
reduces to a NLSM by taking the large $\phi^4$ coupling limit while holding
the ratio of the $\phi^4$ and $\phi^2$ couplings constant.

To proceed with the path-integral, we can linearize the action by introducing the auxiliary field $\rho$:
\beq
S = \frac{N\rho_s}{2}\int d^3 x \bigg( (\partial \vec{\Psi})^2  -\sigma \vec{\Psi}^2
+  (\partial \vec{\Phi})^2- (g +\rho + \sigma)/\lambda \vec{\Phi}^2 - \frac{\rho^2}{4\mu} + \sigma
\bigg)
\ee
Note that we have rescaled $\Phi$ to obtain a canonical kinetic term for $\Phi$ above. $\rho$ is
defined accordingly.

Integrating out $\vec{\Phi}, \vec{\Psi}$, gives an effective action in the remaining path integral
$\int D\sigma \, D\rho \,e^{i N/2 S_{\rm eff}}$
\beq
S_{\rm eff} = \frac{i}{3} \tr \ln (\Box + \sigma) + \frac{2i}{3} \tr \ln (\Box + (g+ \rho+\sigma)/\lambda) + \int d^3 x ( -\frac{\rho_s\rho^2}{4\mu} + \rho_s \sigma ).
\ee
The corresponding gap equations are
\bea\label{gap0}
\rho_s &&= \frac{1}{3} \int \frac{d^{2}k}{(2\pi)^{2}} \left(G_\Psi (k,t) +2 /\lambda G_\Phi (k,t) \right), \nonumber \\
\rho(t) &&= \frac{4\mu}{3\rho_s \lambda} \int \frac{d^{2}k}{(2\pi)^{2}} G_\Phi (k,t),
\eea
where $G_\Psi, G_\Phi$ are spatially Fourier transformed equal time correlation functions. We have
suppressed the details of the  Schwinger-Keldysh time contour, whose only consequence in the leading large $N$ calculation is
to determine the boundary conditions of the Green's functions  that we will review below.

These Green's functions can be conveniently parametrized by the (spatially Fourier transformed) time dependent field as follows \cite{gubser,das}:
\beq
\Phi_k(t) =  \Phi_k(t_i) \sqrt{\frac{\Omega^{\Phi}_k(t_i)}{\Omega^{\Phi}_k(t)}}\cos(\int \Omega^\Phi_k(t) dt) + \Pi_{\Phi\, k}(t_i) \frac{\sin(\int^t dt' \Omega^\Phi_k(t')))}{\sqrt{\Omega^\Phi_k(t_i) \Omega^\Phi_k(t)}},
\label{paramphi}
\ee
and similarly we can parametrize $\Psi$ using these time dependent functions. 
ie 
\beq
\Psi_k(t) =  \Psi_k(t_i) \sqrt{\frac{\Omega^{\Psi}_k(t_i)}{\Omega^{\Psi}_k(t)}}\cos(\int^t \Omega^\Psi_k(t) dt) + \Pi_{\Psi\, k}(t_i) \frac{\sin(\int^t dt' \Omega^\Psi_k(t')))}{\sqrt{\Omega^\Psi_k(t_i) \Omega^\Psi_k(t)}}.
\label{parampsi}
\ee
$\Pi_{a \, k}$ denotes the conjugate field of $a \in \{\Phi, \Psi\}$, and that $t_i$ is some initial 
time which we could beq taken to approach $-\infty$.

These $\Omega^\Phi_k(t)$ satisfies the equation
\beq \label{Oeom}
\frac{\ddot{\Omega}^\Phi_k}{2\Omega^\Phi_k} -\frac{3}{4} \left(\frac{\dot{\Omega}^\Phi_k}{\Omega^\Phi_k}\right)^2+(\Omega^\Phi_k)^2 = k^2 + m_\Phi^2(t).
\ee
Similarly,
\beq \label{Oeom2}
\frac{\ddot{\Omega}^\Psi_k}{2\Omega^\Psi_k} -\frac{3}{4} \left(\frac{\dot{\Omega}^\Psi_k}{\Omega^\Psi_k}\right)^2+(\Omega^\Psi_k)^2 = k^2 + m_\Psi^2(t).
\ee
The effective mass is given by
\beq
m_\Phi^2(t) =  (g+ \rho+\sigma)/\lambda, \qquad m_\Psi^2(t) = \sigma.
\ee
From now on, we take $\mu=0$ and so $\rho=0$. 

\section{Equilibrium properties}

Before delving into time-dependent scenarios, we review the properties of the theory
at equilibrium. 

At equilibrium, the time ordered Green's function at finite temperature is given by
\beq
G^{T\, a}(k,t_1-t_2)= \frac{e^{-i \omega_k |t_1-t_2|}}{2\omega_k}  
\coth(\frac{\beta \omega_k}{2}),\qquad  \omega_k=\sqrt{k^2 + m^2_{a}}.
\ee

The gap equation (\ref{gap0}) therefore becomes
\beq
6\pi \beta  \rho_s =  \bigg( \log[\frac{\sinh(\frac{\beta}{2}\sqrt{ m_\Psi^2+ \Lambda^2})}{\sinh(\frac{\beta}{2}m_\Psi)}] + 
\frac{2}{\lambda} \log[\frac{\sinh(\frac{\beta}{2}\sqrt{\frac{g+ m_\Psi^2+ \lambda \Lambda^2}{\lambda}})}{\sinh(\frac{\beta}{2}\sqrt{\frac{g+ m_\Psi^2}{\lambda}})}]\bigg)
\ee
where $\Lambda$ is the UV cutoff. We have substituted $\lambda m^2_{\Phi} = m^2_{\Psi} + g$.

This expression can be compared with the classical case of Ref.~\onlinecite{o6}, where only the zero Matsubara frequency is kept in the
thermal Green's function for each species $a$ {\em i. e.}
\beq
G^a(\omega_n,k) \sim \frac{T}{ k^2 + m_a^2} .
\ee
The gap equation in this case reduces to
\beq
6\pi \beta  \rho_s =  \bigg( \log[\frac{\sqrt{ m_\Psi^2+ \Lambda^2}}{m_\Psi}] + 
\frac{2}{\lambda} \log[\frac{\sqrt{g+ m_\Psi^2+ \lambda \Lambda^2}}{\sqrt{g+ m_\Psi^2}}]\bigg)
\ee

We compare the Green's function $G_{\Phi}(k=0)$ plotted against temperature at fixed $\rho_s$.
In the low temperature limit the quantum $G_{\Phi}$ falls off much slower than linearly in $T$ because of the behavior of $\cot(\beta m_{\Phi} /2)$ in the Green's function. We recall that $m^2_{\Phi} = (m^2_{\Psi}+g)/\lambda$.

Note that in both the quantum calculation and the classical approximation, $G_{\Phi}(k=0)$ exhibits
a maximum at some temperature $T_p$. The peak marks the change between the low temperature behavior   
where fluctuations are dominated by the superconductivity $\Psi$ component, and the high temperature
behavior which is characterized by fluctuations exploring all directions \cite{o6}.
\begin{figure}
\begin{tabular}{cc}
\includegraphics[width=0.5\textwidth]{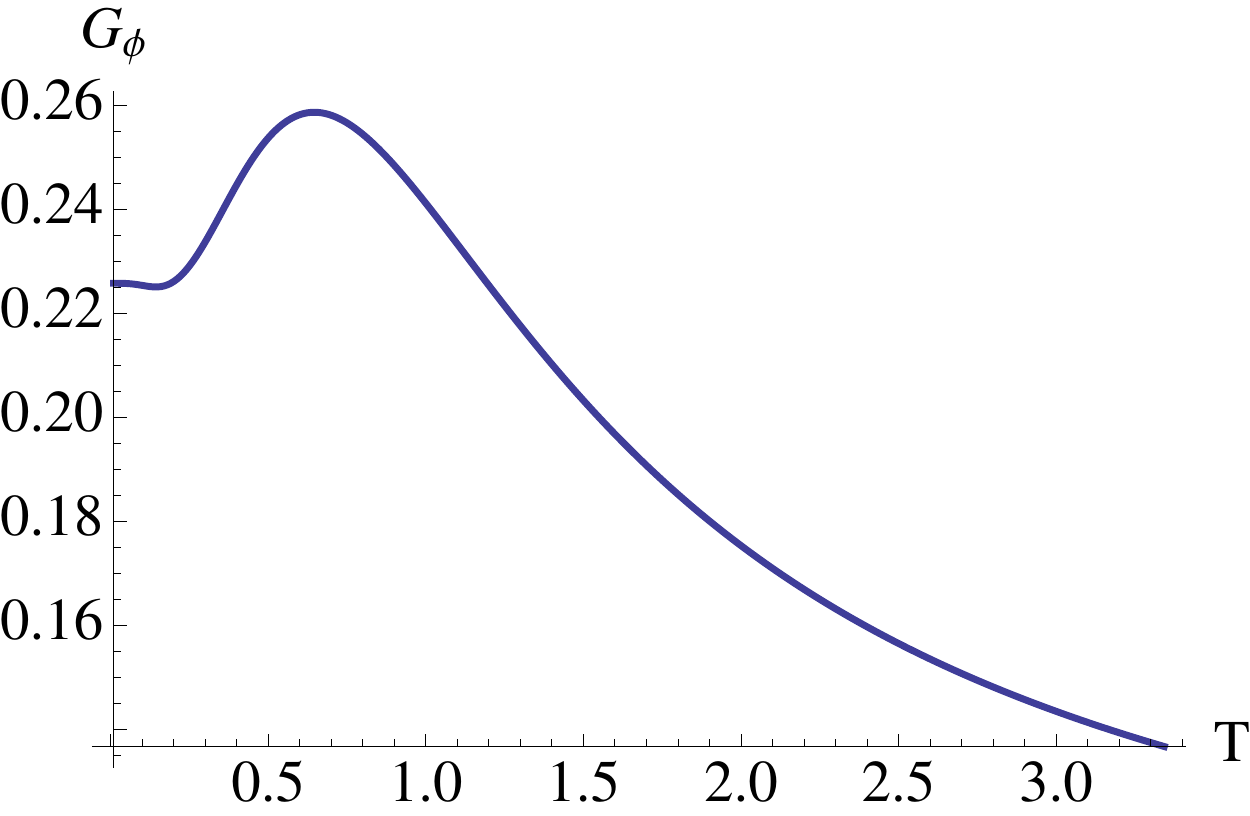} & \includegraphics[width=0.5\textwidth]{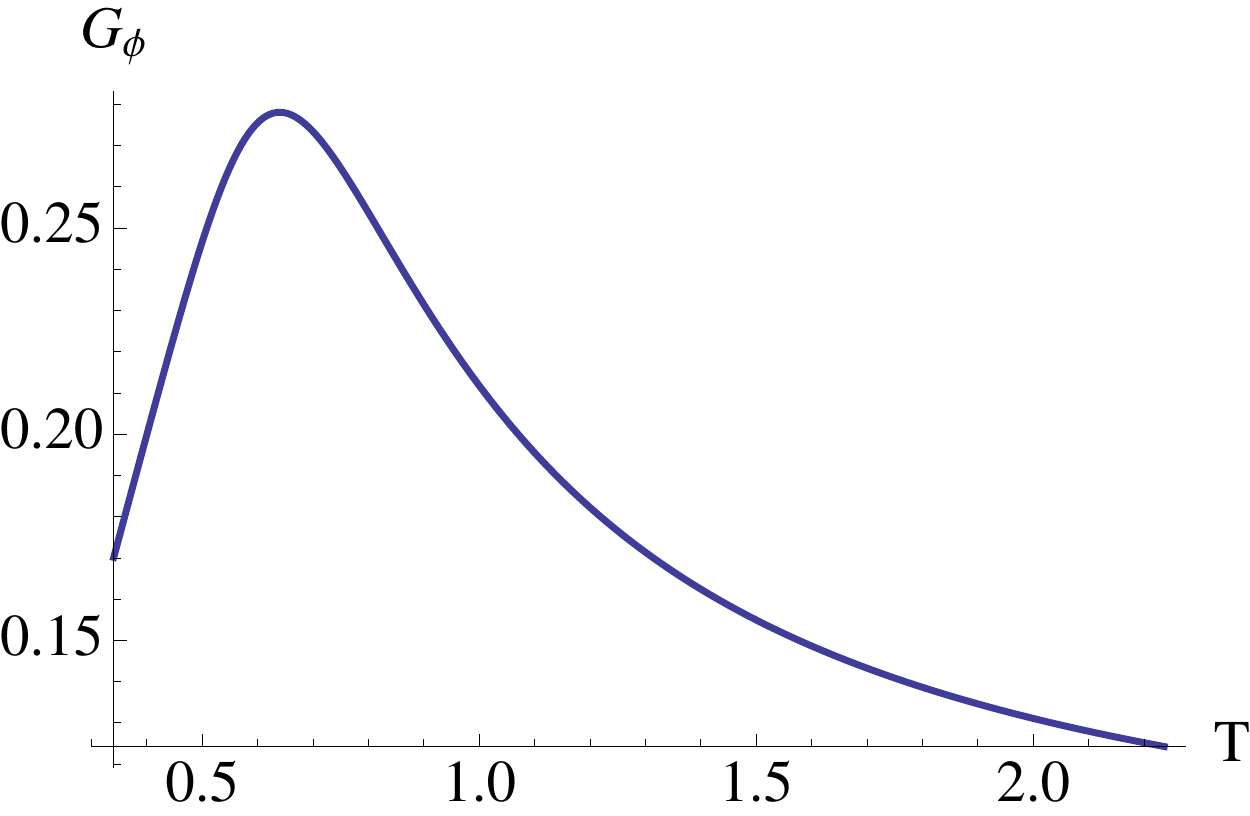}
\end{tabular}
\caption{Equal-time two point function of the charge order
$\Phi$ at vanishing momentum $k$ plotted against temperature at constant $\rho_s$, at cutoff $\Lambda=5$ and $\lambda=g=1$. 
Left: the quantum Green's function. Right: the classical Green's function. The peak position is presumed to be near the onset
of superconductivity: this onset suppresses the charge order fluctuations, leading to a peak in $G_\Phi$ with decreasing temperature.}
\end{figure}


\section{Pulse-like disturbance in $\rho_s$}
\label{sec:pulseII}
In the experiments reported in Ref.~\onlinecite{Orenstein,Gedik,andrea1,andrea2}, the system
is perturbed by pulses of lasers over a short duration of the order of tens of femtoseconds.
As a first brush, to mimic the effect of such a disturbance, we consider perturbing the
system by a time dependent $\rho_s$.
To be precise, we take
\beq
\rho_s = \rho_0 + \delta\rho ( \tanh(v t)^2-1).
\ee
To compute using the saddle point approximation, we should rescale $\Phi,\Psi$:
\beq
\tilde{\Phi}= \sqrt{\rho_s}\Phi,\qquad \tilde{\Psi}=\sqrt{\rho_s}\Psi.
\ee
This leads to a change of the
expression for the effective mass:
\beq
m^2_{\Phi} = \left( \frac{\sigma + g}{\lambda} - (\Box K + (\partial K)^2)  \right)\, \qquad 
m^2_{\Psi}=(\sigma- (\Box K + (\partial K)^2) ),
\ee
where 
\beq
K= \frac{1}{2}\ln \rho_s.
\ee
When $\lambda=1$, these functions $K$ can be absorbed in
the definition of $m^2_{\Psi} = \sigma - (\Box K + (\partial K)^2)$.

We can evolve the system beginning at $v t<-1$, where the time dependence is
negligible, and allow the system to react to the shaking. 
The equal time Green's function in this case, using also the parameterizations
(\ref{paramphi},\ref{parampsi}),  and the initial conditions
\beq
\langle \phi^a(t_i) \phi^a (t_i) \rangle = \frac{1}{2 \Omega^a(t_i)}, \qquad \langle \Pi^a(t_i) \Pi^a(t_i) \rangle = 
\frac{\Omega^a(t_i)}{2},
\ee
takes the form\cite{das}
\be \label{dymG}
G^a(k,t) = \frac{1}{2\Omega^a_k(t)} \coth(\frac{\beta_0  \Omega^a(t_i) }{2}), \qquad a\in \{\Psi,\Phi\}
\ee
and $ \Omega^a(t_i \to -\infty) = \sqrt{k^2 + m_{a}(-\infty)^2}$
where $ m_{a}(-\infty)$
is the initial mass of each field before the application of the disturbance, and that it is set by the initial
temperature $\beta_0$ and $\rho_0$ by solving the gap equation self-consistently at $t=-\infty$.
The gap equation (\ref{gap0}) then  becomes
\bea \label{gaprho}
&&\lambda m^2_{\Phi}(t) - m^2_{\Psi}(t) = g, \nonumber \\
&&m^2_{\Psi}(t) = \frac{A}{2B}, \nonumber \\
&&A=-3\ddot{\rho_s} + \int \frac{dk}{2\pi}  \, k  \bigg(   \big(2 (\Omega^\Psi_k(t)^2-k^2)+\frac{1}{2}\, (\frac{\dot{\Omega}^{\Psi}_k(t)}{\Omega^\Psi(t)})^2\big) G^\Psi(k,t)+\nonumber \\
&&\bigg[\frac{4}{\lambda} \bigg(\Omega^\Phi_k(t)^2-k^2- \frac{g}{\lambda}\bigg)+\frac{1}{\lambda} (\frac{\dot{\Omega}^\Phi_k(t)}{\Omega^\Phi_k(t)})^2\bigg] \,G^\Phi(k,t) \bigg), \nonumber \\
&& B= \int \frac{dk}{2\pi} \, k \bigg(G^{\Psi}(k,t)+ \frac{2}{\lambda^2}G^\Phi(k,t)\bigg)
\eea
The above is obtained by replacing the Green's function in the gap equation
by the explicit forms (\ref{dymG}), and then differentiating the gap equation with respect to time twice.

A natural regularization scheme would be to place the system on a lattice 
with lattice constant $a$. To do so, we make the replacement
\be
k^2 \to \frac{(4-2 \cos(a k_x) -2\cos(a k_y))}{a^2},
\ee
and that $k_x, k_y$ take values between $-\pi/a$ to $\pi/a$.

\subsection{Numerical Results}

\begin{figure}[h]
\begin{tabular}{cc}
\includegraphics[width=0.50\textwidth]{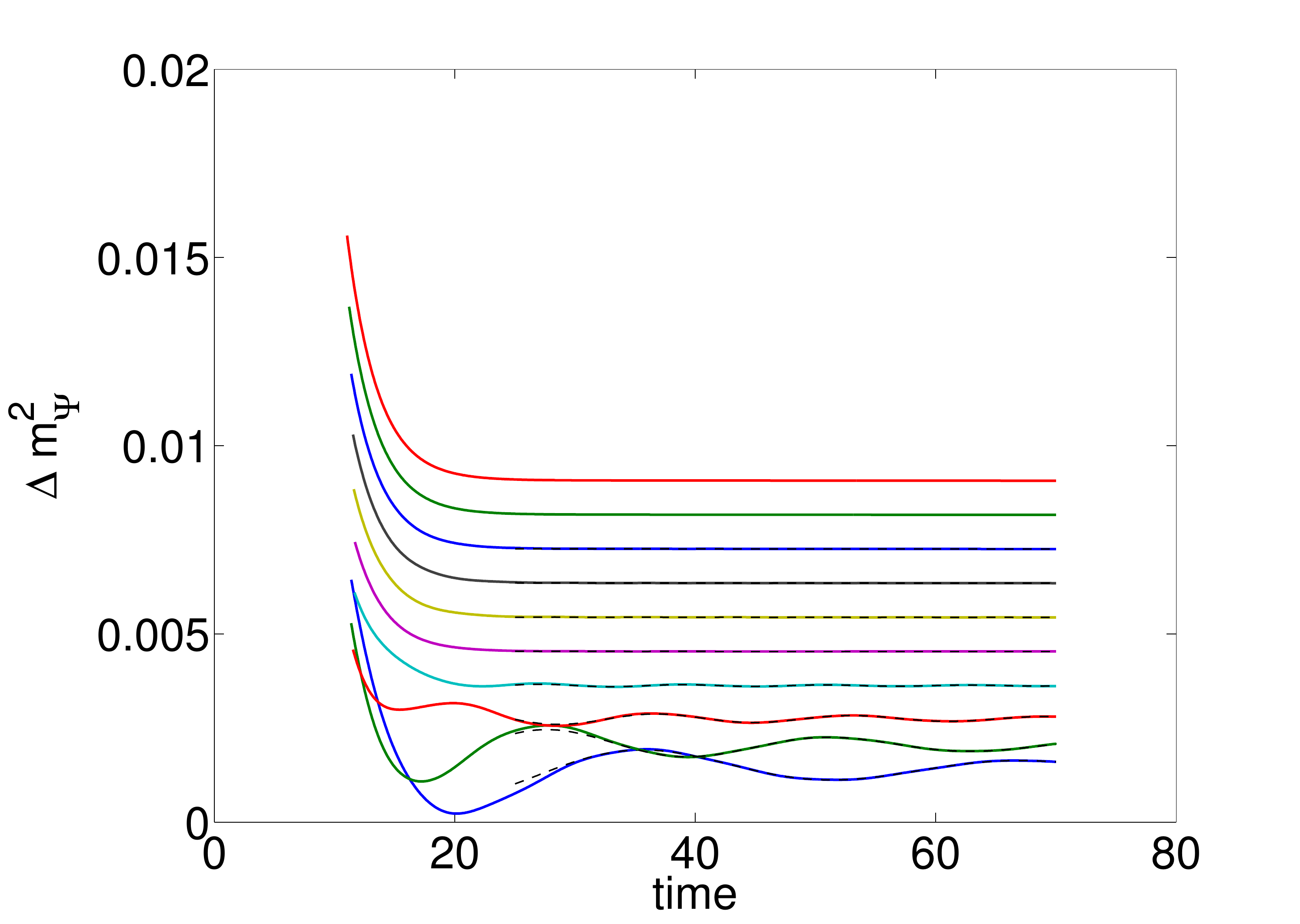} & \includegraphics[width=0.50\textwidth]{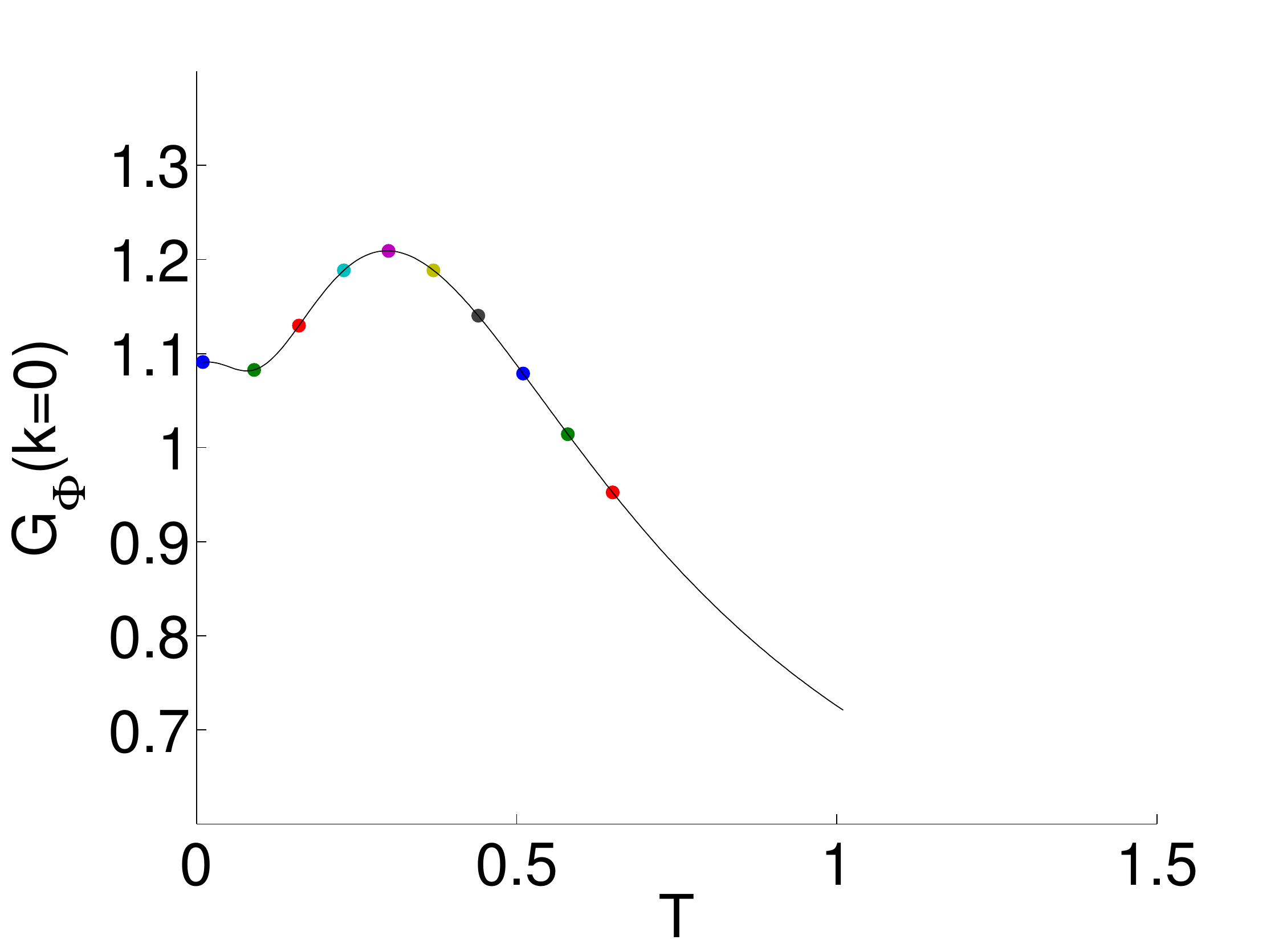} 
\end{tabular}
\caption{Left Panel: Oscillations of $m^2_{\Psi}$ as a function of time at 10 different initial temperatures, from
low temperatures at the bottom of the picture to high temperatures  at the top, at constant
$\rho_0 = 0.0756$  (corresponding to choosing $m_{\Psi}=1/10$ at $T=1/100$, $\lambda=1, g=0.2$, and $a=1$) .  Integral along $k_x$ and $k_y$ is each
divided into 90 steps. The pulse parameter is taken as $v=1/5,\delta\rho= 1/500$. These 10 initial temperatures correspond to 10 points on the equilibrium plot of $G_{\Phi}(k=0)$ against $T$, as shown on the right panel. The color of the markers match the color of the curves on the left.  
Note that variation of the mass before $t=10$ is a huge peak resulting from the disturbance which is not shown in the picture.}\label{10plots}
\end{figure}

\begin{figure}[h]
\begin{tabular}{cc}
\includegraphics[width=0.50\textwidth]{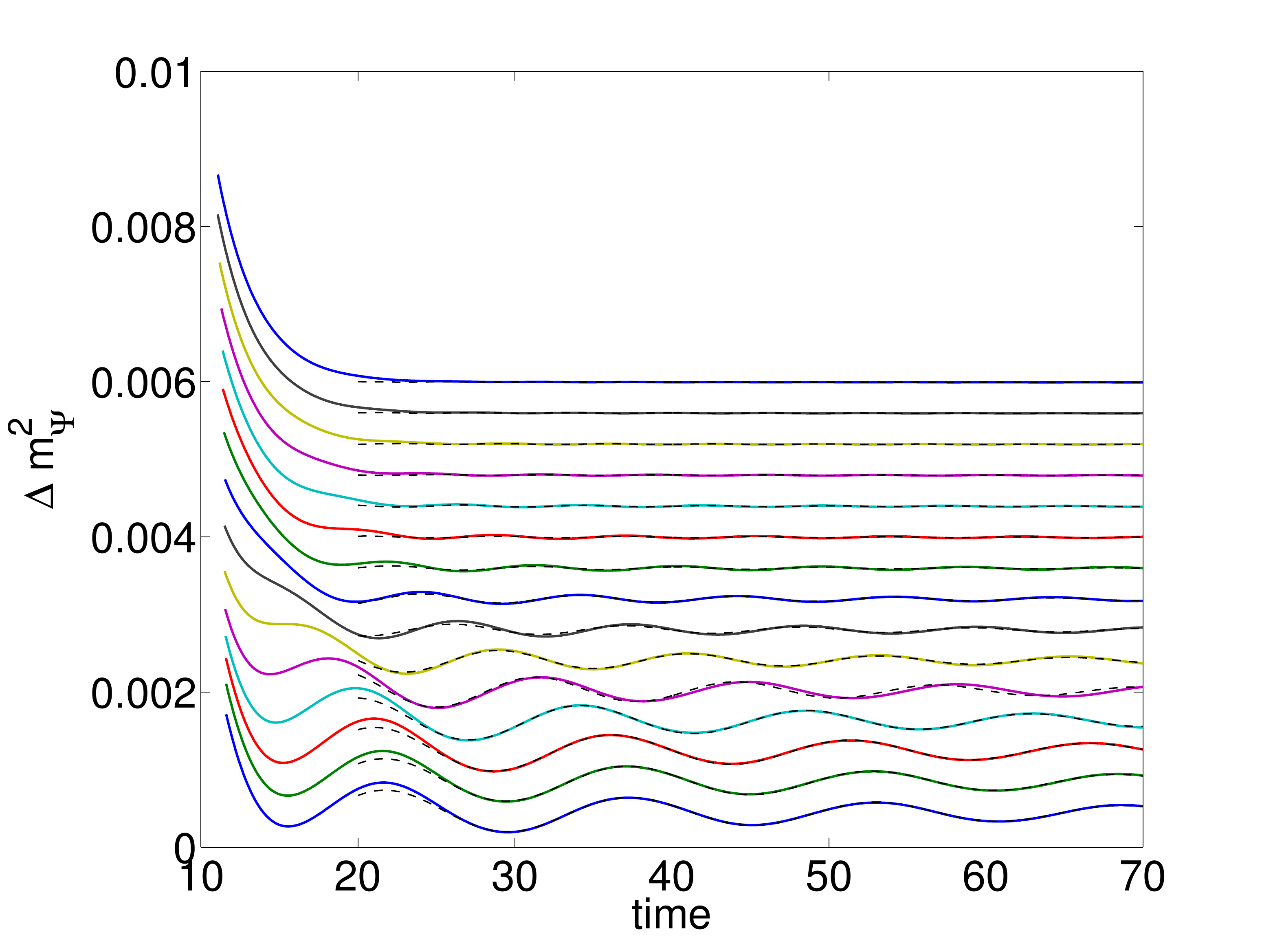} & \includegraphics[width=0.50\textwidth]{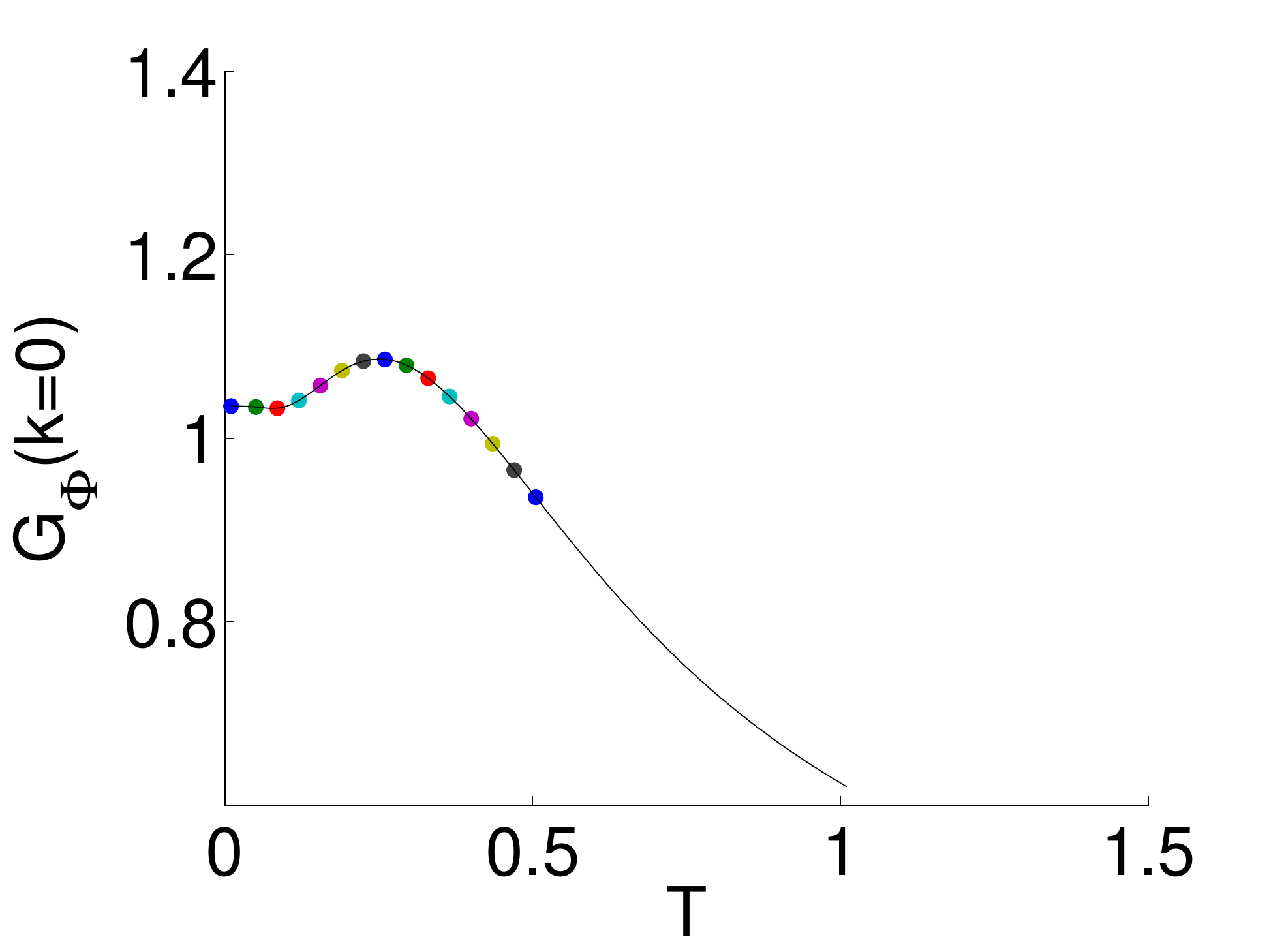} 
\end{tabular}
\caption{Left Panel: Oscillations of $m^2_{\Psi}$ as a function of time at 15 different initial temperatures, from
low temperatures at the bottom of the picture to high temeperatures  at the top, at constant
$\rho_0 =   0.1067$ (corresponding to $m_{\Psi} = 1/5$ at $T=1/100$, $g=0.1, 
\lambda = 6/10$, and $a=1$). Integral along $k_x$ and $k_y$ is each
divided into 90 steps. The pulse parameter is taken as $v=1/5,\delta\rho= 1/500$.
These 15 initial temperatures correspond to 15 points on the equilibrium plot of $G_{\Phi}(k=0)$ against $T$, 
as shown on the right panel. The color of the markers match the color of the curves on the left.  
Note that variation of the mass before $t=10$ is a huge peak resulting from the disturbance which is not shown in the picture.}
\label{10plotslamo6go1}
\end{figure}

\begin{figure}[h!]
\begin{tabular}{cc}
\includegraphics[width=0.50\textwidth]{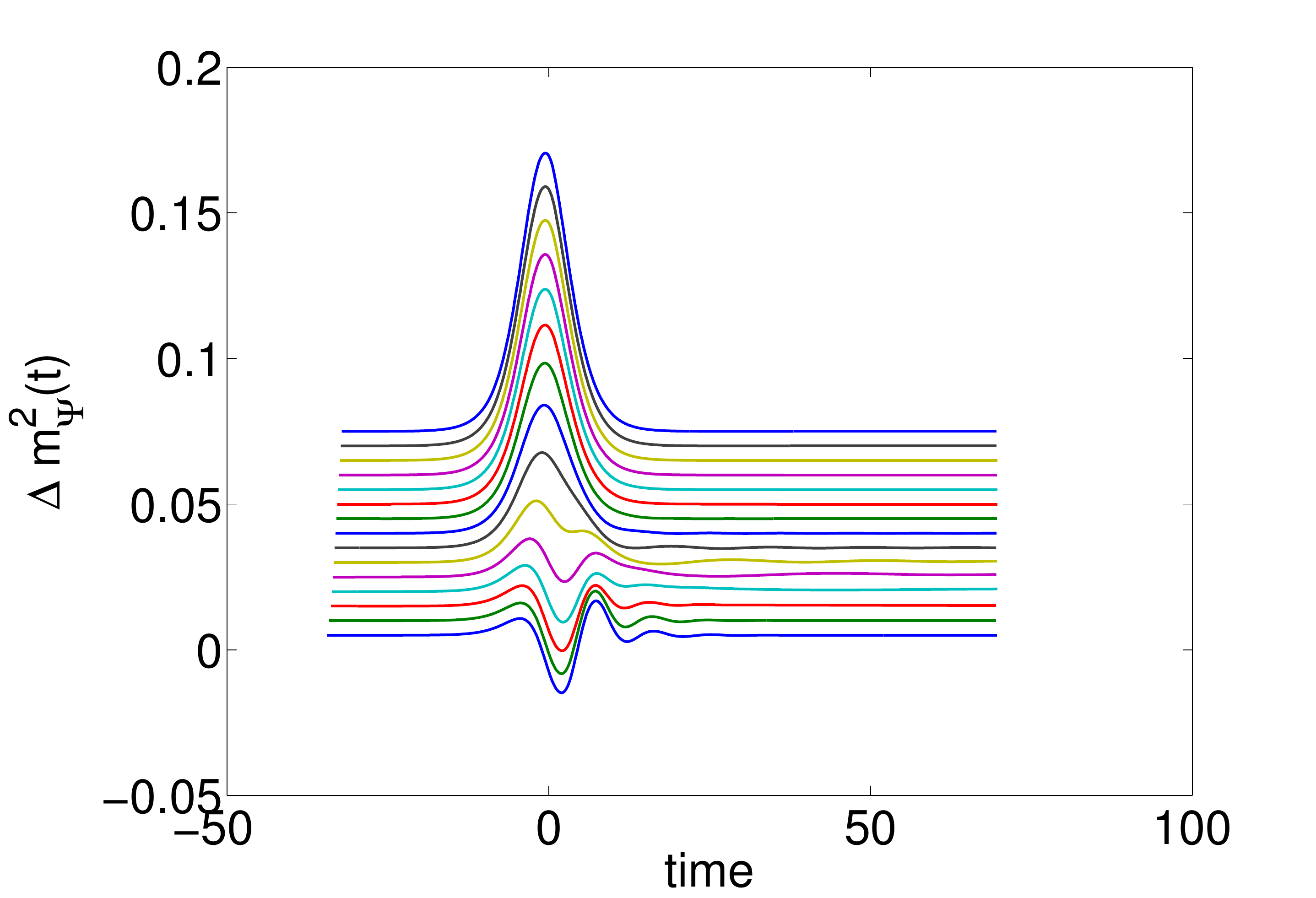} & \includegraphics[width=0.50\textwidth]{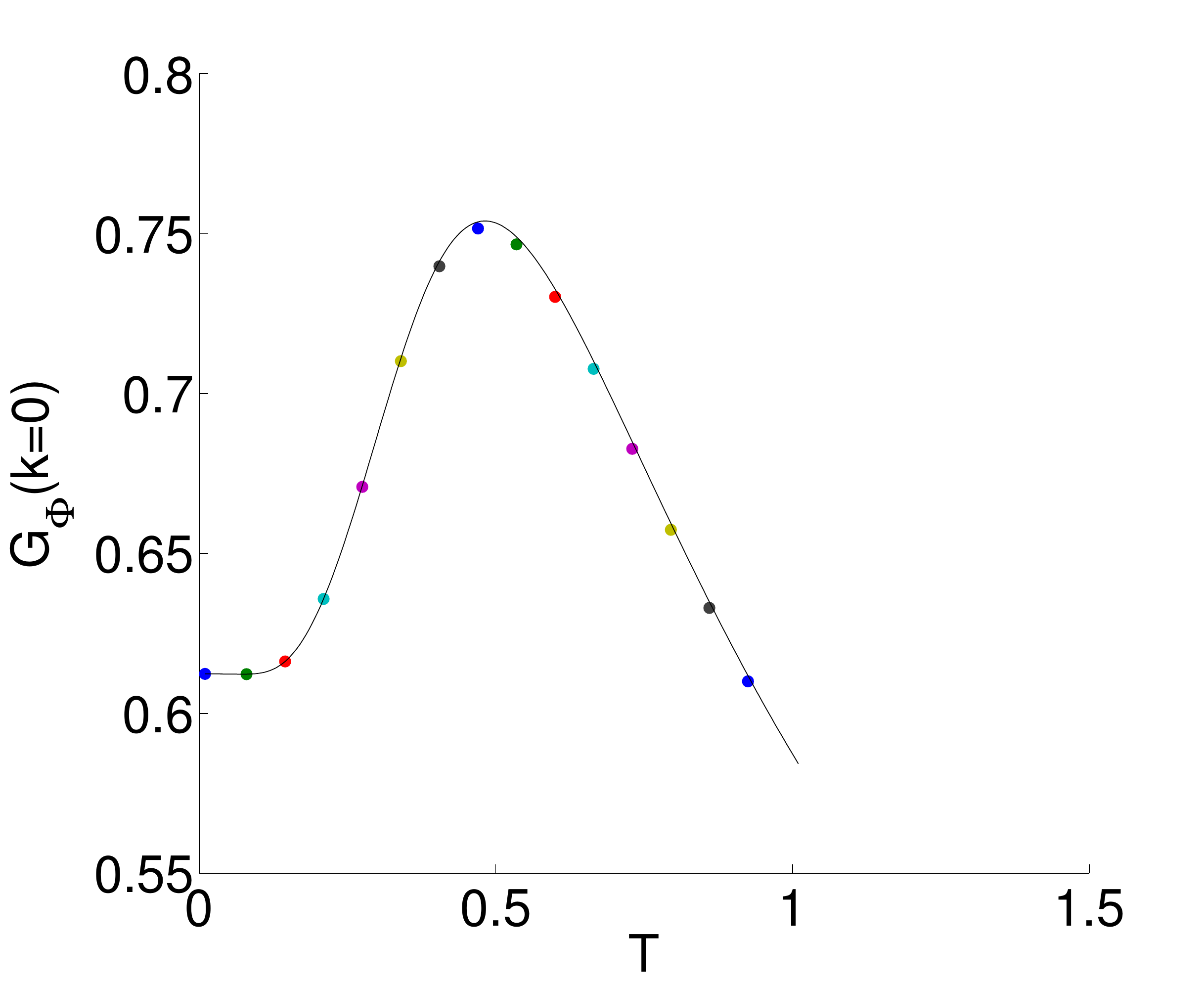} 
\end{tabular}
\caption{Left Panel: Oscillations of $m^2_{\Psi}$ as a function of time at 15 different initial temperatures, from
low temperatures at the bottom of the picture to high temeperatures  at the top, at constant
$\rho_0 =   0.1067$ (corresponding to $m_{\Psi} = 0.0041$ at $T=1/100$, $g=0.4, 
\lambda = 6/10$, and $a=1$). Integral along $k_x$ and $k_y$ is each
divided into 90 steps. The pulse parameter is taken as $v=1/5,\delta\rho= 1/500$.
These 15 initial temperatures correspond to 15 points on the equilibrium plot of $G_{\Phi}(k=0)$ against $T$, 
as shown on the right panel. The color of the markers match the color of the curves on the left.  
Note that in this picture we display the entire oscillations including the large peak. This is because at sufficiently
low temperatures there are extra higher frequency oscillations that begin  earlier.}
\label{10plotslamg4}
\end{figure}

\begin{figure}[h]
\begin{tabular}{cc}
\includegraphics[width=0.5\textwidth]{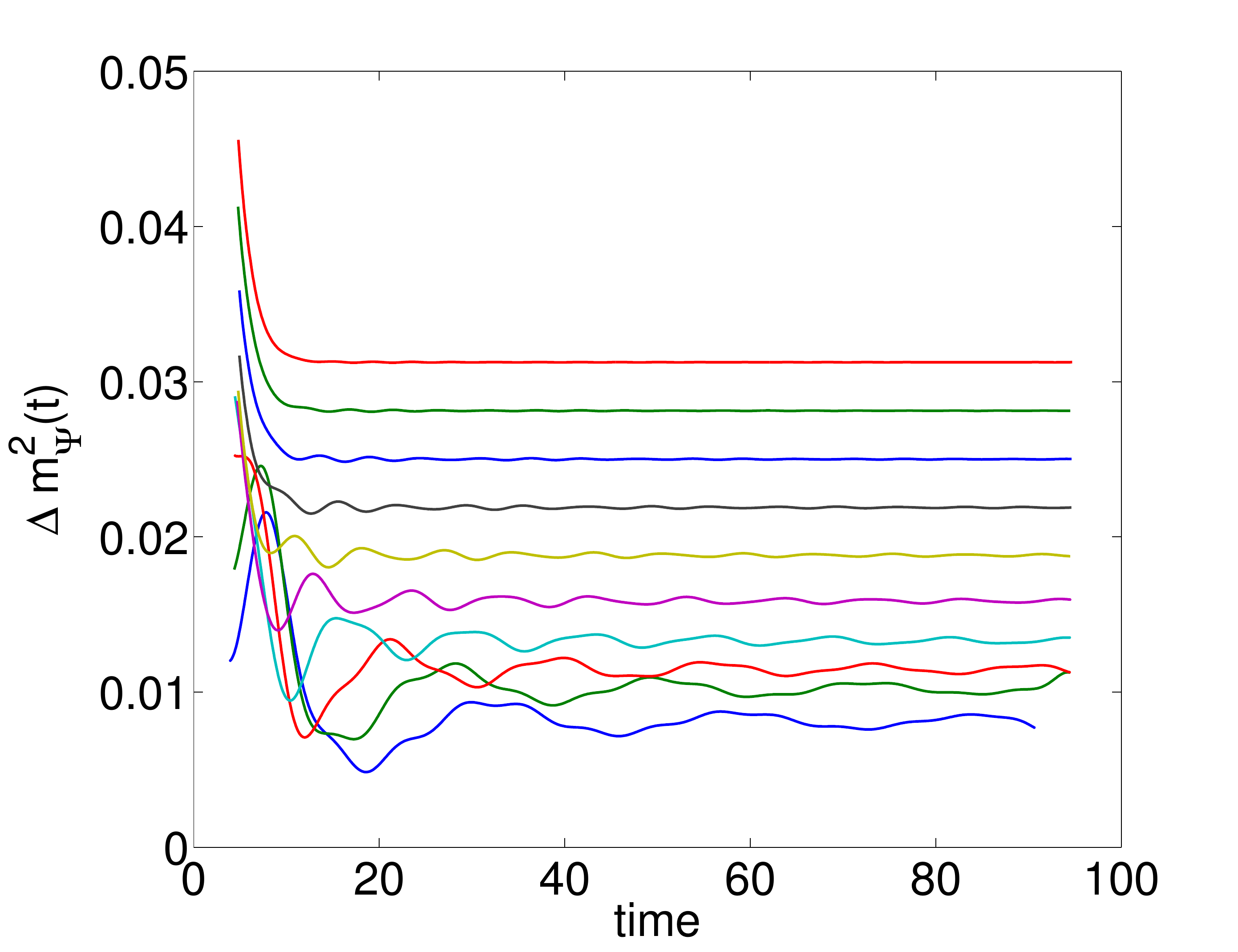} & \includegraphics[width=0.5\textwidth]{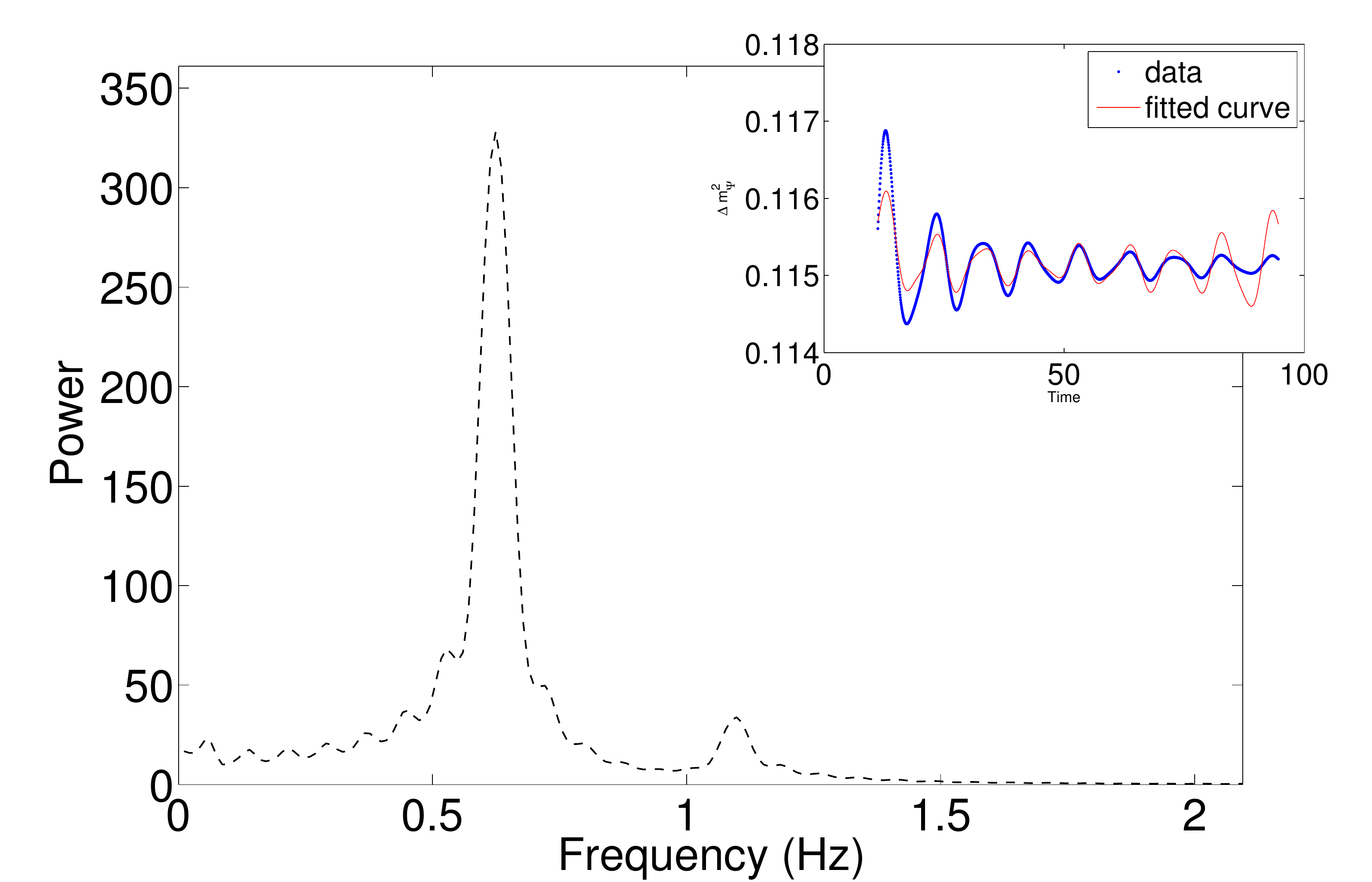}
\end{tabular}
\caption{Left Panel: 10 oscillations taking the same initial conditions as in figure 2 
except that the pulse is more abrupt, with $v=1/3$. At low temperatures corresponding to a regime
$m_{\Psi}\ll v$ the oscillation waveforms look much less regular. Right Panel: a typical plot of the power spectrum of the oscillations, obtained using the magenta curve on the left. The fit makes use of 11 significant frequencies in the power spectrum. }\label{fastg2o10lam110plots}
\end{figure}

We consider dynamical oscillations of the system at various different
choice of parameters $g,\lambda$ and initial temperatures $T=1/\beta_0$, subjected to
different disturbances applied for different durations. 
We plot the oscillations of the self-consistent effective mass $m^2_{\Psi}(t)$
as a function of time. These results are presented in Figs.~\ref{10plots},
\ref{10plotslamo6go1}, \ref{10plotslamg4} and \ref{fastg2o10lam110plots}.
For each set of parameters $g,\lambda$ we obtain the time evolution at 10 different
initial temperatures, and we indicate the position of these initial conditions in the
 equilibrium $G_{\Phi}(k=0)- T$ plot . We look particularly at the vicinity of the peak,
and observe the changes in the oscillations as temperature is increased across the peak.

In all these cases, the self-consistent mass $m_{\Psi}(t)$ displays a large peak
while the disturbance is applied, and exhibits oscillatory behavior after the
time-dependent disturbance is withdrawn. 

The disturbances applied in the cases in figures \ref{10plots} and \ref{10plotslamo6go1}
are relatively slow compared to the initial values of $m_{\Psi}$ and that $m_{\Psi}(t=-\infty)\sim
m_{\Phi}(t=-\infty)$. In these cases, the subsequent oscillations are sinusoidal with
a distinct frequency and a  decaying amplitude. In these cases, we fit the oscillations by the function
\be
f(t) = \exp(-a t) b \sin(c t + 2\pi d) +  e  + f t.
\label{fitfunc}
\ee

\begin{figure}[h]

\includegraphics[width=0.8\textwidth]{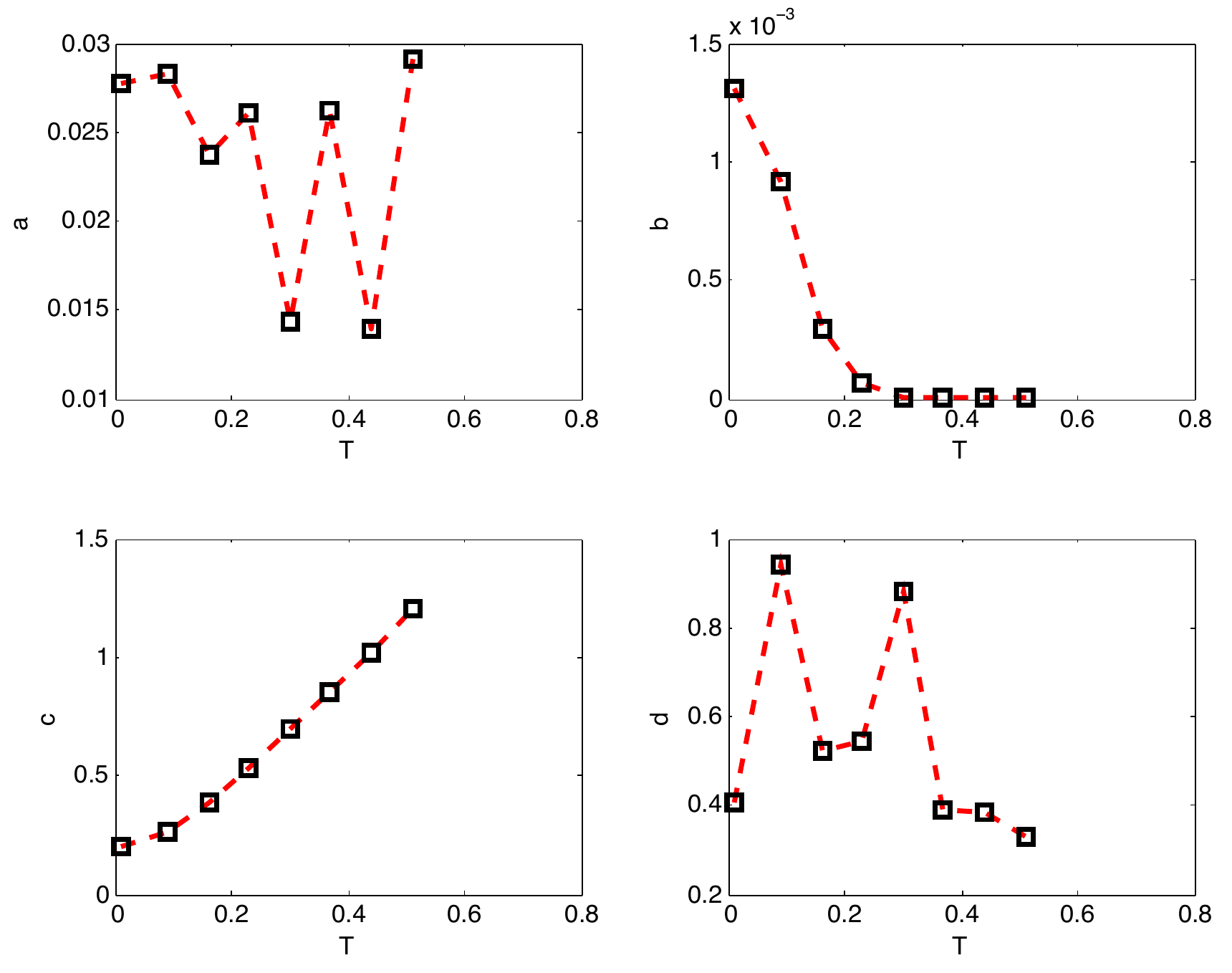}
\caption{Coefficients $a,|b|,c,d$ of the fit function $f(t) = \exp(-a t) b \sin(c t + 2\pi d) + e+ ft$ fitting the data presented in figure \ref{10plots} 
are plotted 
against temperature.}.\label{fitting}
\end{figure}

\begin{figure}[h]
\begin{center}
\includegraphics[width=0.8\textwidth]{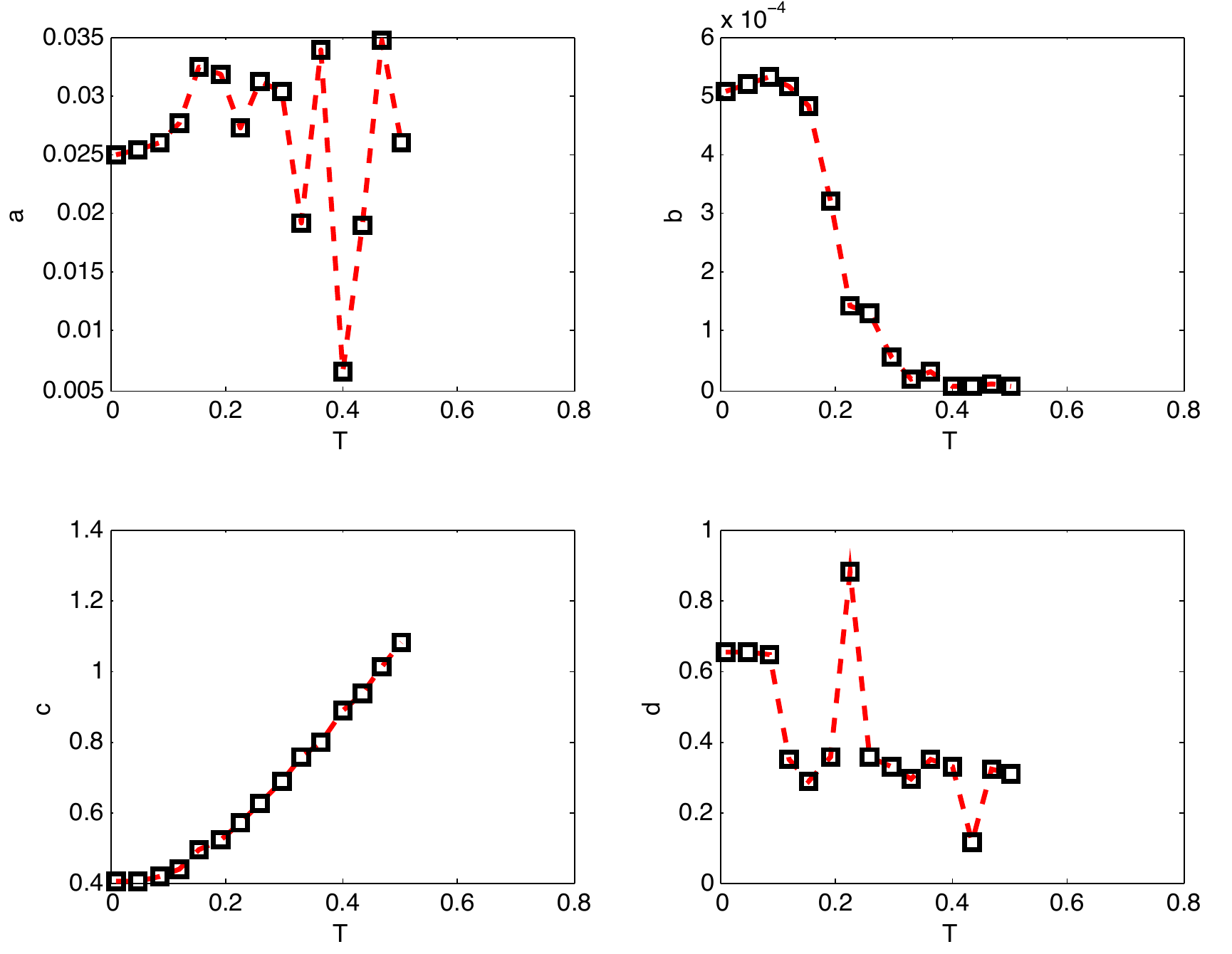}
\end{center}
\caption{Coefficients $a,|b|,c,d$ of the fit function $f(t) = \exp(-a t) b \sin(c t + 2\pi d) + e  + ft$ fitting the data presented in figure \ref{10plotslamo6go1} are plotted 
against temperature.}.\label{fittinglam}
\end{figure}

An infinitesimal, approximately linear downward drift of the oscillations appears at sufficiently high temperatures (i.e. $f< 10^{-3}$ at $a=1$).
This is most likely numerical error since the oscillation amplitudes are also extremely small. We include in the fit function the linear term in order to remove the effect of this drift to improve accuracy for the frequency fit. 
 In all cases,
$e$ is very close to the original value of $m_{\Psi}(t=-\infty)$.
The results of the fit for $a,b,c,d$ corresponding to the data presented in figures \ref{10plots} and \ref{10plotslamo6go1} are presented in figures \ref{fitting} and \ref{fittinglam} respectively. 

A most distinctive feature is that there is a large suppression in the oscillation amplitude as the temperature
$T$ increases across $T_p$, the temperature corresponding to maximal $G_{\Phi}(k=0)$ at equilibrium. This strongly resembles the experimental results \cite{Orenstein} where
oscillations are enhanced below the critical temperature $T_c$ of superconductivity, which is also observed in the electron `hotspot' model considered in the previous section. 
A second feature is that the characteristic frequency of the oscillations increase with
temperature at a rate faster than linearly, and the rate of increase does not appear to level off at high temperatures. One can inspect the ratio of the oscillation frequencies and peak temperature $T_p$. Consider say the results from figure \ref{10plotslamo6go1}. The oscillation frequency is roughly $\omega \sim 0.3-1$, whereas $T_p \sim 0.3$, giving a ratio of order  between 1 -- 3. 
In the experiments, the oscillation frequencies are of order $2\pi \times 2/\textrm{ps}$ and the critical temperature
is of order $\sim$ 50K, giving a ratio of 
\be
\hbar \omega/ (k_B T) \sim 2\pi \times 0.3,
\ee
which is very close to our data.  

 
We note that the accuracy we can achieve for the value of $a$, the rate of exponential
decay of the amplitude is much lower than frequency $c$ and the amplitude $b$ itself. This
is particularly true at higher temperatures, where the oscillation amplitudes are very small, which explains
the apparent larger fluctuations. However, it is clear that the point at which oscillatory behavior begins 
is shifted toward later times as temperature increases, a trend most apparent as we inspect figure
\ref{10plotslamo6go1}, where the first trough has a reducing depth until it disappears altogether as initial temperature is increased. For that matter, there is not an obvious definition of a relative phase between oscillations with different initial temperatures, although simply by eye-balling the oscillations, it is very suggestive of a phase shift with temperatures.

The data presented in figure \ref{10plotslamg4} corresponds to parameters chosen at $g=0.4, \lambda=0.6, \rho_0=0.1067$. At very low temperatures,  the onset of oscillatory behavior appears to begin even before the time dependent disturbance is withdrawn, a feature that eventually disappears as temperature is increased. 

Another feature demonstrated most clearly in figure \ref{10plotslamg4} is that the height of the first peak responding to the time dependent disturbance
increases as temperature is increased. We have checked that this is true for all our data sets.

\begin{figure}[h]
\includegraphics[width=0.7\textwidth]{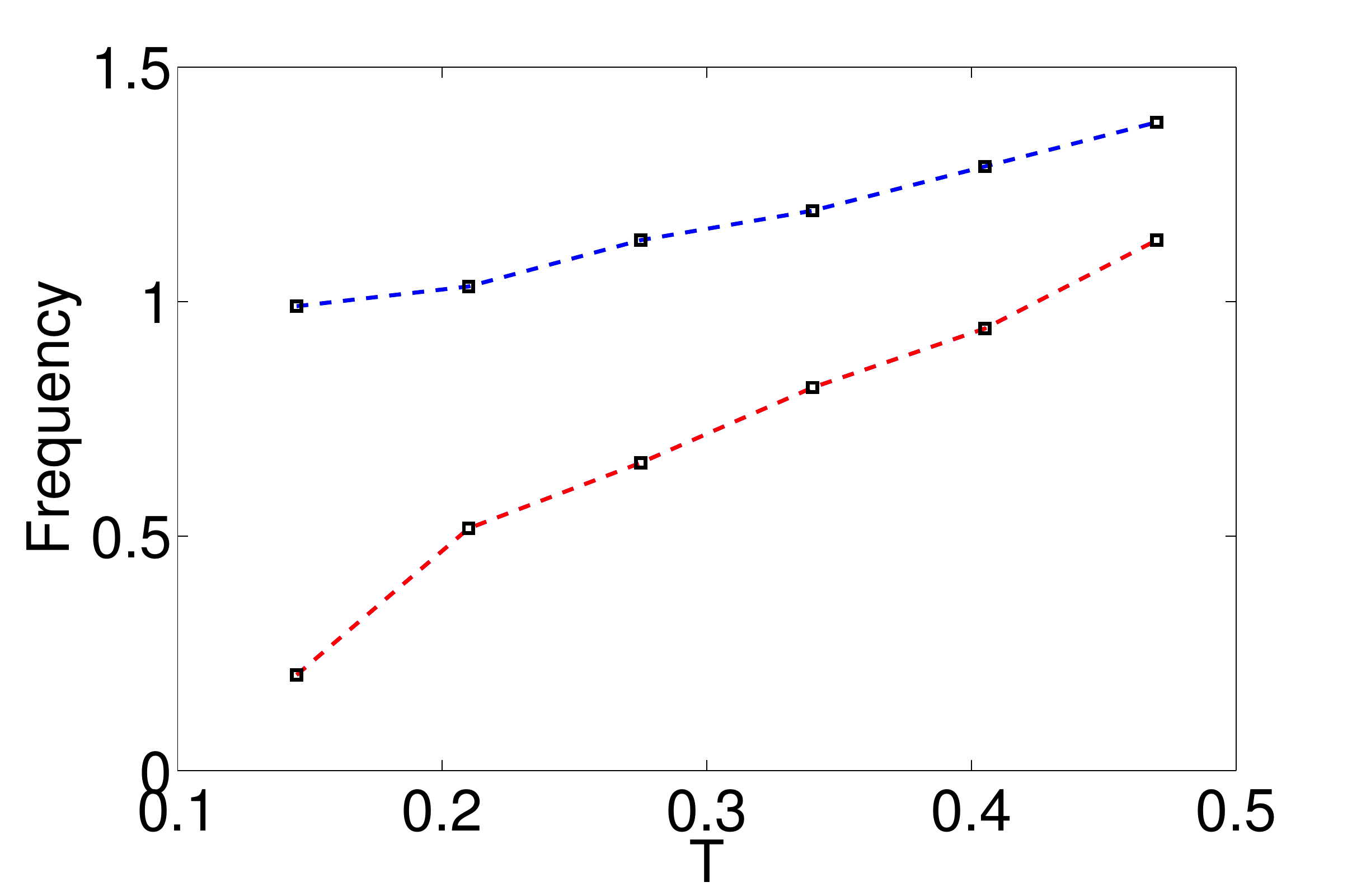} 
\caption{Frequency variation of the higher and lower frequency peaks in the power spectrum with initial temperature.}\label{2freqsT}
\end{figure}

In  the left panel of figure \ref{fastg2o10lam110plots} we show oscillations at 10 differerent temperatures
with the same initial conditions
as in  figure \ref{10plots}, except that the pulse disturbance is more abrupt, set at $v=1/3$. 
We note that the waveform looks much less regular at low temperatures
corresponding to a regime in which the mass scale $m_{\Psi}\ll v$. At higher temperatures corresponding to
higher $m_{\Psi}$ the waveform returns to sinusoidal. Suspecting that there are more than one 
Fourier component with significant amplitude, we inspect the power spectrum of these oscillations. 
A plot of the power spectrum of data obtained using initial temperature $T$ close to $T_p$ 
is shown in the right panel of figure \ref{fastg2o10lam110plots}. (i.e. The set of data corresponding to the magenta curve in figure \ref{fastg2o10lam110plots}). There is a distinct second peak in all the Fourier transforms, and we can track the variation of the
frequencies of the two significant peaks with initial temperatures, shown in figure \ref{2freqsT}.

\begin{figure}[ch!]
\includegraphics[width=0.7\textwidth]{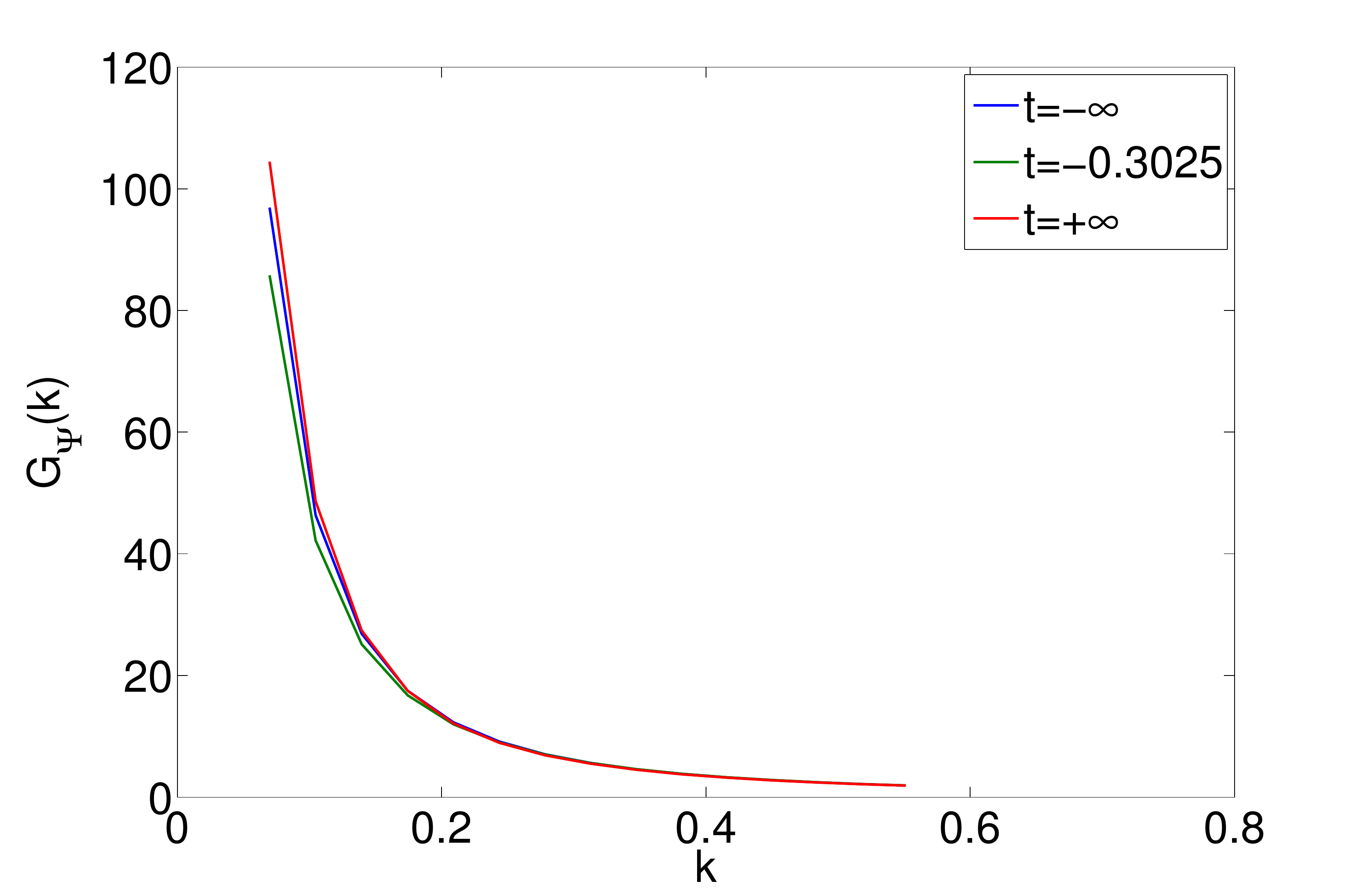} 
\caption{A plot of $G_{\Psi}(k)$ as a function of $k$ at different
times. The parameters take values $g=0.4, \lambda=0.6, \rho_0=0.1067$, and the 
pulse is characterized by $v=1/5, \delta\rho=1/5000$. In fact
it corresponds to the pulse leading to oscillations displayed in the red curve (the third curve
from the bottom of the left panel in figure \ref{10plotslamg4}). }\label{Gpsi}
\end{figure}

It is also of interest to inspect  the Green's function $G_{\Psi}(k)$ at different times.
A typical plot is shown in figure \ref{Gpsi}. At $t=-\infty$ the Green's function
is a thermal Green's function. As the time-dependent disturbance sets in, one can see that
all the departure from the thermal Green's function occurs at low momenta. 
At late times long after the withdrawal of the disturbance, the Green's function
appears to approach the original thermal value. 

\subsection{Remark: pulses in $g$}

 Let us also remark on the response of the system upon shaking the parameter $g$.

Consider $g$ as a function of time given by
\be
g(t) = g_0  + \delta g (\tanh^2(v t)-1).
\ee

The  gap equation is then modified to
\bea
&&\lambda m^2_{\Phi}(t) - m^2_{\Psi}(t) = g(t), \nonumber \\
&&m^2_{\Psi}(t) = \frac{A}{2B}, \nonumber \\
&&A= \int \frac{dk}{2\pi}  \, k  \bigg(   \big(2 (\Omega^\Psi_k(t)^2-k^2)+\frac{1}{2}\, (\frac{\dot{\Omega}^{\Psi}_k(t)}{\Omega^\Psi(t)})^2\big) G^\Psi(k,t)+\nonumber \\
&&\frac{4}{\lambda} \bigg(\Omega^\Phi_k(t)^2-k^2- \frac{g(t)}{\lambda}+\frac{1}{\lambda} (\frac{\dot{\Omega}^\Phi_k(t)}{\Omega^\Phi_k(t)})^2\bigg) \,G^\Phi(k,t), \nonumber \\
&& B= \int \frac{dk}{2\pi} \, k \bigg(G^{\Psi}(k,t)+ \frac{2}{\lambda^2}G^\Phi(k,t)\bigg) \label{gappulse}
\eea

\begin{figure}[h!]
\begin{tabular}{cc}
\includegraphics[width=0.7\textwidth]{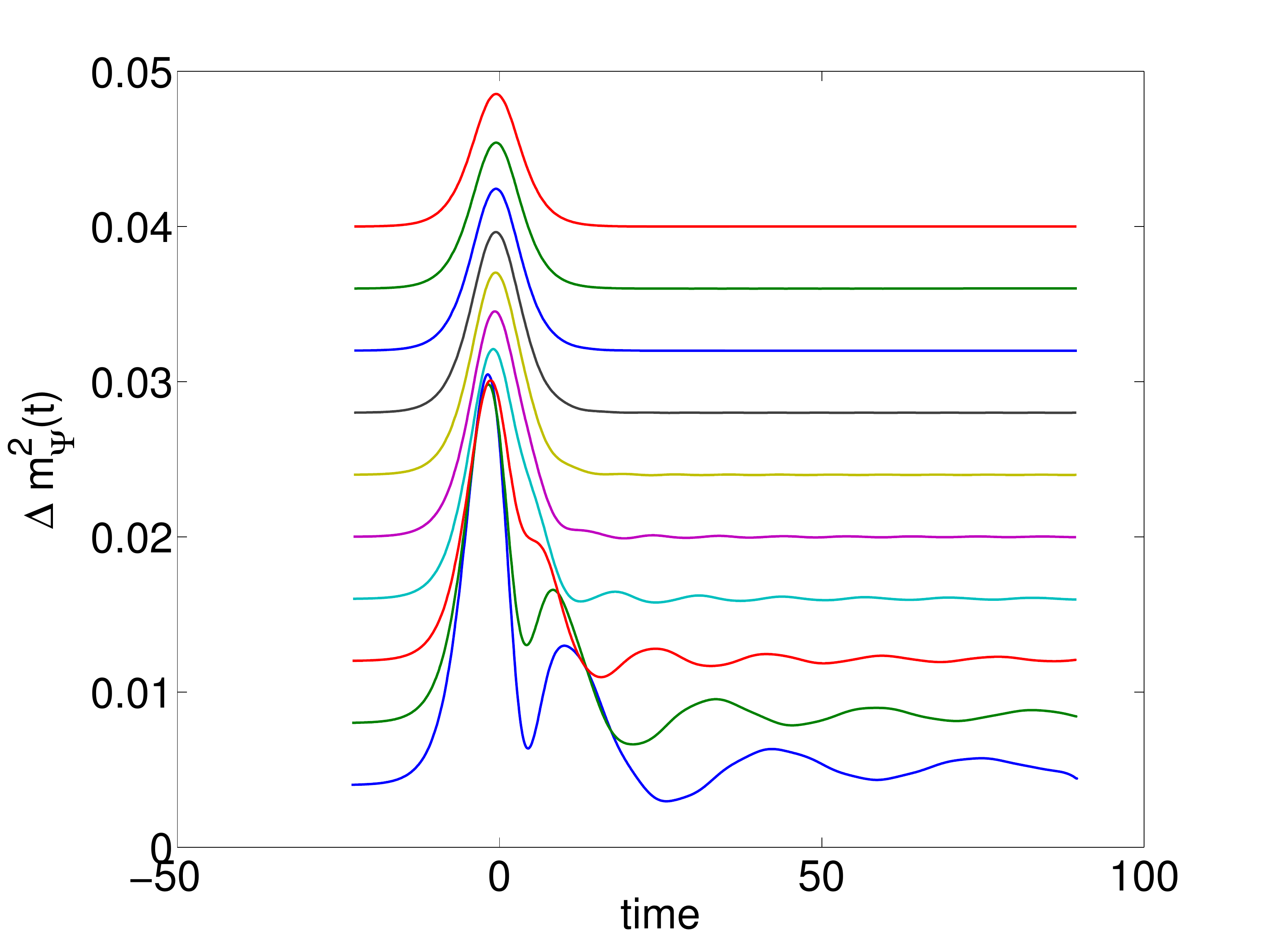}  
\end{tabular}
\caption{Oscillations of $m^2_{\Psi}$ as a function of time at 10 different initial temperatures.
The parameters $g_0,\lambda$ and $T$ take the same values as the 10 plots in figure \ref{10plots}.
 Here $v=1/5$ and $\delta g = 1/50$.}
\label{10Tshakeglam1}
\end{figure}

Below we find the plots at different $T$ in figure \ref{10Tshakeglam1}.  The oscillations following a
kick in $g$
are qualitatively the same as what happens when $\rho_s$ is taken as the time-dependent
disturbance instead. In particular, the oscillations are sinusoidal with a decaying amplitude, and
that the amplitude falls off as temperature increases. What is interesting however is that
contrary to a time-dependent $\rho_s$, the initial large response to the disturbance before oscillatory behavior sets in has a large amplitude at low temperatures which decreases with increasing temperatures.

\section{Conclusions}
\label{sec:conc}

In the first part if the paper, we studied the non-equilibrium dynamics of SC and 
CDW order parameters in the t-J-V model\cite{jay}, using a time-dependent Hartree-Fock computation. We examined two setups: quench and pulse in the interaction parameters, and compared with a recent optical experiment.\cite{Orenstein} 
We used a decaying sinusoidal function to fit the oscillation which gives qualitatively good fitting parameters compared with the spectral analysis, other than the fact that near $T_c$, there are relatively large side peaks in the spectrum. When perturbing with the nearest neighbor Coulomb interaction $V$, we found an enhanced oscillation amplitude of the CDW order below the superconducting critical temperature in both setups.
We interpret this enhancement as a competition between the charge order and superconducting order. The oscillation frequency is of the order of terahertz if we plug in the energy scale from the information of $T_c$. The frequency of the oscillations
has a strong dependence on temperature, which is however different from that in the experiments. 
The change in frequency makes it subtle to define the relative phase between oscillations at different temperatures. 
But in the pulse case crossing $T_c$, if we choose a particular phase starting point in the fit function, we find a nearly $\pi$ phase shift, 
as is evident by eye from Fig.~\ref{pv01}.
When perturbing the exchange interaction $J$, in both setups, there is no obvious enhancement in the oscillation 
amplitude crossing $T_c$ because of the similar enhancement effects of $J$ on both SC condensation and charge density waves.

In the second part of the paper, we considered the dynamical evolution of a large $N$ NLSM inspired by Ref.~\onlinecite{o6} 
subjected to a short pulse like disturbance to mimic the effect of a pulse of laser radiation on the underdoped cuprates. We considered in detail oscillations of the system upon sending a short pulse in 
the coupling $\rho_s$ at different initial temperature $T$, helicity moduli $\lambda$ and relative energetic cost of superconductivity and charge density wave order $g$. We find that for disturbance with a rate of change $v$ taken at the same order as the effective mass scales of the system, the subsequent oscillation after the withdrawal of the disturbance is, to a very good approximation, sinusoidal with an exponentially decaying amplitude. The amplitude is greatly suppressed as initial temperature is increased, particularly beyond the maximal point in the equilibrium $G_{\Phi}(k=0) -T$ plot.  The characteristic oscillation frequency increases faster than linearly with increasing temperature.  The onset of oscillatory behavior begins at a later time at large temperatures making a comparison of relative phases between different oscillations ambiguous, even though the data is suggestive of a phase shift as temperature changes. The increase in oscillation frequency,
 the presence of a phase shift and a strongly suppressed oscillation amplitude across the critical temperatures are features qualitatively consistent with the experiments \cite{Orenstein}.
For more abrupt disturbances or at very low temperatures such that $m_{\Psi}$ is several orders of magnitudes less than $m_{\Phi}$ however, the oscillatory behavior is characterized by more than one frequency.

\acknowledgments

We thank A.~Cavelleri, N.~Gedik, F.~Mahmood, J.~Orenstein, and A.~Vishwanath for valuable discussions.
The research was supported by the U.S.\ National Science Foundation under grant DMR-1103860,
and by the Templeton Foundation. L. H. was supported by the Croucher foundation. This research was also supported in part by Perimeter Institute for Theoretical Physics; research at Perimeter Institute is supported by the Government of Canada through Industry Canada and by the Province of Ontario through the Ministry of Research and Innovation.

\newpage

\appendix

\section{Equations of motion}
\label{app:motion}

It is a simple matter to evaluate the commutators of the operators in Eq.~(\ref{su4}):
\bea
\left[ N_1 (\bk), \Delta_1 (\bk) \right] &=& \Delta_1 (\bk) \nn
    \left[ N_1 (\bk ), \Delta_1^\dagger (\bk) \right]  &=& -\Delta_1^\dagger (\bk) \nn
\left[ N_1 (\bk), \Pi_1 (\bk) \right] &=& \Pi_1 (\bk) \nn
    \left[ N_1 (\bk ), \Pi_1^\dagger (\bk) \right]  &=& -\Pi_1^\dagger (\bk) \nn
\left[ N_1 (\bk), P_1 (\bk) \right] &=&  P_1 (\bk) \nn
    \left[ N_1 (\bk ), P_1^\dagger (\bk) \right]  &=& -  P_1^\dagger (\bk) \nn
\left[ N_3 (\bk), \Delta_1 (-\bk) \right] &=& \Delta_1 (-\bk) \nn
    \left[ N_3 (\bk ), \Delta_1^\dagger (-\bk) \right]  &=& -\Delta_1^\dagger (-\bk) \nn
\left[ N_3 (\bk), \Pi_1 (\bk) \right] &=& -\Pi_1 (\bk) \nn
    \left[ N_3 (\bk ), \Pi_1^\dagger (\bk) \right]  &=& \Pi_1^\dagger (\bk) \nn
\left[ N_3 (\bk), P_3 (\bk) \right] &=&  P_3 (\bk) \nn
    \left[ N_3 (\bk ), P_3^\dagger (\bk) \right]  &=& -  P_3^\dagger (\bk) \nn
 \left[ \Delta_1 (\bk), \Delta_1^\dagger (\bk) \right] &=& N_1 (\bk) + N_3 (-\bk) \nn
 \left[ \Delta_1 (\bk), \Pi_1 (- \bk) \right] &=& - P_1 (\bk) \nn   
 \left[ \Delta_1 (\bk), \Pi_1^\dagger ( \bk) \right] &=& - P_3 (\bk)   \nn
 \left[ \Delta_1 (\bk), P_1^\dagger (\bk) \right] &=& \Pi_1^\dagger (-\bk) \nn
 \left[ \Delta_1 (\bk), P_3^\dagger (\bk) \right] &=& \Pi_1 (\bk) \nn
 \left[ \Pi_1 (\bk), \Pi_1^\dagger (\bk)  \right] &=& N_1 (\bk) - N_3 (\bk) \nn
 \left[ \Pi_1 (\bk), P_1^\dagger (\bk) \right] &=& - \Delta_1^\dagger (-\bk) \nn
 \left[ \Pi_1 (\bk), P_3 (\bk) \right] &=& \Delta_1 (\bk) \nn
 \left[ P_1 (\bk), P_1^\dagger (\bk) \right] &=& N_1 (\bk) + N_1 (-\bk) -1 \nn
 \left[ P_3 (\bk), P_3^\dagger (\bk) \right] &=& N_3 (\bk) + N_3 (-\bk) -1
 \label{commutation}
\eea
and some others that follow under $\bk \rightarrow -\bk$ and/or Hermitian conjugates.

A similar set of relations follow from $1 \rightarrow 2$ and $3 \rightarrow 4$, yielding a second SU(4) algebra. However, we will
not need these because we will always assume $\Delta_2 = -\Delta_1$ and $\Pi_2 = -\Pi_1$.

Then we can use $H_{MF}$ in Eq.~(\ref{hmf}) to obtain the equations of motion of the average values of the operators
in Eq.~(\ref{su4}):
\begin{footnotesize}
\begin{align}
\frac{d\Delta_1(\bk)}{dt}&=-i \left[ -\epsilon_1(\bk)\Delta_1(\bk)-\epsilon_1(\bk)\Delta_1(\bk)+\frac{(3J-V)}{2}\Delta_2(N_1(\bk)+N_3(-\bk))+\frac{3J+V}{2}(-\Pi_2^\ast P_1(\bk)-\Pi_2 P_3(\bk))\right]  \nn
\frac{d\Delta_1(-\bk)}{dt}&=-i \left[ -\epsilon_1(-\bk)\Delta_1(-\bk)-\epsilon_1(-\bk)\Delta_1(-\bk)+\frac{(3J-V)}{2}\Delta_2(N_1(-\bk)+N_3(\bk))+\frac{3J+V}{2}(-\Pi_2^\ast P_1(\bk)-\Pi_2 P_3(\bk))\right]  \nn
\frac{d\Delta_1^{\dagger}(\bk)}{dt}&=-i \left[  \epsilon_1(\bk)\Delta_1^\dagger(\bk)+\epsilon_1(\bk)\Delta_1^\dagger(\bk)-\frac{(3J-V)}{2}\Delta_2^\ast(N_1(\bk)+N_3(-\bk))-\frac{3J+V}{2}(-\Pi_2 P_1^{\dagger}(\bk)-\Pi_2^\ast P_3^{\dagger}(\bk))\right]  \nn
\frac{d\Delta_1^\dagger(-\bk)}{dt}&=-i \left[ \epsilon_1(-\bk)\Delta_1^\dagger(-\bk)+\epsilon_1(-\bk)\Delta_1^\dagger(-\bk)-\frac{(3J-V)}{2}\Delta_2^\ast(N_1(-\bk)+N_3(\bk))-\frac{3J+V}{2}(-\Pi_2 P_1^\dagger(\bk)-\Pi_2^\ast P_3^{\dagger}(\bk))\right]  \nn
\frac{d\Pi_1(\bk)}{dt}&=-i \left[ -\epsilon_1(\bk) \Pi_1(\bk)+\epsilon_1(-\bk)\Pi_1(\bk)+\frac{(3J-V)}{2}(-\Delta_2P_3^{\dagger}(\bk)+\Delta_2^\ast P_1(\bk))+\frac{3J+V}{2}\Pi_2(N_1(\bk)-N_3(\bk))\right]  \nn
\frac{d\Pi_1(-\bk)}{dt}&=-i \left[ -\epsilon_1(-\bk) \Pi_1(-\bk)+\epsilon_1(\bk)\Pi_1(-\bk)+\frac{(3J-V)}{2}(-\Delta_2P_3^{\dagger}(\bk)+\Delta_2^\ast P_1(\bk))+\frac{3J+V}{2}\Pi_2(N_1(-\bk)-N_3(-\bk))\right]  \nn
\frac{d\Pi_1^\dagger(\bk)}{dt}&=-i \left[ \epsilon_1(\bk) \Pi_1^\dagger(\bk)-\epsilon_1(-\bk)\Pi_1^\dagger(\bk)-\frac{(3J-V)}{2}(-\Delta_2^\ast P_3(\bk)+\Delta_2 P_1^\dagger(\bk))-\frac{3J+V}{2}\Pi_2^\ast(N_1(\bk)-N_3(\bk))\right]  \nn
\frac{d\Pi_1^\dagger(-\bk)}{dt}&=-i \left[\epsilon_1(-\bk) \Pi_1^\dagger(-\bk)-\epsilon_1(\bk)\Pi_1^\dagger(-\bk)-\frac{(3J-V)}{2}(-\Delta_2^\ast P_3(\bk)+\Delta_2 P_1^\dagger(\bk))-\frac{3J+V}{2}\Pi_2^\ast(N_1(-\bk)-N_3(-\bk))\right]  \nn
\frac{dP_1(\bk)}{dt}&=-i \left[ -(\epsilon_1(\bk)+\epsilon_1(-\bk)) P_1(\bk)+\frac{(3J-V)}{2}\Delta_2(\Pi_1(-\bk)+\Pi_1(\bk))-\frac{3J+V}{2}\Pi_2(\Delta_1(-\bk)+\Delta_1(\bk))\right]  \nn
\frac{dP_1^\dagger(\bk)}{dt}&=-i \left[ (\epsilon_1(\bk)+\epsilon_1(-\bk)) P_1^\dagger(\bk)-\frac{(3J-V)}{2}\Delta_2^\ast(\Pi_1^\dagger(-\bk)+\Pi_1^\dagger(\bk))+\frac{3J+V}{2}\Pi_2^\ast (\Delta_1^\dagger(-\bk)+\Delta_1^\dagger(\bk))\right]  \nn
\frac{dP_3(\bk)}{dt}&=-i \left[ -(\epsilon_1(-\bk)+\epsilon_1(\bk)) P_3(\bk)+\frac{(3J-V)}{2}\Delta_2(\Pi_1^\dagger(\bk)+\Pi_1^\dagger(-\bk))-\frac{3J+V}{2}\Pi_2^\ast(\Delta_1(\bk)+\Delta_1(-\bk))\right]  \nn
\frac{dP_3^\dagger(\bk)}{dt}&=-i \left[ (\epsilon_1(-\bk)+\epsilon_1(\bk)) P_3^\dagger(\bk)-\frac{(3J-V)}{2}\Delta_2^\ast(\Pi_1(\bk)+\Pi_1(-\bk))+\frac{3J+V}{2}\Pi_2(\Delta_1^\dagger(\bk)+\Delta_1^\dagger(-\bk))\right] \nn
\frac{dN_1(\bk)}{dt}&=-i\left[\frac{3J-V}{2}(\Delta_2^\ast\Delta_1(\bk)-\Delta_2\Delta_1^{\dagger}(\bk))+\frac{3J+V}{2}(\Pi_2^\ast\Pi_1(\bk)-\Pi_2\Pi_1^{\dagger}(\bk))\right] \nn
\frac{dN_1(-\bk)}{dt}&=-i\left[\frac{3J-V}{2}(\Delta_2^\ast\Delta_1(-\bk)-\Delta_2\Delta_1^{\dagger}(-\bk))+\frac{3J+V}{2}(\Pi_2^\ast\Pi_1(-\bk)-\Pi_2\Pi_1^{\dagger}(-\bk))\right] \nn
\frac{dN_3(\bk)}{dt}&=-i\left[\frac{3J-V}{2}(\Delta_2^\ast\Delta_1(-\bk)-\Delta_2\Delta_1^{\dagger}(-\bk))+\frac{3J+V}{2}(-\Pi_2^\ast\Pi_1(\bk)+\Pi_2\Pi_1^{\dagger}(\bk))\right] \nn
\frac{dN_3(-\bk)}{dt}&=-i\left[\frac{3J-V}{2}(\Delta_2^\ast\Delta_1(\bk)-\Delta_2\Delta_1^{\dagger}(\bk))+\frac{3J+V}{2}(-\Pi_2^\ast\Pi_1(-\bk)+\Pi_2\Pi_1^{\dagger}(-\bk))\right] 
\label{eom}
\end{align}
\end{footnotesize}

Also note that the operators $P_1+P_2$ and $P_3 + P_4$ (and their Hermitian conjugates) commute with the original Hamiltonian $H$ for $V=0$; these operators generate the pseudospin symmetry between the SC and CDW order parameters.

One quality that remains constant during the oscillation is the mean field energy $\left\langle H_{MF}\right\rangle$, if the interaction parameters are constant with respect to time, like during the time after the quench. This can be easily verified using  Eq.~(\ref{eom}). And this can be used to check the validity of the numerics.

\bea
&&\left\langle H_{MF}\right\rangle=\sum_{\bk}\Biggl[ \epsilon_1(\bk)N_1(\bk)+\epsilon_1(-\bk)N_3(\bk)+\epsilon_2(\bk)N_2(\bk)+\epsilon_2(-\bk)N_4(\bk)\Biggr]\nn
&&+\frac{3J-V}{2}\left(\Delta_1\Delta_2^\ast+\Delta_2\Delta_1^\ast\right)+\frac{3J+V}{2}\left(\Pi_1\Pi_2^\ast+\Pi_1^\ast\Pi_2\right)
\eea

Notice that we have added back some subtractions terms to Eq.~(\ref{hmf}). Then the time derivative of $\left\langle H_{MF}\right\rangle$ becomes
\begin{footnotesize}
\bea
\frac{1}{2}\frac{d\left\langle H_{MF}\right\rangle}{dt}&=&\sum_{\bk}-i\epsilon_1(\bk)\left[\frac{3J-V}{2}(\Delta_2^\ast\Delta_1(-\bk)-\Delta_2\Delta_1^{\dagger}(-\bk))+\frac{3J+V}{2}(\Pi_2^\ast\Pi_1(-\bk)-\Pi_2\Pi_1^{\dagger}(-\bk))\right] \nn
&-&i\epsilon_1(-\bk)\left[\frac{3J-V}{2}(\Delta_2^\ast\Delta_1(-\bk)-\Delta_2\Delta_1^{\dagger}(-\bk))+\frac{3J+V}{2}(-\Pi_2^\ast\Pi_1(\bk)+\Pi_2\Pi_1^{\dagger}(\bk))\right] \nn
&-&i\frac{3J-V}{2}\Delta_2^\ast\left[ -\epsilon_1(\bk)\Delta_1(\bk)-\epsilon_1(\bk)\Delta_1(\bk)+\frac{(3J-V)}{2}\Delta_2(N_1(\bk)+N_3(-\bk))+\frac{3J+V}{2}(-\Pi_2^\ast P_1(\bk)-\Pi_2 P_3(\bk))\right]  \nn
&-&i\frac{3J-V}{2}\Delta_2\left[  \epsilon_1(\bk)\Delta_1^\ast(\bk)+\epsilon_1(\bk)\Delta_1^\ast(\bk)-\frac{(3J-V)}{2}\Delta_2^\ast(N_1(\bk)+N_3(-\bk))-\frac{3J+V}{2}(-\Pi_2 P_1^{\ast}(\bk)-\Pi_2^\ast P_3^{\ast}(\bk))\right]  \nn
&-&i\frac{3J+V}{2}\Pi_2^\ast\left[ -\epsilon_1(\bk) \Pi_1(\bk)+\epsilon_1(-\bk)\Pi_1(\bk)+\frac{(3J-V)}{2}(-\Delta_2P_3^{\ast}(\bk)+\Delta_2^\ast P_1(\bk))+\frac{3J+V}{2}\Pi_2(N_1(\bk)-N_3(\bk))\right]  \nn
&-&i\frac{3J+V}{2}\Pi_2\left[ -\epsilon_1(\bk) \Pi_1^\ast(\bk)+\epsilon_1(-\bk)\Pi_1^\ast(\bk)-\frac{(3J-V)}{2}(-\Delta_2^\ast P_3(\bk)+\Delta_2 P_1^\ast(\bk))-\frac{3J+V}{2}\Pi_2^\ast(N_1(\bk)-N_3(\bk))\right]\nn
&=&0
\eea
\end{footnotesize}
where we have assumed time-independence of $J$ and $V$, which is true in the quench case.

\newpage

\section{More simulations of the hot spot model}
\label{app:simulation}

Here we describe multiple additional initial conditions for the hot spot model as a supplement to the main text. 

\begin{figure}[H]
\begin{center}
\includegraphics[width=3in]{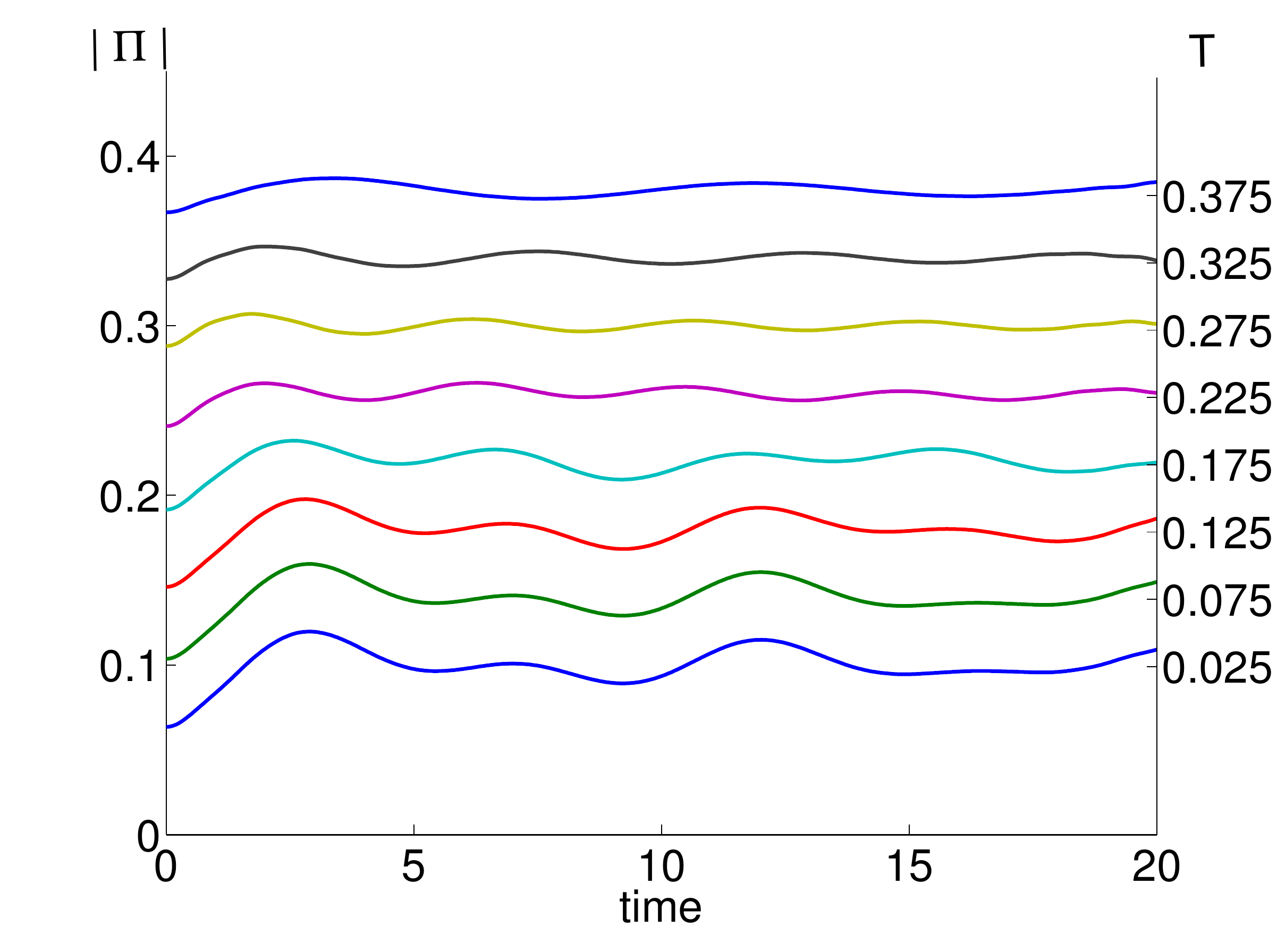}
\includegraphics[width=3in]{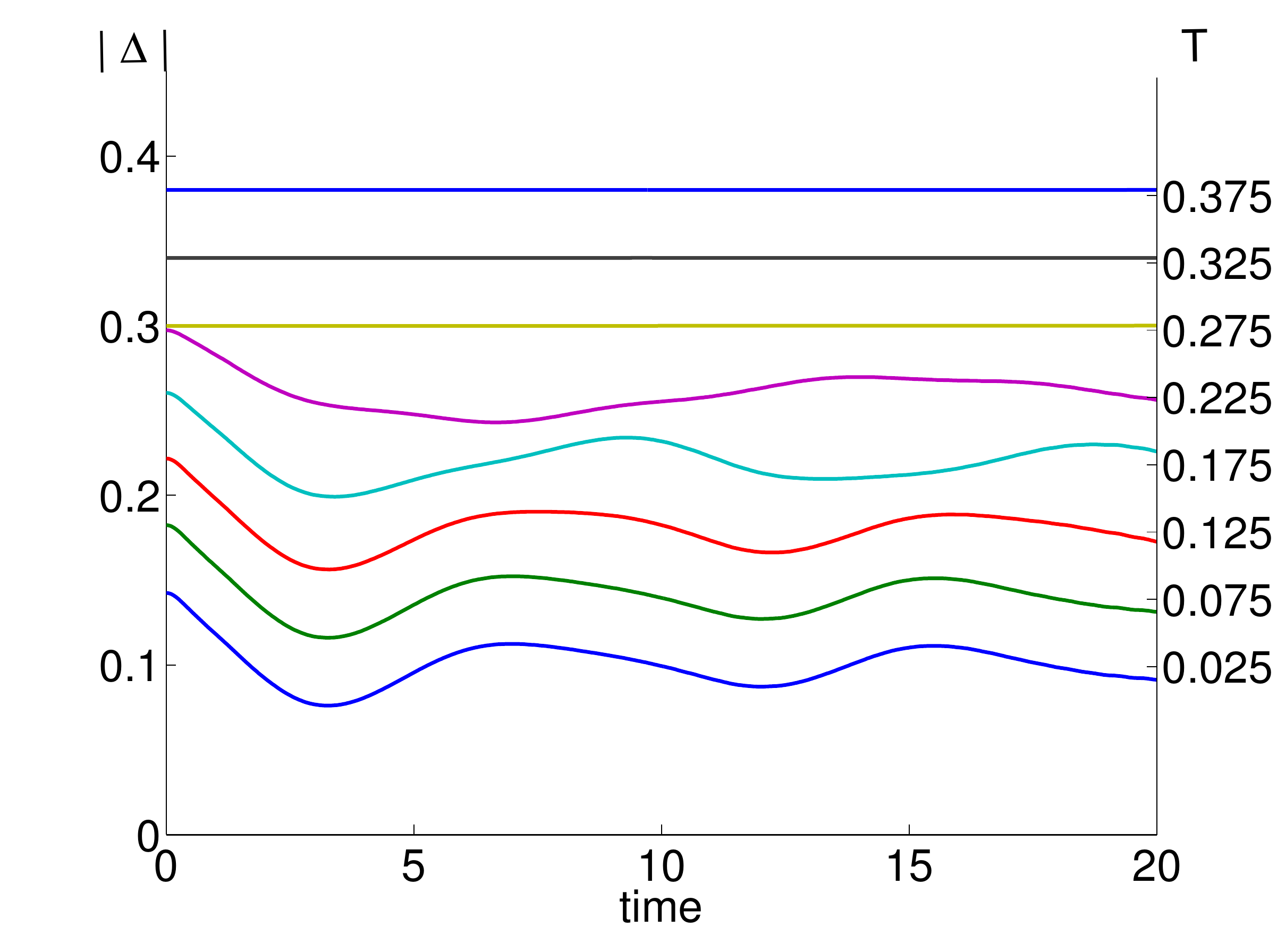}
\caption{Oscillation of CDW order parameter $\Pi$(left) and SC order parameter $\Delta$(right) as a function of time in the quench case, from low temperature at the bottom to high temperature at the top, temperatures are taken from 0.025 to 0.375 with 0.05 step. The initial value  $J_0=1.2$, $V_0=0.9$, the quench is taken as $\Delta J=0, \Delta V=0.1$.  The initial $T_c$ at equilibrium can be computed to be 0.25, the final $T_c$ to be 0.2.}
\label{qv01}
\end{center}
\end{figure}

In Fig.~\ref{qv01}, we considered the quench case, where $\Delta V=0.1$,$\Delta J=0$. We still have amplitude enhancement below $T_c$, although the enhancement is not that big. Besides, there is also a phase shift upon crossing $T_c$.  

\begin{figure}[H]
\begin{center}
\includegraphics[width=3in]{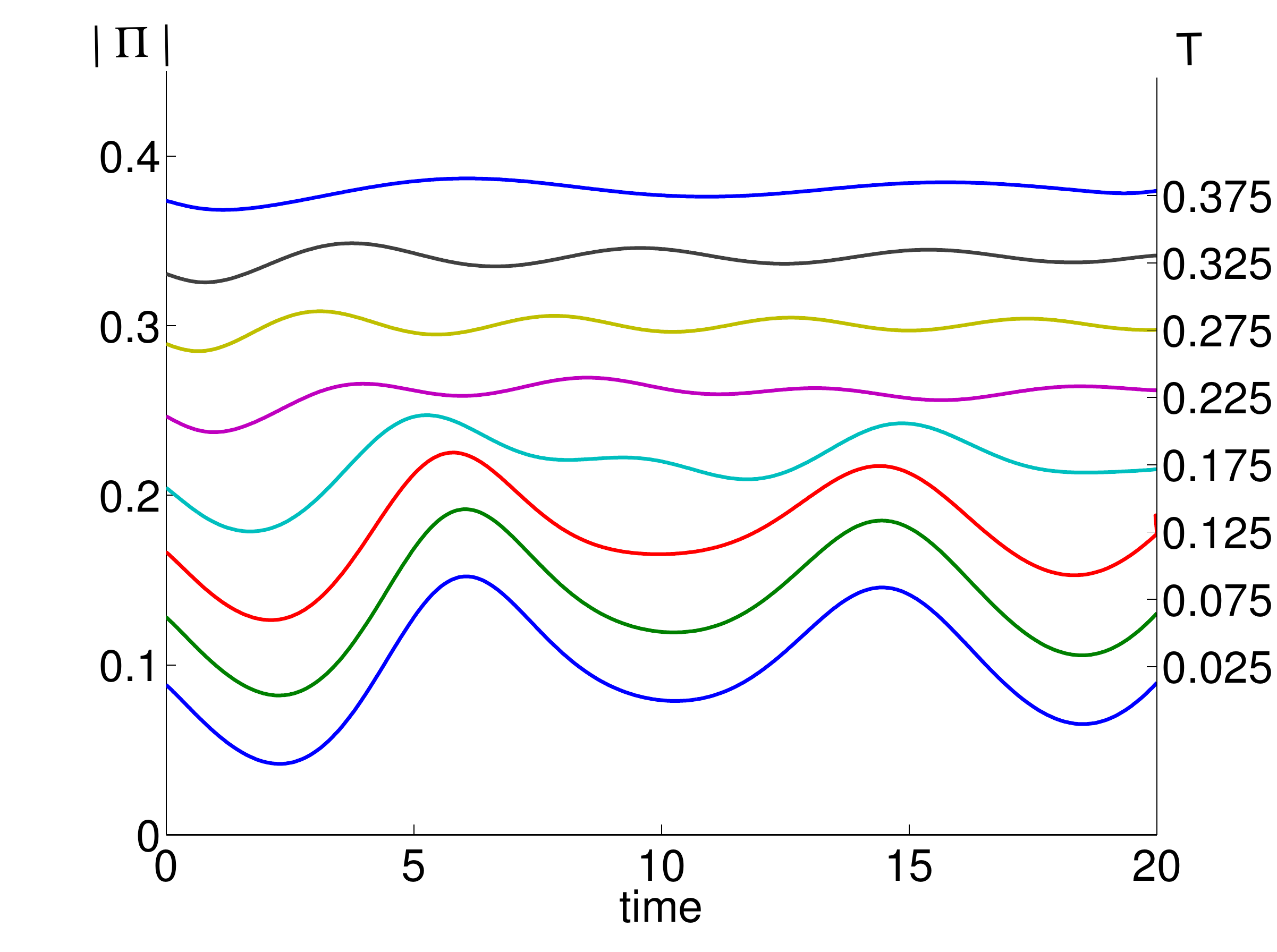}
\includegraphics[width=3in]{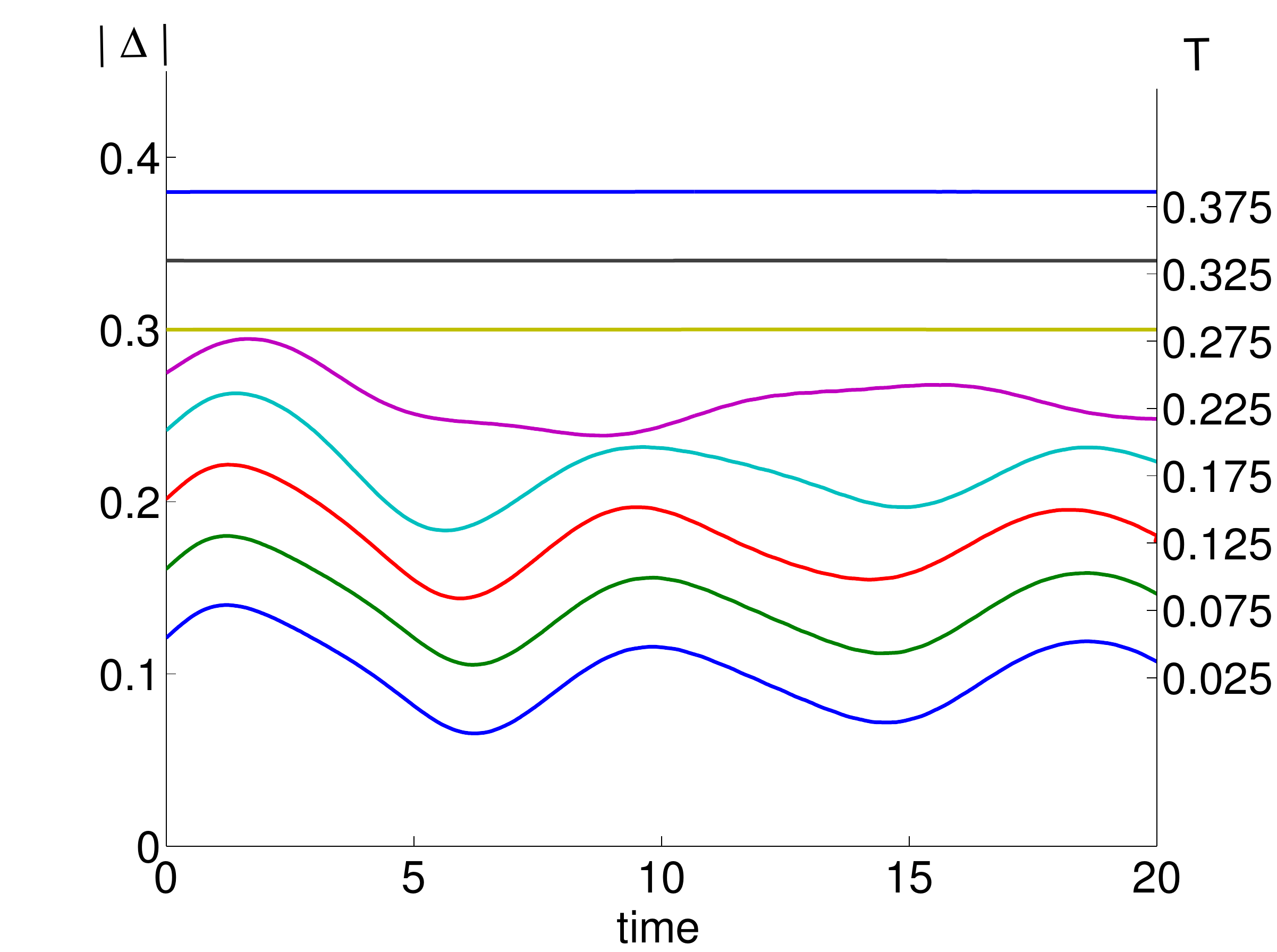}
\caption{Oscillation of CDW order parameter $\Pi$(left) and SC order parameter $\Delta$(right) as a function of time in the pulse case, from low temperature at the bottom to high temperature at the top, temperatures are taken from 0.025 to 0.375 with 0.05 step. The initial value  $J_0=1.2$, $V_0=0.9$, the pulse is taken as $\Delta J=0, \Delta V=-0.1$, $\omega=1$.  The initial $T_c$ at equilibrium can be computed to be 0.25, at the largest derivation $V=V_0+\Delta V$, the corresponding equilibrium $T_c$ to be 0.33.}
\label{pv-01}
\end{center}
\end{figure}

\begin{figure}[H]
\begin{center}
\includegraphics[width=2in]{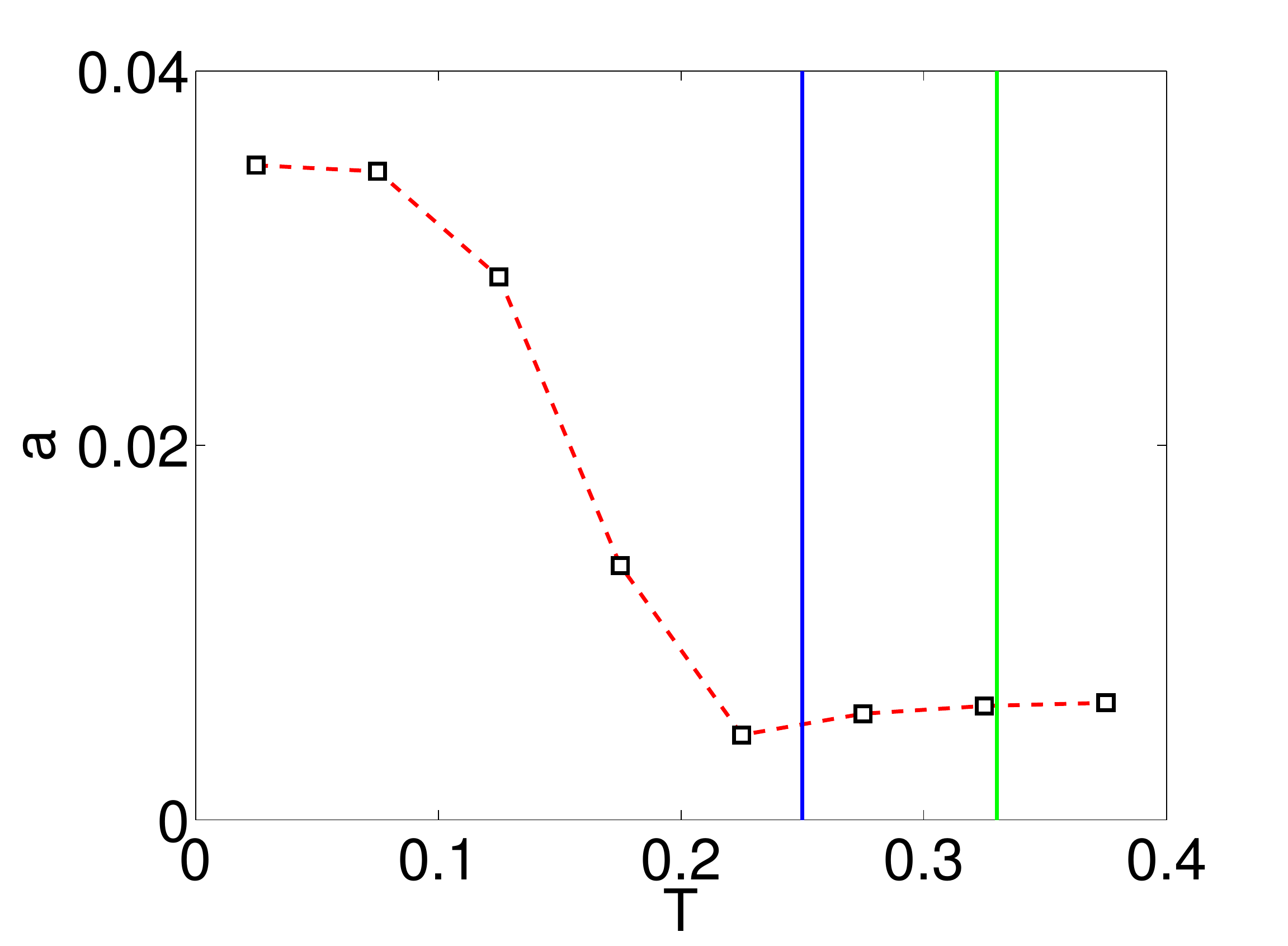}
\includegraphics[width=2in]{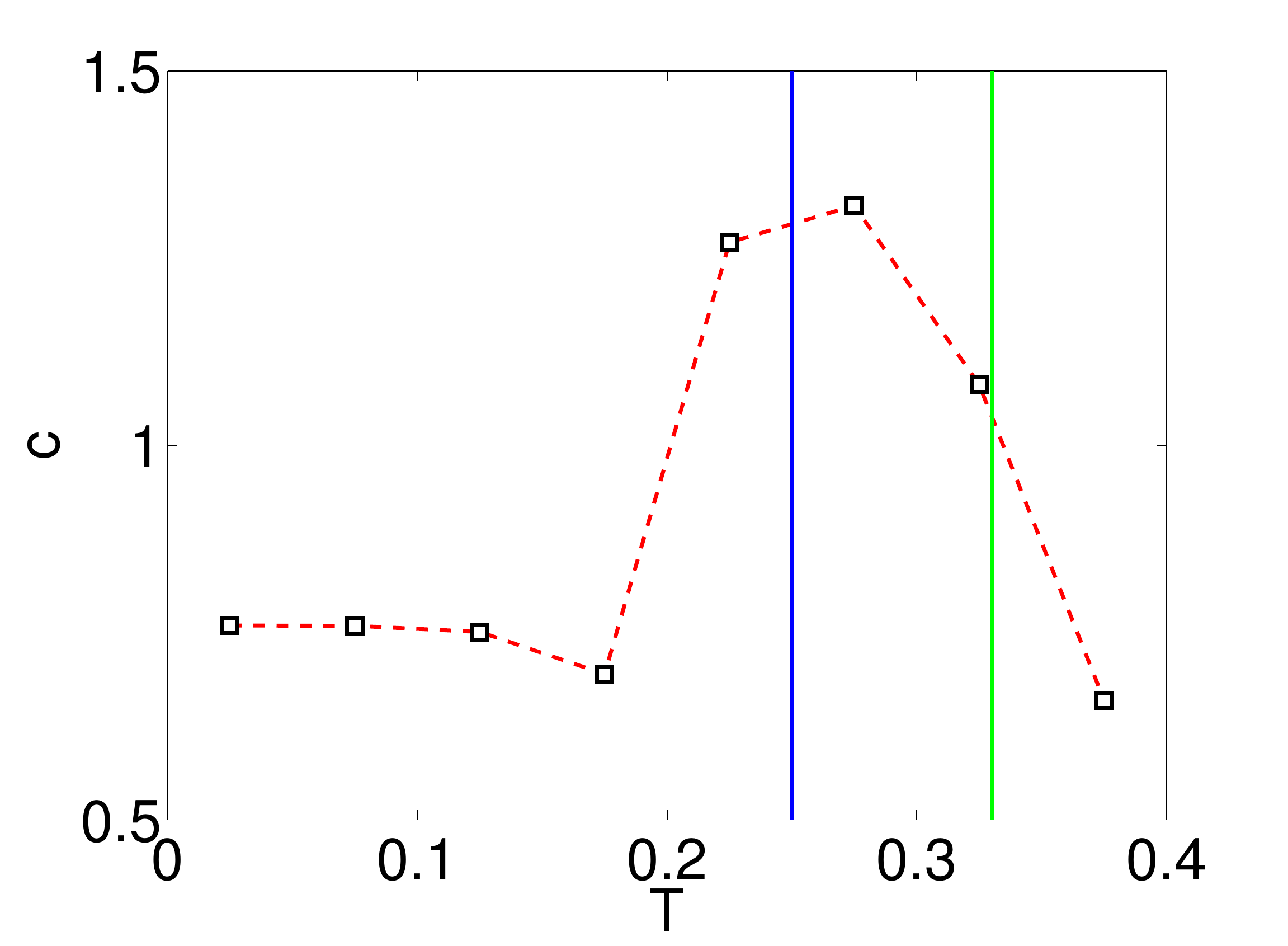}
\includegraphics[width=2in]{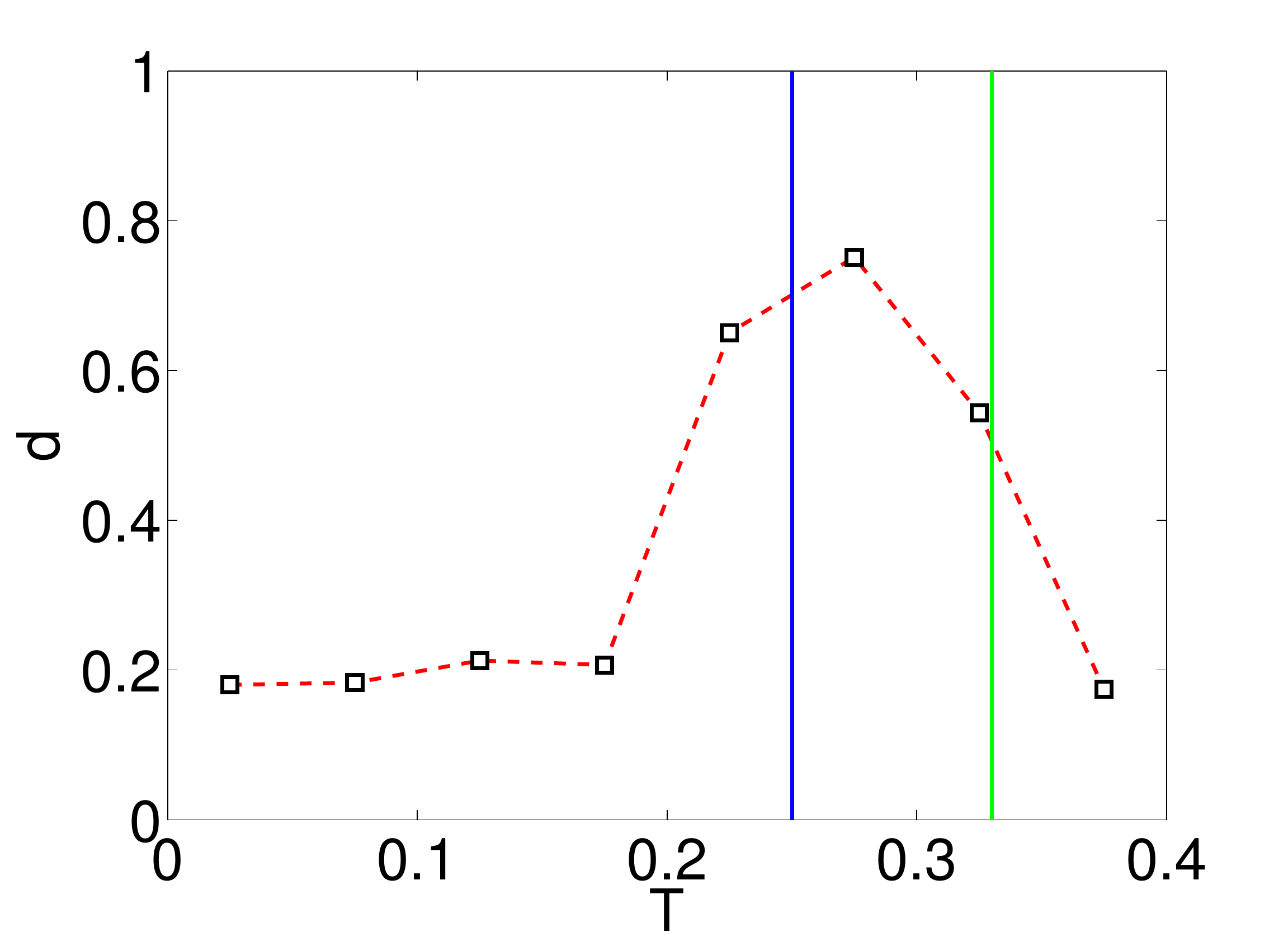}
\caption{Left to right: amplitude a, frequency c and phase d of the fit Eq.~\ref{fit} fitting the data in Fig.~\ref{pv-01} as a function of temperature. The blue line denotes initial equilibrium $T_c=0.25$,the green line denotes at the largest derivation $V=V_0+\Delta V$, the equilibrium $T_c=0.33$.}
\label{pv-01fitpara}
\end{center}
\end{figure}

\begin{figure}[H]
\begin{center}
\includegraphics[width=3in]{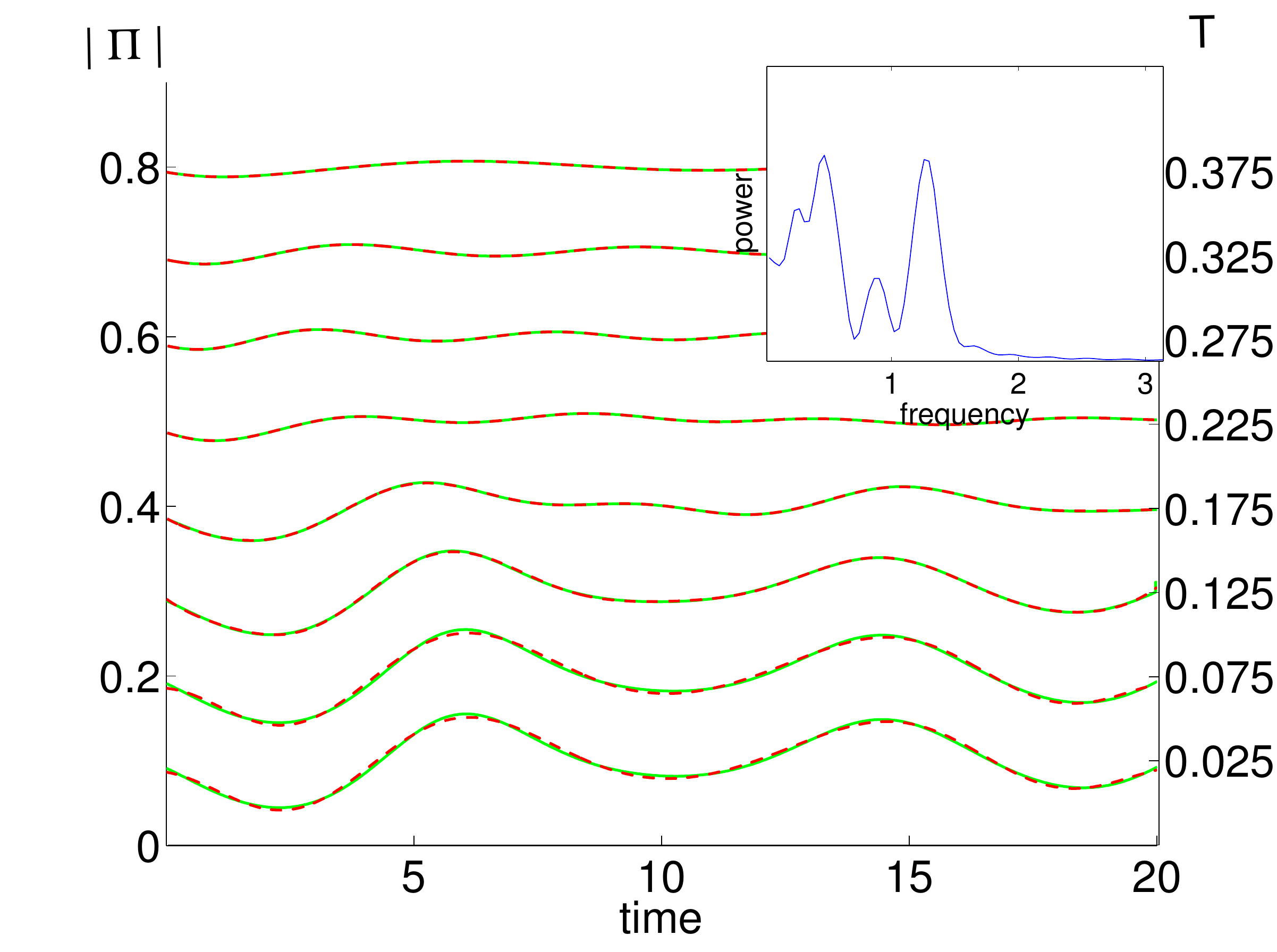}
\includegraphics[width=3in]{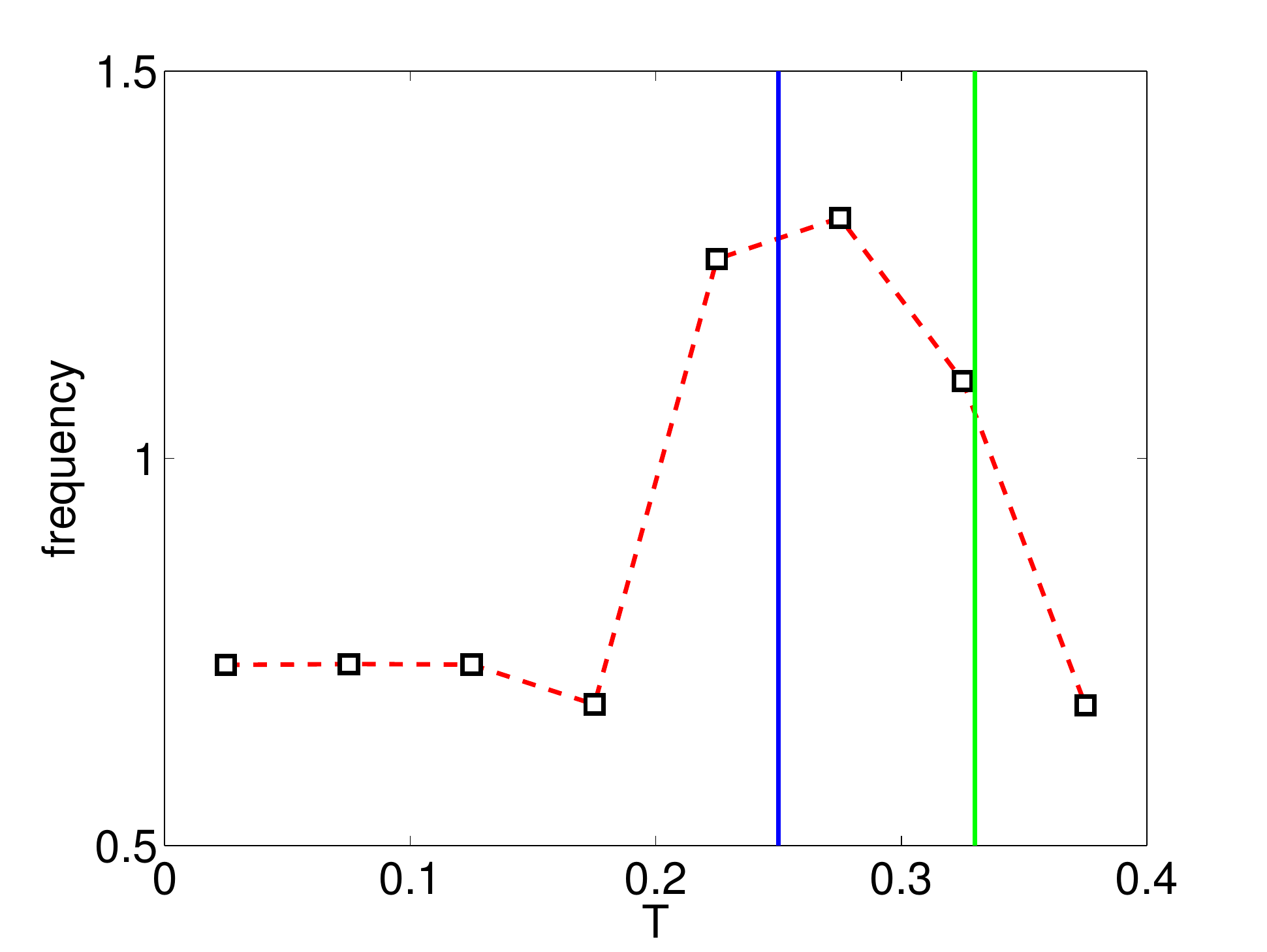}
\caption{Left: spectral fitting of the left panel in Fig.\ref{pv-01}. The inserted figure is the spectral power at $T=0.225$ near $T_c$. Right:  peak frequency from the left panel as a function of temperature.
 }
\label{pv-01fitspec}
\end{center}
\end{figure}

In Figs.~\ref{pv-01}--\ref{pv-01fitspec}, we showed the pulse case, where $\Delta V=-0.1$, $\Delta J=0$. We have used both decayed sinusoidal fitting and spectral analysis to characterize oscillation properties. Fitting parameters are found to be similar to those in Figs.~\ref{pv01}--\ref{pv01fitspec}, although the phase shift has the opposite sign. In the spectral analysis, near $T_c$ the side peaks gets larger and become the main peak: that is why we observe 
frequency change against temperature in this region.

\begin{figure}[H]
\begin{center}
\includegraphics[width=3in]{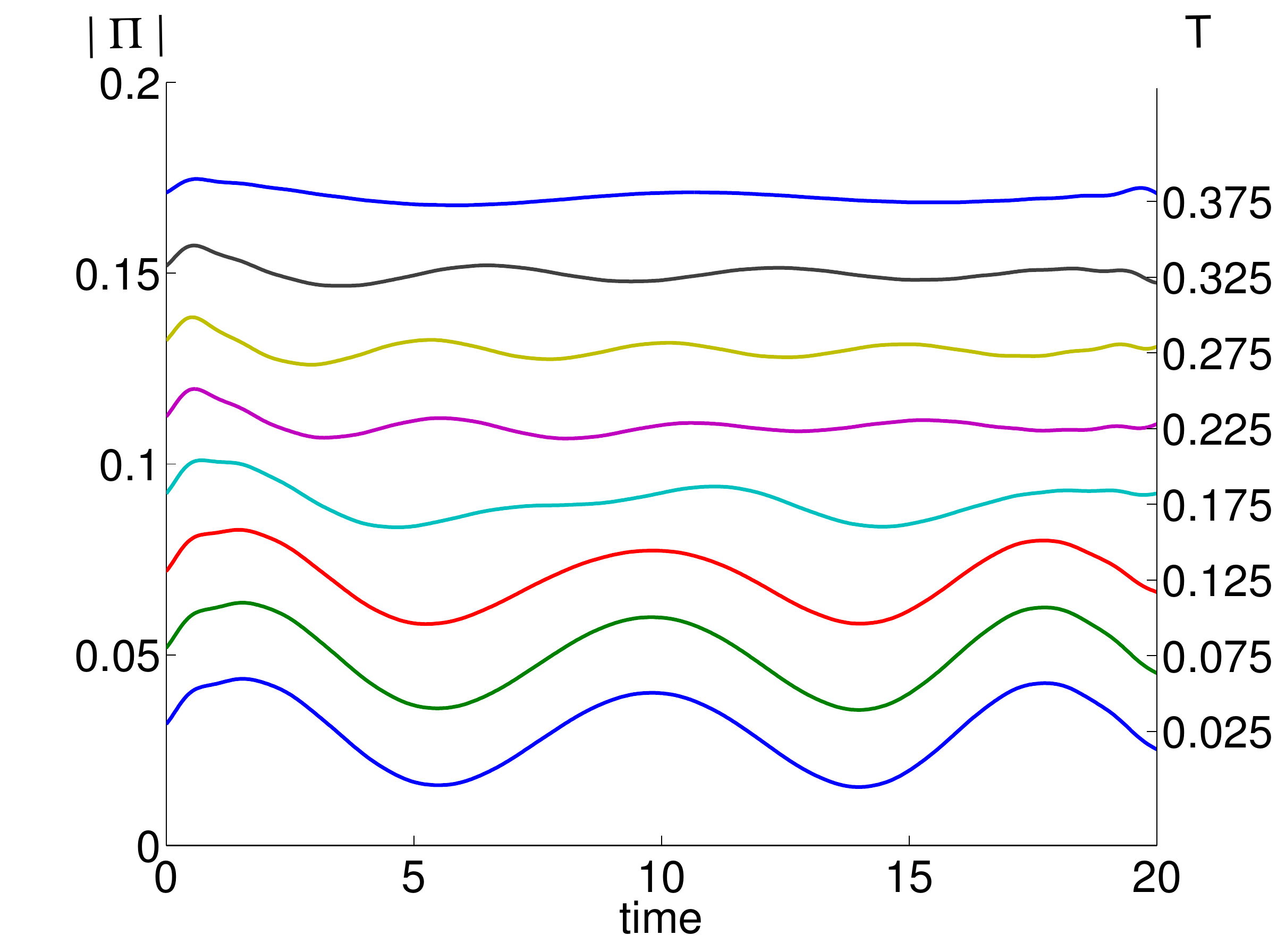}
\includegraphics[width=3in]{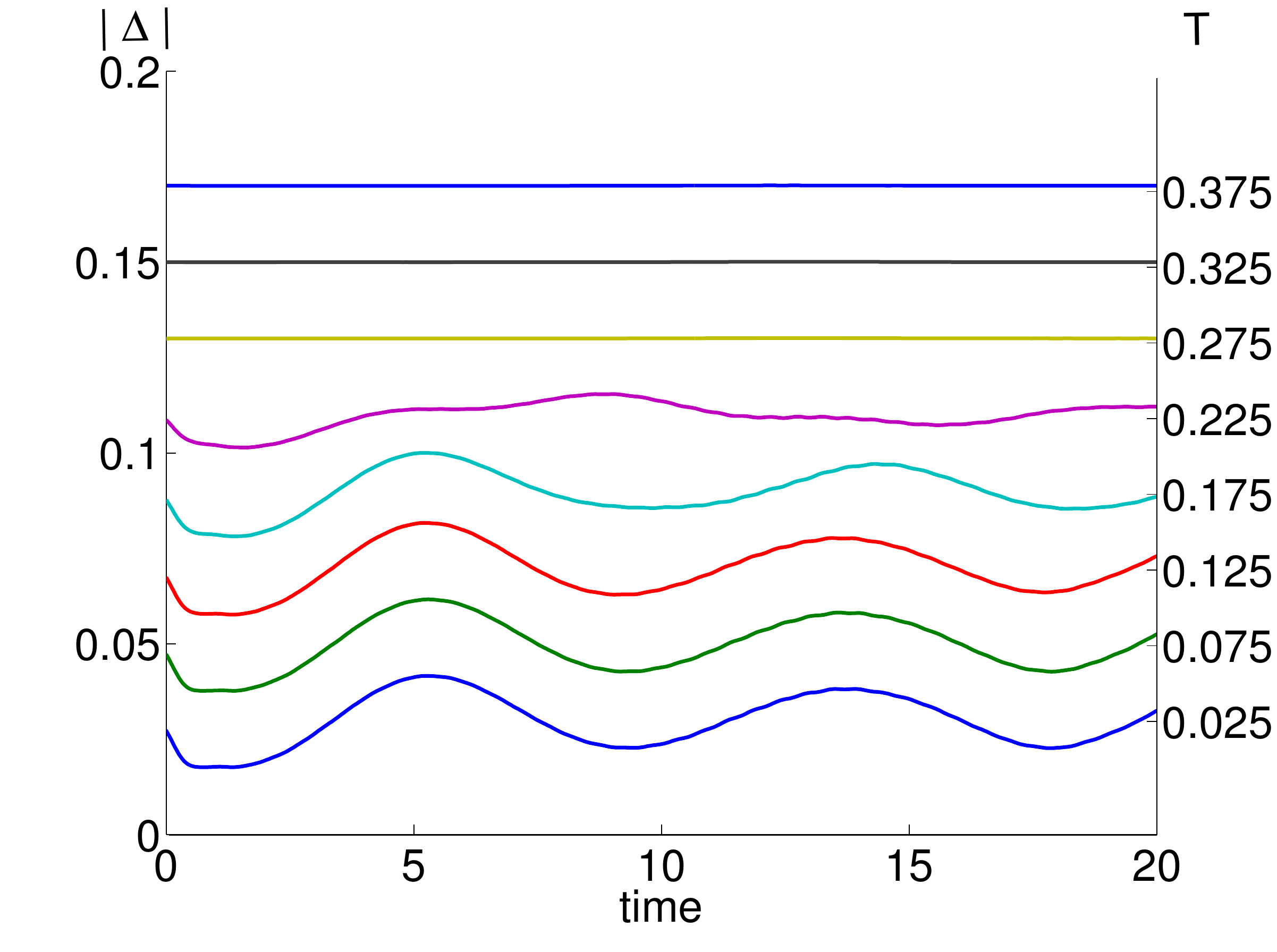}
\caption{Oscillation of CDW order parameter $\Pi$(left) and SC order parameter $\Delta$(right) as a function of time in the pulse case, from low temperature at the bottom to high temperature at the top, temperatures are taken from 0.025 to 0.375 with 0.05 step. The initial value  $J_0=1.2$, $V_0=0.9$, the pulse is taken as $\Delta J=0, \Delta V=0.1$, $\omega=4$. The initial $T_c$ at equilibrium can be computed to be 0.25, at the largest derivation $V=V_0+\Delta V$, the corresponding equilibrium $T_c$ to be 0.2.}
\label{pv01fre4}
\end{center}
\end{figure}

\begin{figure}[H]
\begin{center}
\includegraphics[width=2in]{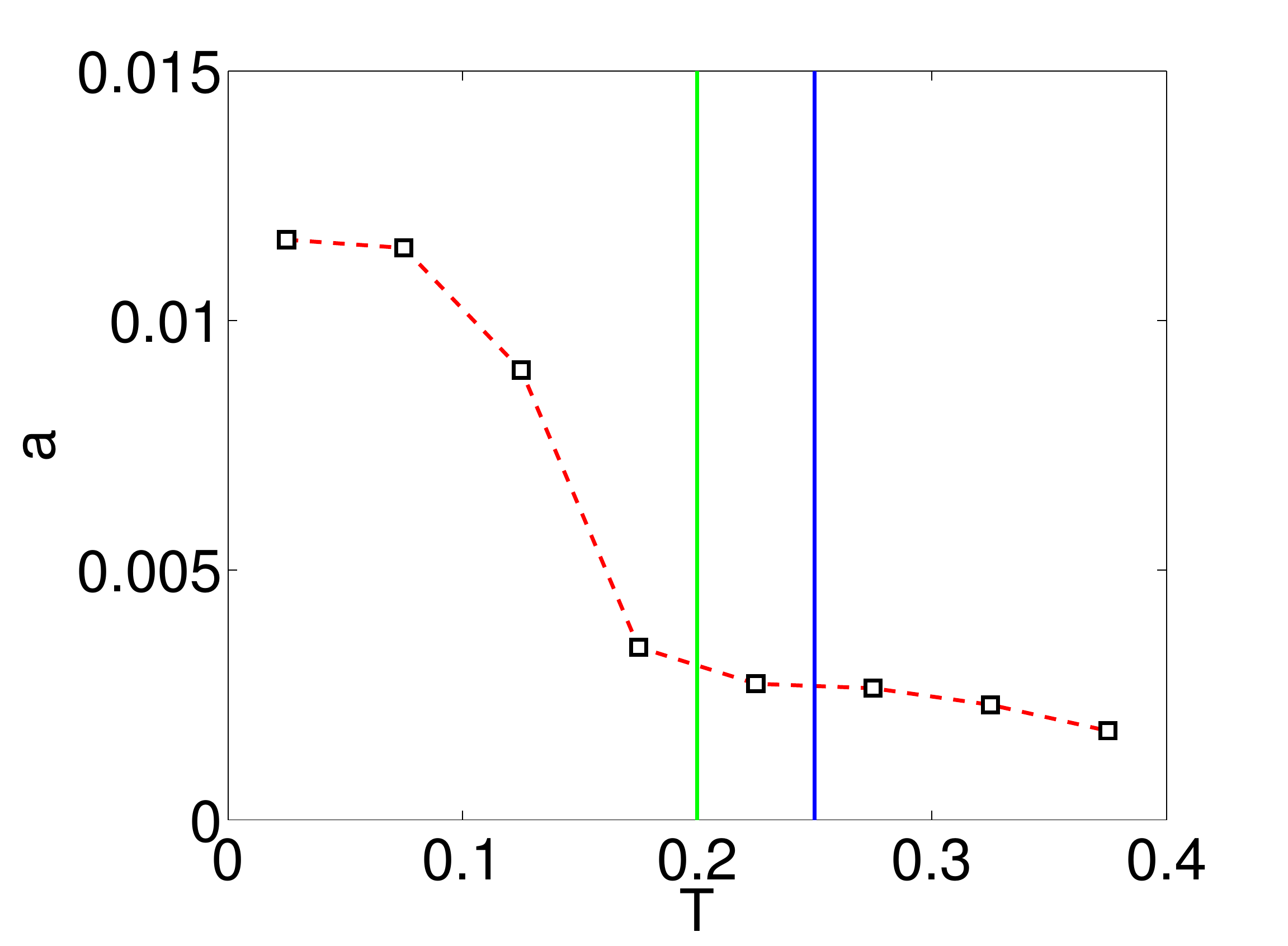}
\includegraphics[width=2in]{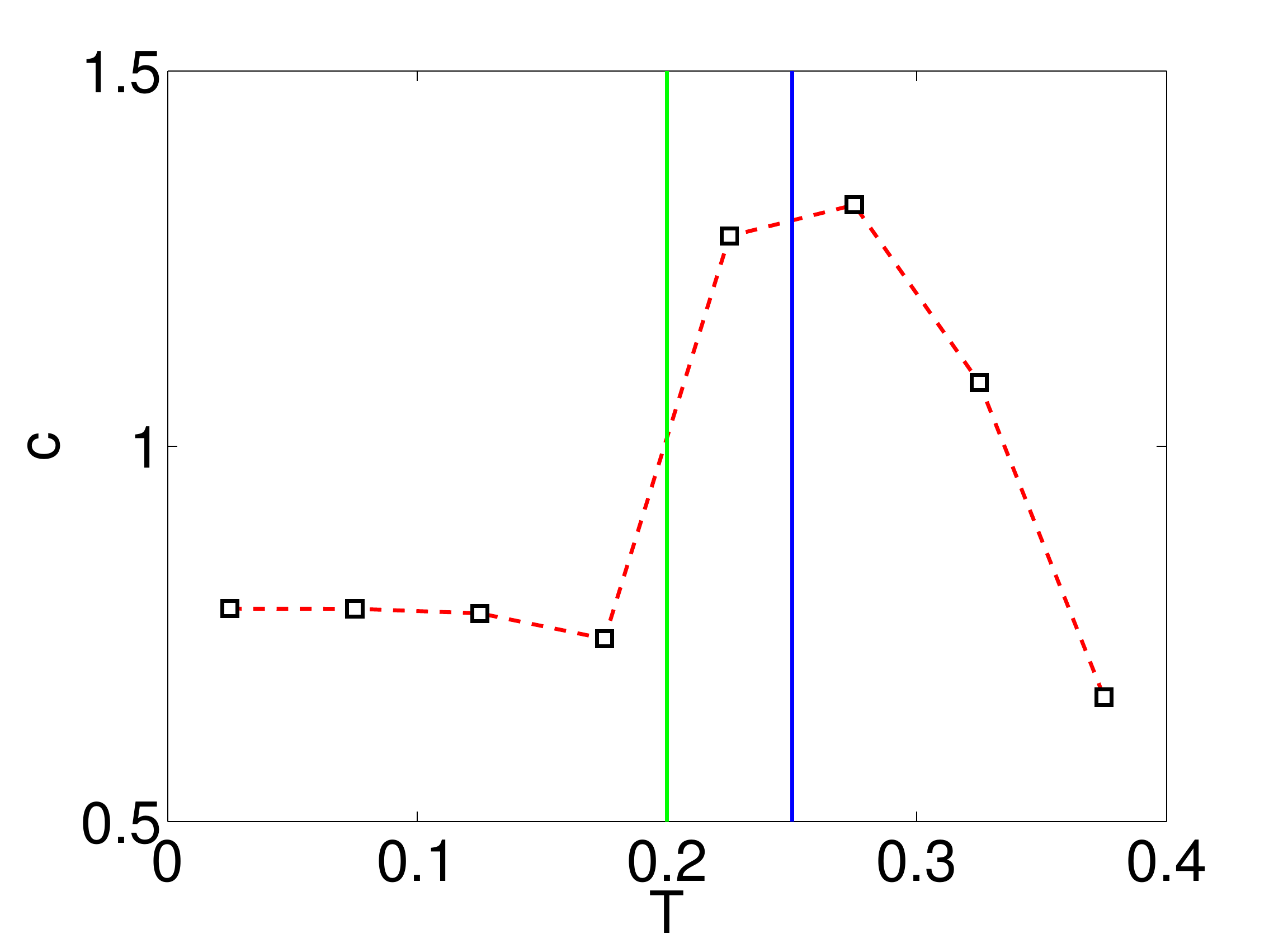}
\includegraphics[width=2in]{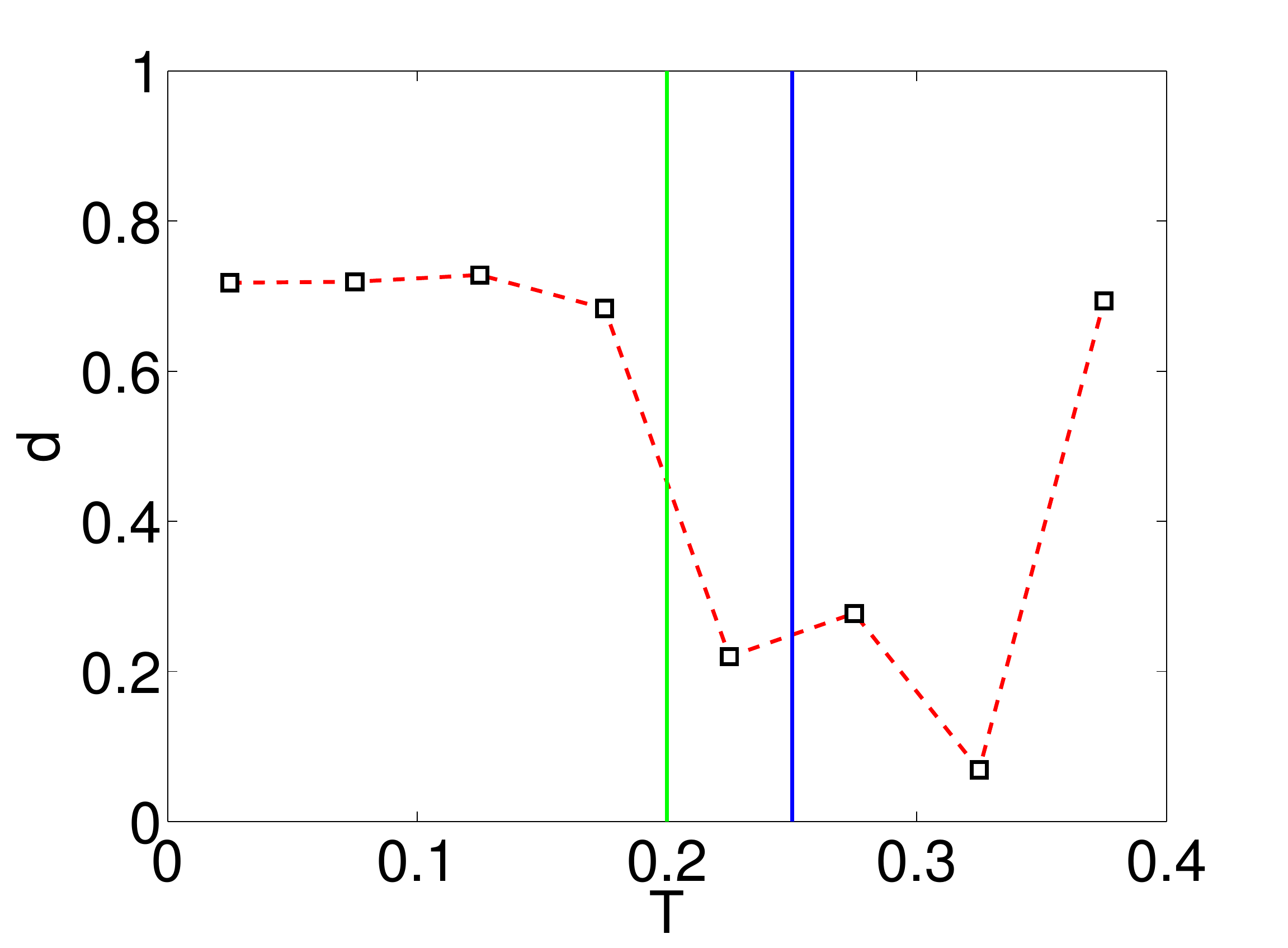}
\caption{Left to right: amplitude a, frequency c and phase d of the fit Eq.~\ref{fit} fitting the data in Fig.~\ref{pv01fre4} as a function of temperature. The blue line denotes initial equilibrium $T_c=0.25$,the green line denotes at the largest derivation $V=V_0+\Delta V$, the equilibrium $T_c=0.2$.}
\label{pv01fre4fitpara}
\end{center}
\end{figure}

\begin{figure}[H]
\begin{center}
\includegraphics[width=3in]{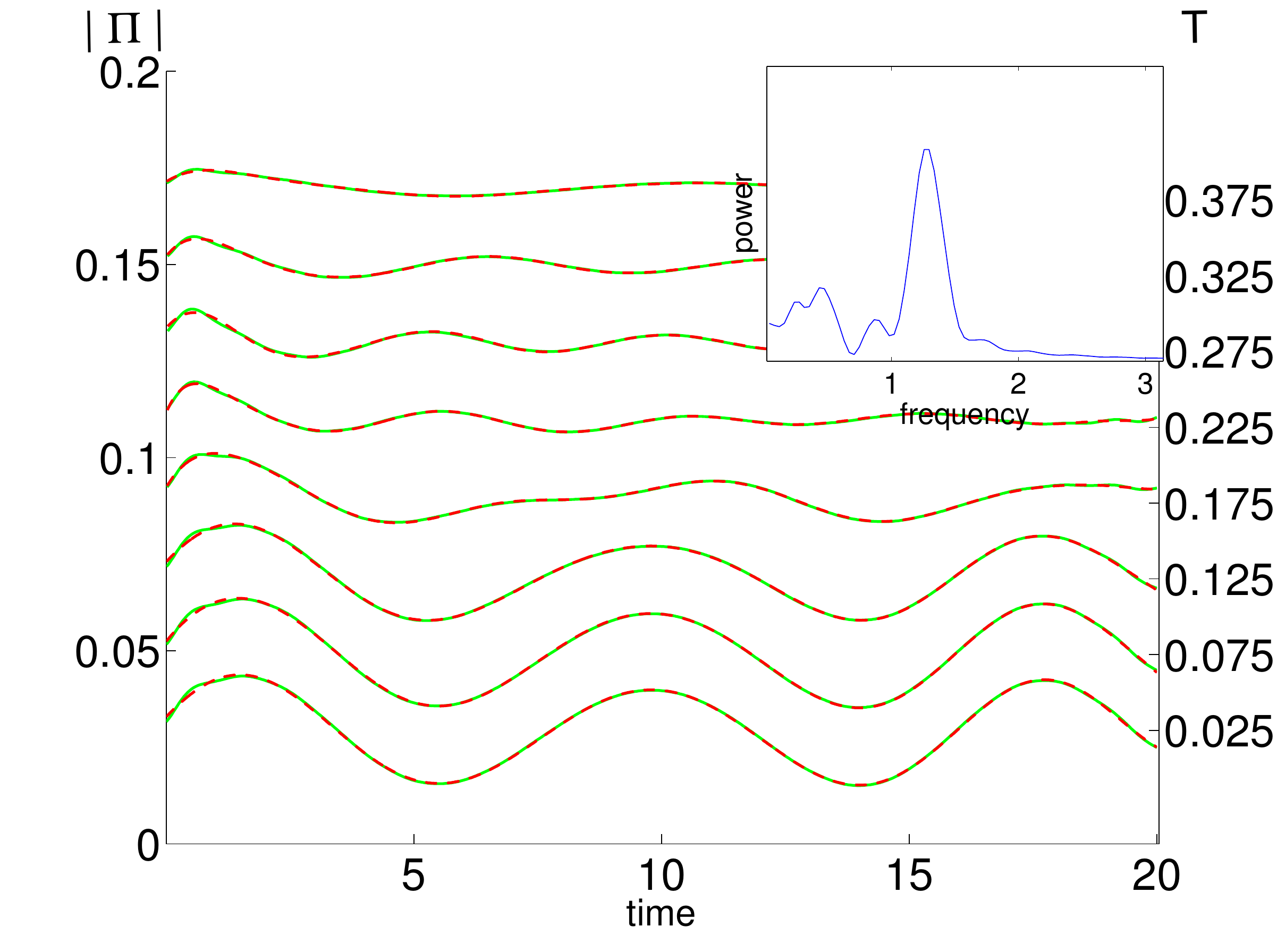}
\includegraphics[width=3in]{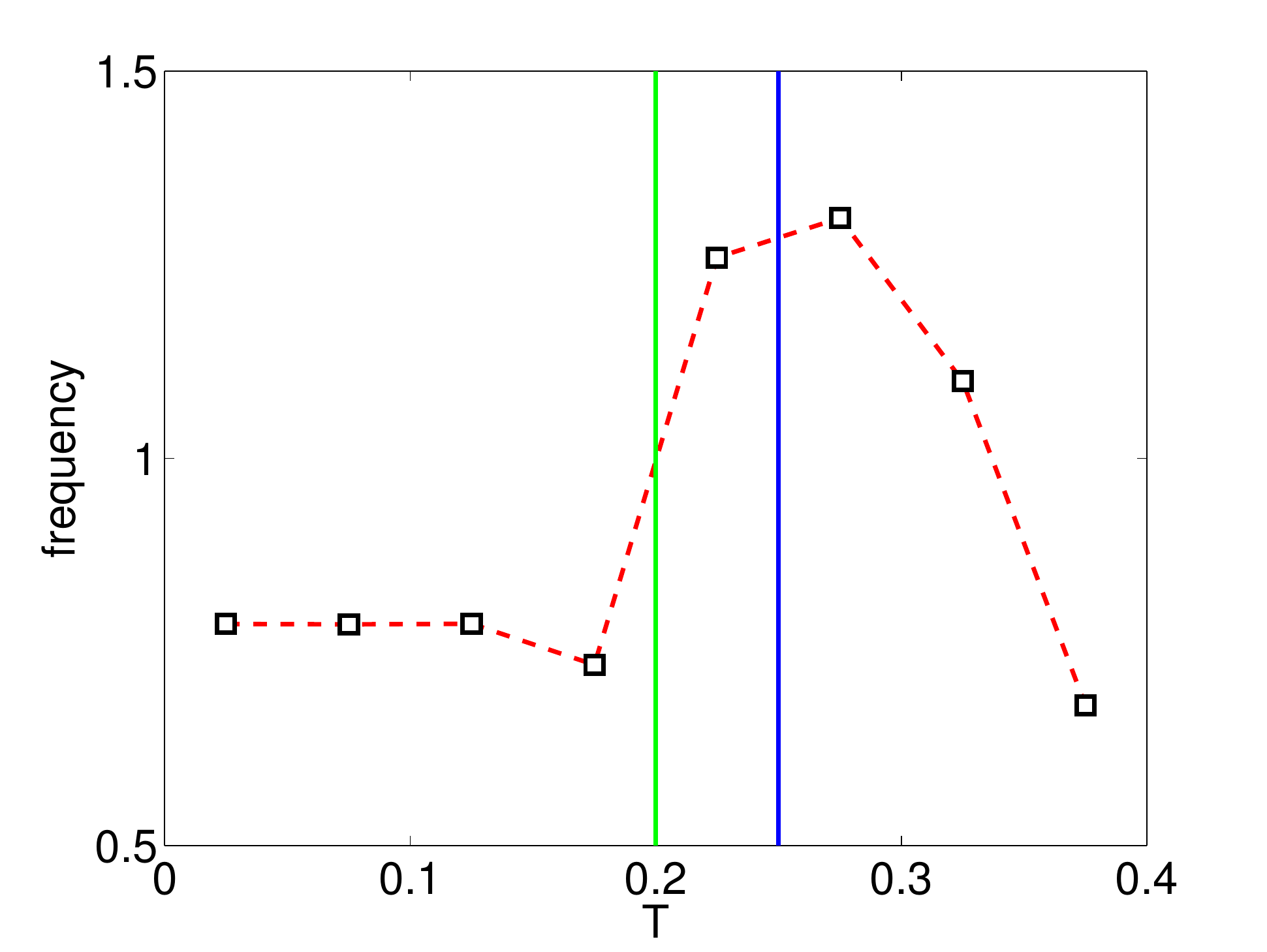}
\caption{Left : Fitting Fig.~\ref{pv01fre4} using spectral analysis. The inserted figure is the power spectrum at $T=0.225$ near $T_c$. Right: Peak frequency as a function of temperature. }
\label{pv01fre4fitspec}
\end{center}
\end{figure}

In Fig~.\ref{pv01fre4} -~\ref{pv01fre4fitspec}, we made the pulse sharper with $\omega=4$. If we put energy scale $T_c\approx 75K$ in the simulation, this gives the pulse duration is about $25$ fs. We also found similar amplitude enhancement and phase shift crossing $T_c$, except that here the overall oscillation amplitude gets smaller because of the shorter disturbance. With more simulations, we found that as long as the pulse duration does not exceed the oscillation period, these effects are quite robust.

\begin{figure}[H]
\begin{center}
\includegraphics[width=3in]{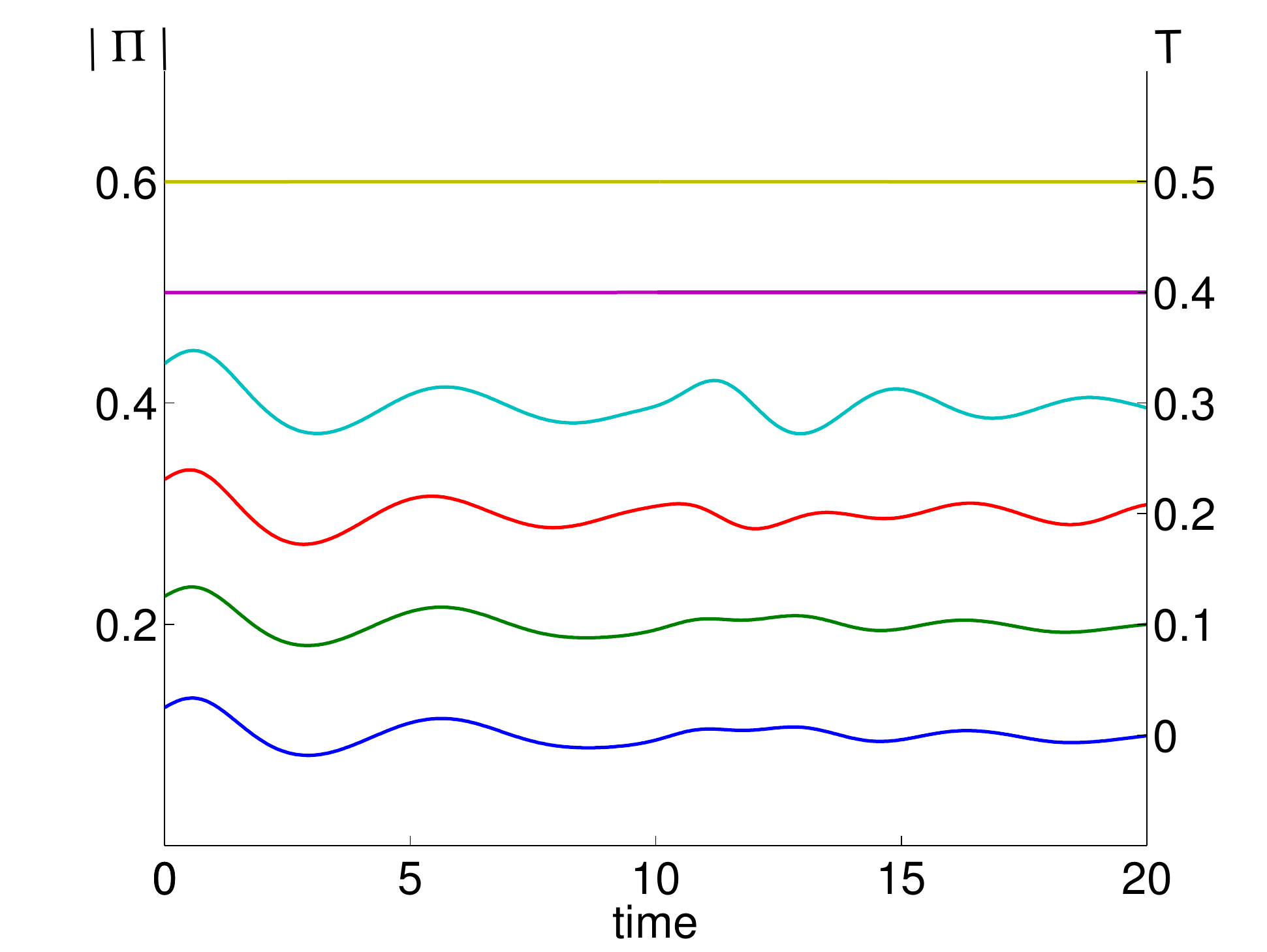}
\includegraphics[width=3in]{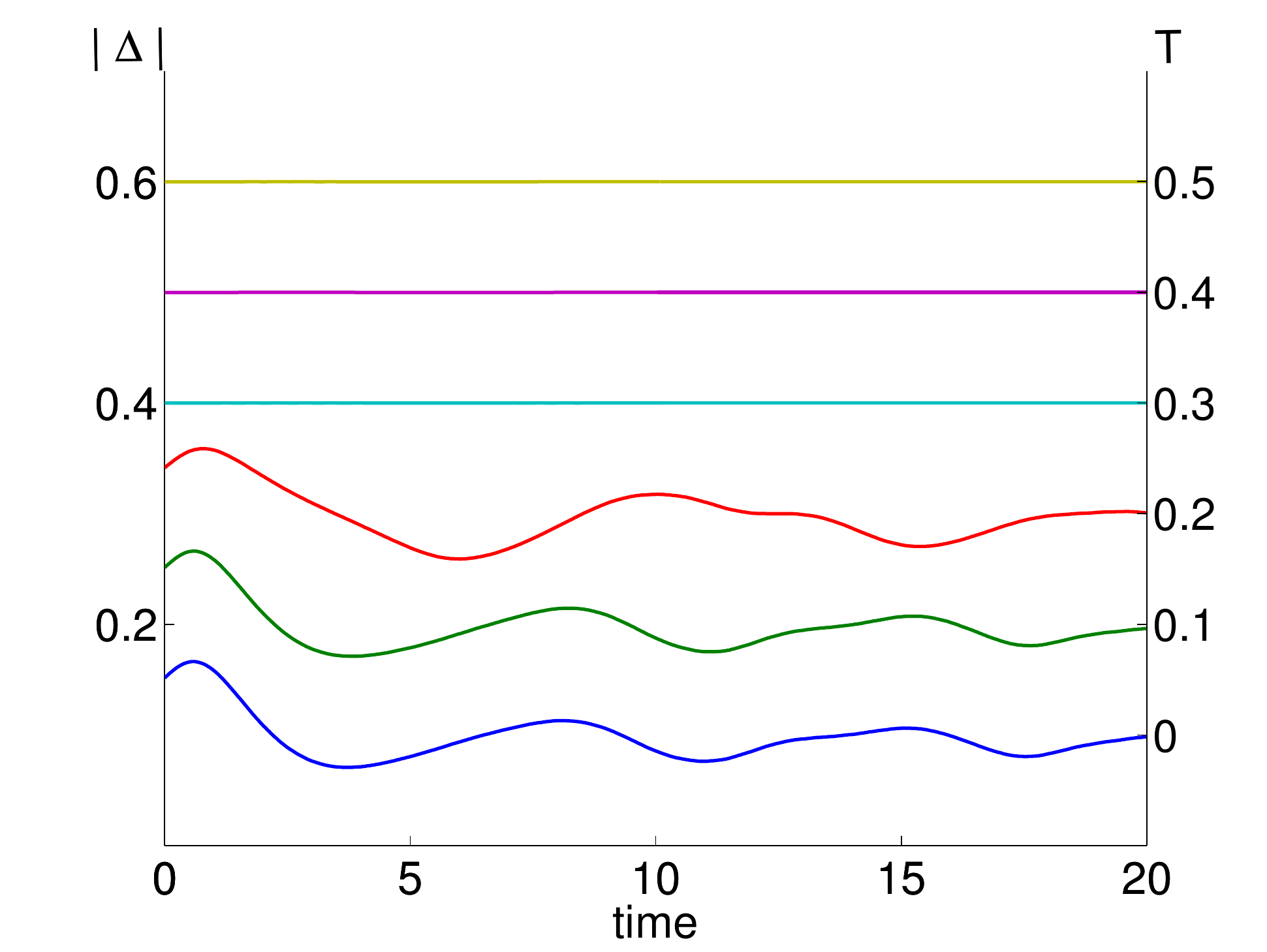}
\caption{Oscillation of CDW order parameter $\Pi$(left) and SC order parameter $\Delta$(right) as a function of time in the quench case, from low temperature at the bottom to high temperature at the top, temperatures are taken from 0 to 0.5 with 0.1 step. The initial value  $J_0=1.2$, $V_0=0.9$, the quench is taken as $\Delta J=0.1, \Delta V=0$. }
\label{pj01}
\end{center}
\end{figure}

Finally, we considered quench $J$ case in Fig.~\ref{pj01}. However, now the behavior is quite different, as we discussed in the main text; $V$ rather than $J$ affects SC and CDW differently, which mimics the effect of the pump in the experiment.

\newpage

\section{A different corner in parameter space for the O(6) model}
\begin{figure}[ch!]
\includegraphics[width=0.7\textwidth]{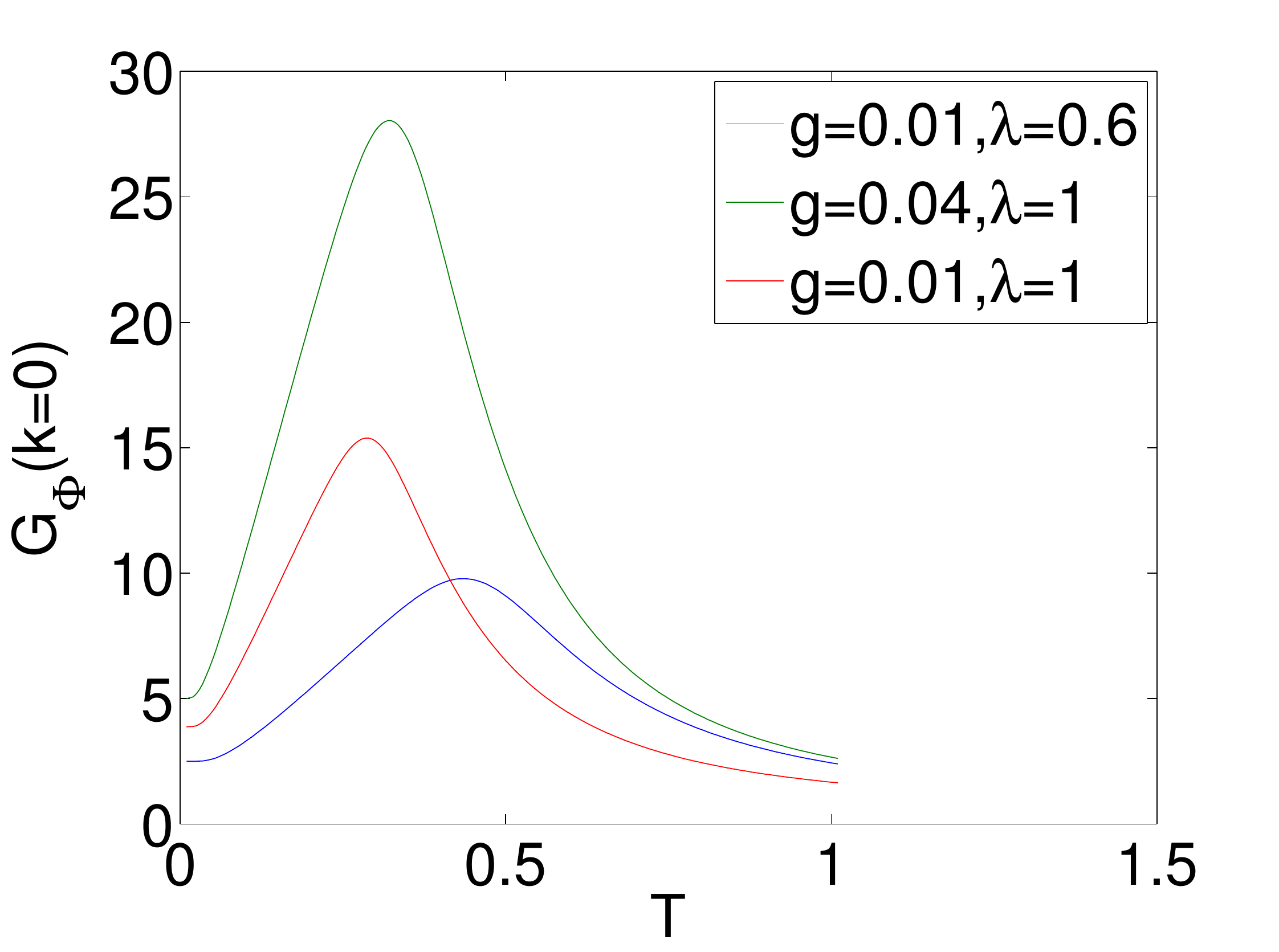} 
\caption{Equilibrium equal time correlation functions of $\Phi$ at zero momentum vs temperatures at three
different values of $\lambda$ and $g$ and $m_\Psi = 1/500$ at $T=1/100$. }\label{3equiplots}
\end{figure}

In the main text, we have been inspecting a region of parameter space where $g$ is of the order of $1/10$
of the lattice cutoff $a$. 
We present here results from a vastly different parameter space, in which
we reduce $g$ by a factor of 10.  At equilibrium, the correlation functions are plotted in figure \ref{3equiplots},
which has a much stronger resemblance to the classical results in \cite{o6}. 

\begin{figure}[h!]
\begin{tabular}{cc}
\includegraphics[width=0.50\textwidth]{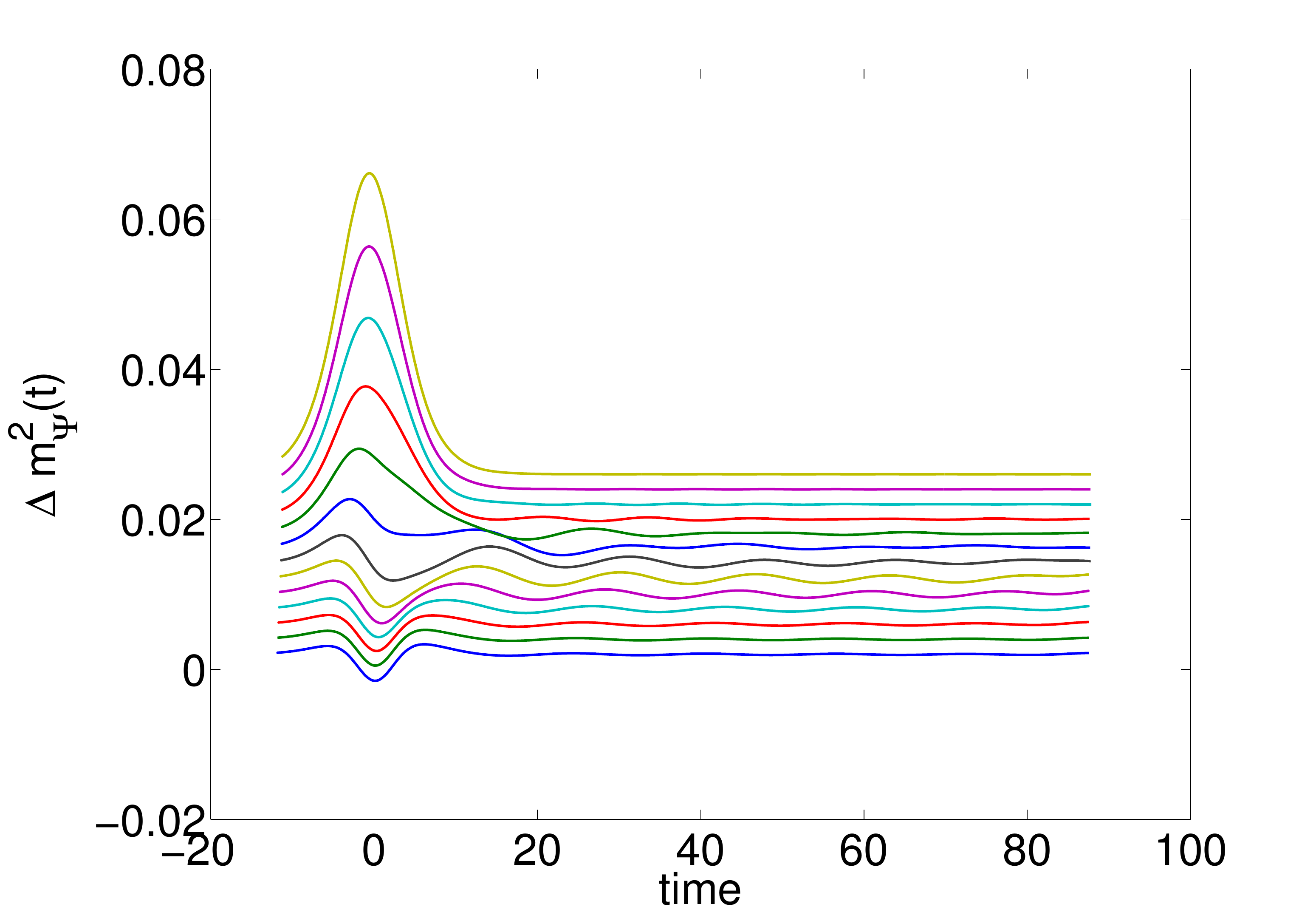} & \includegraphics[width=0.50\textwidth]{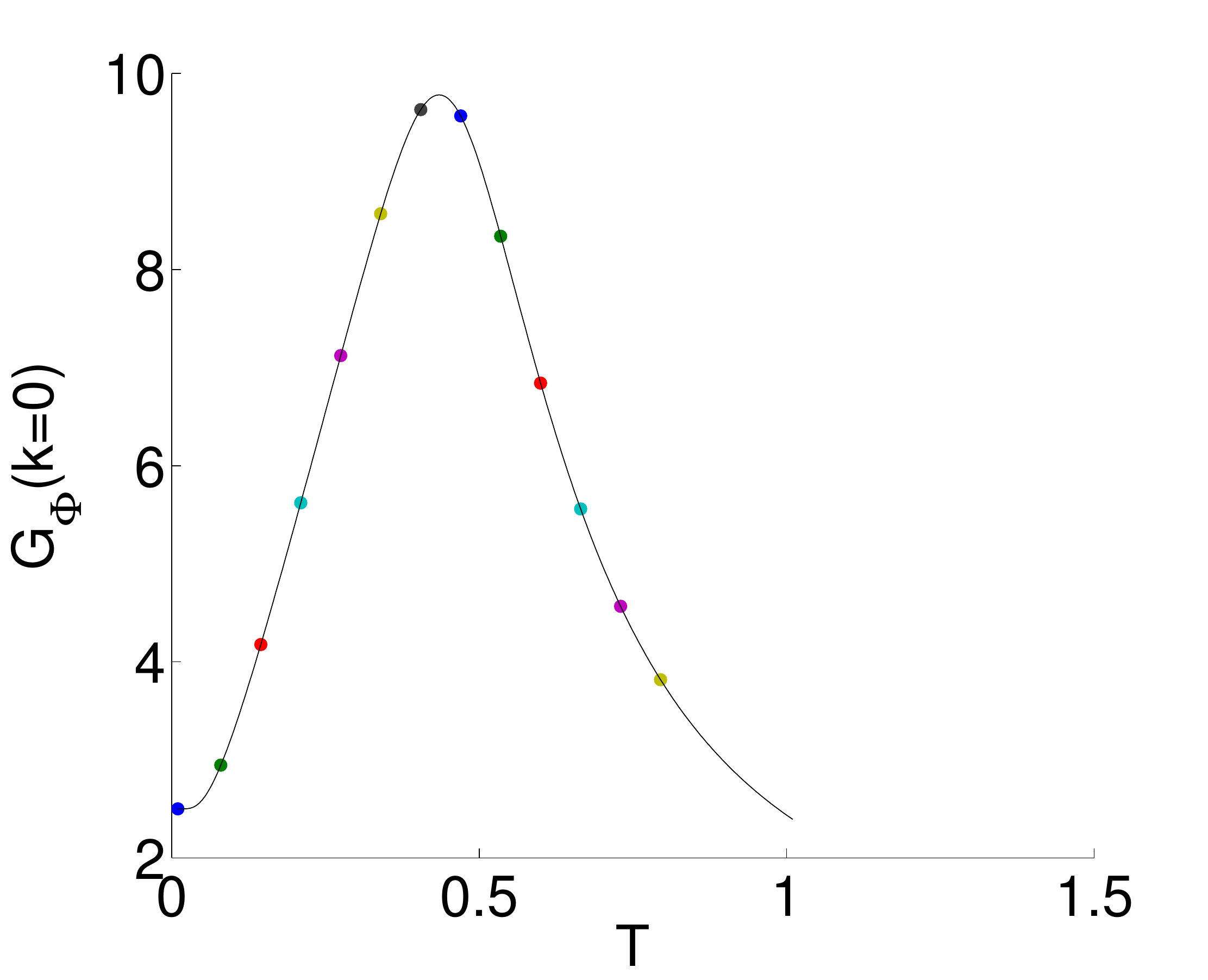} 
\end{tabular}
\caption{Left Panel: Oscillations of $m^2_{\Psi}$ as a function of time at 13 different initial temperatures, from
low temperatures at the bottom of the picture to high temeperatures  at the top, at constant
$\rho_0 =   0.1058$ (corresponding to $m_{\Psi} = 0.002$ at $T=1/100$, $g=0.04, 
\lambda = 1$, and $a=1$). Integral along $k_x$ and $k_y$ is each
divided into 90 steps. The pulse parameter is taken as $v=1/5,\delta\rho= 1/500$.
These 13 initial temperatures correspond to 13 points on the equilibrium plot of $G_{\Phi}(k=0)$ against $T$, 
as shown on the right panel. The color of the markers match the color of the curves on the left.  
Note that in this picture we display the entire oscillations including the large peak. }
\label{10plotsnewg4}
\end{figure}

\begin{figure}[h!]
\begin{tabular}{cc}
\includegraphics[width=0.80\textwidth]{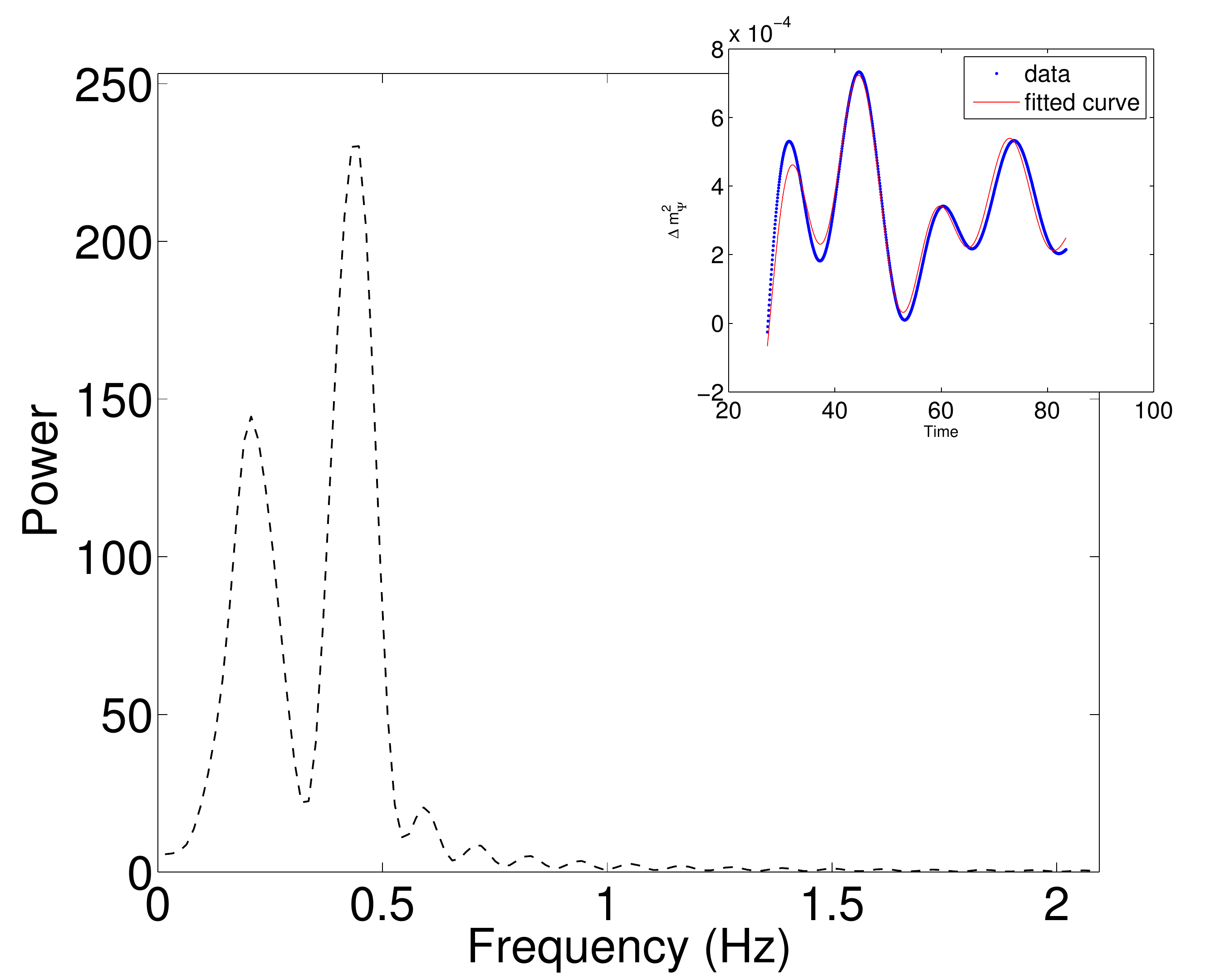} 
\end{tabular}
\caption{Power spectrum at $T=0.47$.}
\label{FT2}
\end{figure}

\begin{figure}[h]

\includegraphics[width=0.8\textwidth]{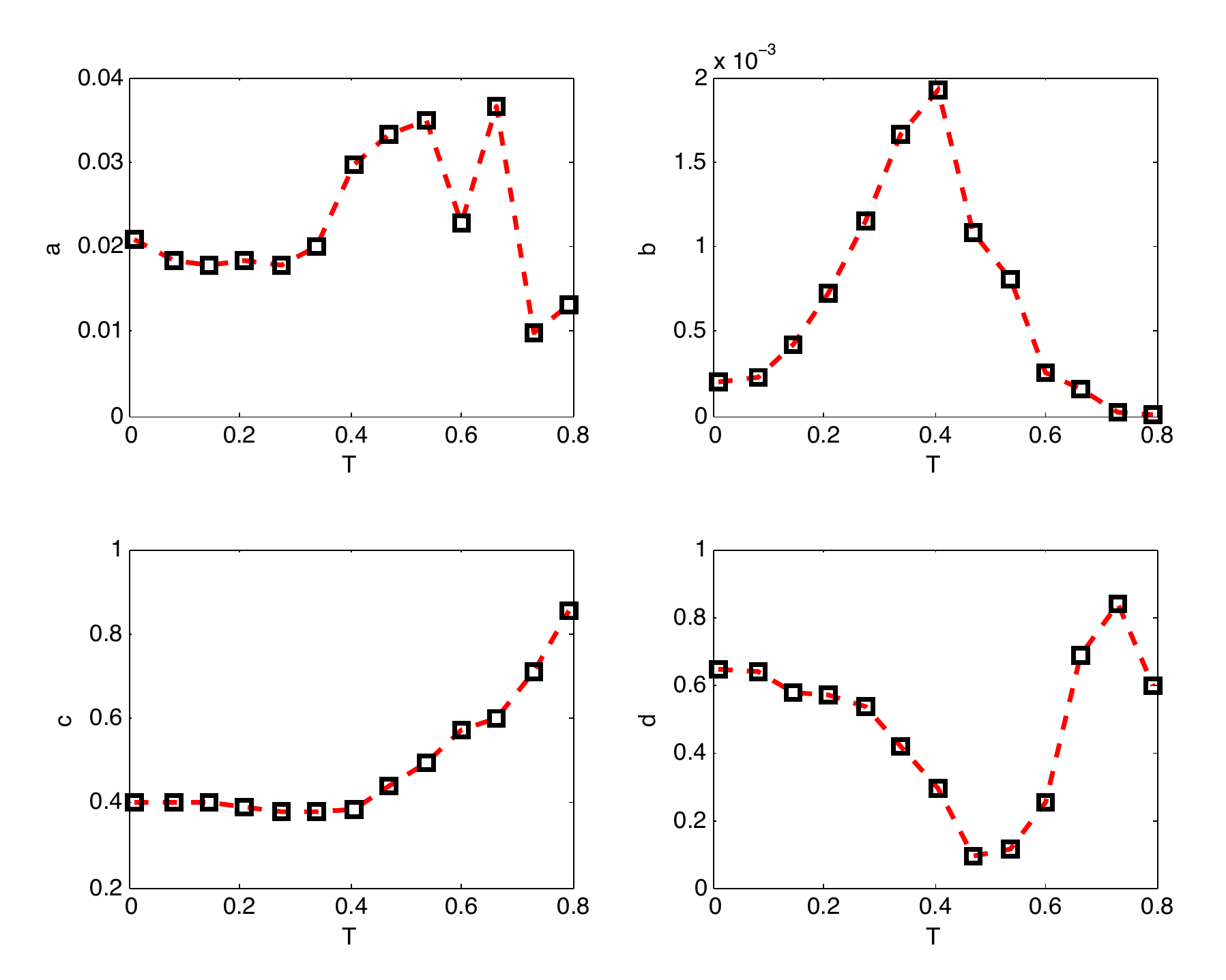}
\caption{Coefficients $a,|b|,c,d$ of the fit function $f(t) = \exp(-a t) b \sin(c t + 2\pi d) + e$ fitting the data presented in figure \ref{10plotsnewg4} 
are plotted 
against temperature.}.\label{fitdatanew}
\end{figure}

In figure \ref{10plotsnewg4} below we present the oscillations at $\lambda=1, g=0.04$.  We note that comparing
with figure \ref{10plotslamg4}, we find that this feature where the initial large response to the disturbance
changes sign close to the peak temperature $T_p$ becomes a completely generic feature in this regime
of parameter space. Note that right across the peak temperature (see blue and green curves in the middle of
the right panel) the initial trough disappears, and this skipped trough leads to a more convincing $\pi$ phase
shift actually happening independently how the data is fitted. We note that in all our data sets collected this always happens close to $T_p$, in this case and also $g=0.01$, just above $T_p$,
and whereas in the case in figure \ref{10plotslamg4}, just below. Another feature is that ultimately the frequency still
stays at roughly the scale set by $T_p$ despite dramatically reducing $g$ while keeping $T_p$ roughly the same.
The oscillation amplitude is highest close to $T_p$ but decreases again as temperature is further lowered,
although not as steeply as in the high temperature regime. The fit of the exponential decay
at very high temperature is a gross underestimate, because the oscillation amplitude is extremely small and that
it is strongly affected by the starting point of the fit. 
We also present in figure \ref{fitdatanew} results of numerical fits of the oscillations by the fit function in equation
\ref{fitfunc}. To improve the results of the fits, we note that close to the peak temperature, there are in fact
two major Fourier components with similar amplitudes (see figure \ref{FT2}). Therefore, between $T=0.5-0.6$ we modify the fit function to 
\be
f_{\textrm{modified}}(t)= e^{-a t} (b \sin(\omega_1 t + d_1) + b_2 \sin(\omega_2 t + d_2)) 
\ee 
We take the higher frequency peak and its amplitude in the plot of frequencies and amplitudes verses temperature in figure \ref{fitdatanew} .

\newpage




\begin{thebibliography}{}

\bibitem{Orenstein} J.~P.~Hinton, J.~D.~Koralek, Y.~M.~Lu, A.~Vishwanath, J.~Orenstein, D.~A.~Bonn, W.~N.~Hardy, and Ruixing Liang, Phys. Rev. B {\bf 88}, 060508 (2013).

\bibitem{Gedik} D.~H.~Torchinsky, F.~Mahmood, A.~T.~Bollinger, Ivan Bo\v{z}ovi\'c, and N.~Gedik, Nature Materials {\bf 12}, 
387 (2013).

\bibitem{andrea1} D. Fausti, R. Tobey, N. Dean, S. Kaiser, A. Dienst, M. Hoffmann, S. Pyon, T. Takayama, H. Takagi, and A. Cavalleri, Science {\bf 331}, 189 (2011).

\bibitem{andrea2} S. Kaiser, D. Nicoletti, C. R. Hunt, W. Hu, I. Gierz, H. Y. Liu, M. Le Tacon, T. Loew, D. Haug, B.~Keimer, 
and A. Cavalleri, arXiv:1205.4661.

\bibitem{keimer} G.~Ghiringhelli,  M. Le Tacon, M. Minola, S. Blanco-Canosa, C. Mazzoli, N. B. Brookes, 
G.~M.~De Luca, A.~Frano, D. G. Hawthorn, F. He, T. Loew, M. Moretti Sala, D. C. Peets, M.~Salluzzo, 
E. Schierle, R. Sutarto, G.~A.~Sawatzky, E. Weschke, B. Keimer, and L. Braicovich,
Science {\bf 337}, 821 (2012).

\bibitem{chang} J.~Chang, E. Blackburn, A. T. Holmes, 
N. B. Christensen, J. Larsen, J. Mesot, Ruixing Liang, D. A. Bonn, W.~N.~Hardy,
A. Watenphul, M. v. Zimmermann, E.~M.~Forgan, and S. M. Hayden,
Nature Phys. {\bf 8}, 871 (2012).

\bibitem{hawthorn} A.~J.~Achkar, R. Sutarto, X. Mao, F. He, A. Frano, S. Blanco-Canosa, M. Le Tacon, 
G.~Ghiringhelli, L.~Braicovich, M. Minola, M. Moretti Sala, 
C. Mazzoli, Ruixing Liang, D. A. Bonn, W. N. Hardy, B. Keimer, G.~A.~Sawatzky, 
and D. G. Hawthorn
Phys. Rev. Lett. {\bf 109}, 167001 (2012).


\bibitem{jay}  J. D. Sau and S. Sachdev, Phys. Rev. B {\bf 89}, 075129 (2014).

\bibitem{levitov} R.~A.~Barankov and L.~S.~Levitov, Phys. Rev. Lett. {\bf 96}, 230403 (2006).

\bibitem{o6} L.~E.~Hayward, D.~G.~Hawthorn, R.~G.~Melko, and S.~Sachdev, Science {\bf 343}, 1336 (2014).

\bibitem{metlitski10-2} M. A. Metlitski and S. Sachdev,
Phys. Rev. B {\bf 82}, 075128 (2010).

\bibitem{rolando} S.~Sachdev and R.~La Placa, Phys. Rev. Lett. {\bf 111}, 027202 (2013).




\bibitem{cardy} 
  S.~Sotiriadis and J.~Cardy,
  Phys.\ Rev.\ B {\bf 81}, 134305 (2010).

\bibitem{gubser}
A. ~Chandran, A.~Nanduri, S.~S.~ Gubser and S.~L.~Sondhi,
Phys Rev B, {\bf 88},  024306 (2013).

\bibitem{hung} 
  L.~-Y.~Hung, M.~Smolkin and E.~Sorkin,
  Phys.\ Rev.\ Lett.\  {\bf 109}, 155702 (2012).
 
  

\bibitem{das} 
  S.~R.~Das and K.~Sengupta,
  JHEP {\bf 1209}, 072 (2012).
 








\end{thebibliography}
\end{document}